   \documentclass[final,1p,times]{elsarticle}

\usepackage{graphicx}
\usepackage{epsfig}

\usepackage{epstopdf}
\usepackage{soul}
\usepackage{cancel}

\usepackage{amssymb}
\usepackage{amsmath}
\usepackage{lineno}

\biboptions{sort&compress}

\usepackage{orcidlink}

\newcounter{bla}

\usepackage{mathrsfs}
\usepackage{physics}
\usepackage{tikz-feynman}

\journal{Computer Physics Communications}

\usepackage{tikz}
\tikzfeynmanset{compat=1.0.0}
\usetikzlibrary{shapes.geometric, arrows}
\tikzstyle{startstop} = [rectangle, rounded corners, 
minimum width=2.5cm, 
minimum height=1cm,
text centered, 
draw=black, 
fill=red!30]

\tikzstyle{io} = [trapezium, 
trapezium stretches=true, 
trapezium left angle=70, 
trapezium right angle=110, 
minimum width=2.5cm, 
minimum height=1cm, text centered, 
draw=black, fill=blue!30]

\tikzstyle{process} = [rectangle, 
minimum width=2.5cm, 
minimum height=1cm, 
text centered, 
text width=3cm, 
draw=black, 
fill=orange!30]

\tikzstyle{generate} = [rectangle, 
minimum width=2.5cm, 
minimum height=1cm, 
text centered, 
text width=3cm, 
draw=black, 
fill=yellow!30]

\tikzstyle{decision} = [diamond, 
minimum width=2.5cm, 
minimum height=1cm, 
text centered, 
draw=black, 
aspect=1.8,
fill=green!30]
\tikzstyle{arrow} = [ultra thick,->,>=stealth]

\newcommand{\diff}{\mathop{}\!d}

\newcommand{\HTP}{\textsc{HadroTOPS}~} 

\allowdisplaybreaks

\begin{document}

\begin{frontmatter}



\title{\textsc{HadroTOPS}: A Monte Carlo Event Generator For Hadron Production In Two-Photon Scattering In Electron Positron Collisions}


\author[a,b]{Max Lellmann\corref{lellmann}\,\orcidlink{0000-0002-2154-9292}}
\author[a,b]{Igor Danilkin\,\orcidlink{0000-0001-8950-0770}} 
\author[a,b,c]{Achim Denig\,\orcidlink{0000-0001-7974-5854}}
\author[a,b]{Jan Muskalla\,\orcidlink{0009-0001-5006-370X}}
\author[a,b]{Christoph F. Redmer\,\orcidlink{0000-0002-0845-1290}}
\author[b,c,d]{Xiu-Lei Ren\,\orcidlink{0000-0002-5138-7415}}
\author[a,b,c]{Marc Vanderhaeghen\,\orcidlink{0000-0003-2363-5124}} 

\cortext[lellmann] {\textit{E-mail address:} lellmann@uni-mainz.de}

\address[a]{Institute for Nuclear Physics, Johannes Gutenberg University, D-55099 Mainz}
\address[b]{PRISMA$^\mathrm{+}$ Cluster of Excellence, Johannes Gutenberg University, D-55099 Mainz}
\address[c]{Helmholtz Institute Mainz, Johannes Gutenberg University, D-55099 Mainz}
\address[d]{Shandong Provincial Key Laboratory of Nuclear Science, Nuclear Energy Technology and Comprehensive Utilization \& School of Nuclear Science, Energy and Power Engineering, Shandong University, Jinan 250061, China}

\begin{abstract}
We present a Monte Carlo event generator specifically developed for the study of hadronic two-photon scattering events in two-photon scattering at electron-positron colliders. The code enables the generation of events with exact leading-order QED coupling and a flat phase space decay of the hadronic state into an arbitrary number of final state particles as selected by the user. Thus, this generator is well-suited for the use of partial wave analyses tools to study the two-photon production of higher-multiplicity final states across a wide range of energies and photon virtualities. Furthermore, the code integrates both experimental and theoretical inputs on the two-photon couplings of hadrons to simulate two-photon production processes. Motivated by the investigations of the BESIII collaboration, the final states $\pi^+\pi^-$, $\pi^0\pi^0$, $\pi^0\eta$, $K^+K^-$, $K^0_SK^0_S$, $\eta\eta$, and $f_1(1285)\to \eta\pi^+\pi^-$ via $a_0^\pm(980)\pi^\mp$ and $f_0(500)\eta$ are currently included. The code is sufficiently flexible to easily add additional final states as well as quickly change the already included channels.





\end{abstract}


\end{frontmatter}



{\bf PROGRAM SUMMARY}

\begin{small}
\noindent
{\em Program Title: } HadroTOPS          \\
{\em Developer's repository link:} { \url{http://gitlab.rlp.net/malellma/HadroTOPS}} \\
{\em Licensing provisions:} GPLv3 \\
{\em Programming language:} C++                                   \\
{\em Nature of problem:}\\
The analysis of hadronic two-photon interactions at electron-positron colliders necessitates the use of a Monte Carlo generator capable of modeling the complex hadronic processes involved in the fusion mechanism across a broad kinematic spectrum. In particular, the study of multi-hadron production is hindered by the lack of suitable simulation tools. Partial wave analyses of various hadronic final states require a Monte Carlo simulation that precisely predicts the quantum electrodynamics (QED) components of the process, while simultaneously generating a uniform phase space distribution for two-photon to multi-hadron transitions. This allows for a correct description of the {leading-order QED part,} while providing an easy decay model, which can be utilized in partial wave analyses and allows for the two-photon cross section to be replaced by the relevant observables.

{\em Solution method:}\\
{In this code, the leading-order QED calculation for the inclusive and single-meson production process $e^+e^- \to e^+e^- X$ from Refs.~\cite{Bonneau:1973kg, Budnev:1975poe, Pascalutsa:2012pr} is extended to the fully exclusive two-meson ($M_1 M_2$) production process $e^+e^- \to e^+e^- M_1 M_2$. This is combined with the phase-space generation algorithm outlined in Ref.~\cite{Schuler:1997ex} which, via a specialized mapping detailed below, eliminates the cross section's dependence on the momentum transfers, $(t_1 t_2)^{-1}$. This approach significantly enhances numerical stability, particularly in the critical region of small photon virtualities.}
This combination enables an accurate leading QED description of two-photon production for any hadronic final state, with a uniform decay distribution, making it a suitable tool for partial wave analyses. Being able to incorporate theoretical calculations and experimental results of the cross section of two-photon processes, the generator can simulate the two-photon production of any hadronic final states for which the relevant information is provided. 
For the two-meson production channels, explicit estimates are given based on a dispersive formalism to incorporate the (virtual) two-photon fusion process $\gamma^{(\ast)} \gamma^{(\ast)} \to M_1 M_2$.   
The program is designed to be easily extendable, allowing for the inclusion of additional hadronic final states or more refined hadronic models in future developments.
\end{small}

\section{Physics Case and Theoretical Foundation}
\label{sec:intro}

Two-photon processes in $e^+e^-$ collisions of the form $e^+e^- \to e^+e^- X$ are a well established tool to study hadronic structure in photon-photon interactions. In such processes, each $e^\pm$ emit a quasi-real or virtual photon, and the photon-photon fusion produces a hadronic final state $X$. In this work, we focus on the exclusive production of pion pairs ($X=\pi^+\pi^-,\,\pi^0\pi^0$), $X=\pi^0\eta$, and the production of the axial state $f_1(1285)$ ($f_1(1285)\to\eta\pi^+\pi^-$). Other final states ($K\bar{K}$, $\eta\eta$) are purely described by available experimental measurements. All these reactions proceed via $\gamma^{(\ast)}\gamma^{(\ast)}$ fusion, where $\gamma^*$ denotes an off-shell photon. At $e^+e^-$ colliders, two-photon production of hadrons has been studied in detail to extract 
two-photon couplings and properties of the $f_0(500)$, $f_0(980)$, $a_0(980)$, $f_2(1270)$, $a_2(1320)$ and other resonances in the hadronic system \cite{Belle:2003xlt,Belle:2004bpk,Belle:2008bmg,BaBar:2009rrj,Belle:2010ckn,Kuessner2022,Kussner:2024ryb}. There is renewed interest in accurate theoretical descriptions of $\gamma^*\gamma^* \to X$ type hadronic reactions, driven by the role of the processes in hadronic light-by-light (HLbL) contribution \cite{Colangelo:2015ama,Masjuan:2017tvw,Colangelo:2017fiz,Hoferichter:2018kwz,Eichmann:2019tjk,Bijnens:2019ghy,Leutgeb:2019gbz,Cappiello:2019hwh,Masjuan:2020jsf,Bijnens:2020xnl,Bijnens:2021jqo,Danilkin:2021icn,Stamen:2022uqh,Leutgeb:2022lqw,Hoferichter:2023tgp,Hoferichter:2024fsj,Estrada:2024cfy,Ludtke:2024ase,Deineka:2024mzt,Eichmann:2024glq,Bijnens:2024jgh,Hoferichter:2024bae,Holz:2024diw,Cappiello:2025fyf,Colangelo:2014qya,Blum:2019ugy,Chao:2021tvp,Chao:2022xzg,Blum:2023vlm,Fodor:2024jyn} to the muon anomalous magnetic moment $(g-2)_\mu$ \cite{Aliberti:2025beg}. In particular, an accurate modeling of the $\gamma^*\gamma^* \to \pi\pi$ amplitudes, including the effects of strong final-state interactions (FSI) between pions, is essential for precision HLbL computations \cite{Colangelo:2017qdm, Colangelo:2017fiz, Danilkin:2021icn}. Dispersive approaches have been developed to address this need, providing partial wave (p.w.) amplitudes that incorporate $\pi\pi$ rescattering unitarily and tie into experimental $\pi\pi$ scattering data \cite{Garcia-Martin:2010kyn,Hoferichter:2011wk,Moussallam:2013una,Danilkin:2018qfn,Hoferichter:2019nlq,Danilkin:2019opj}.
{
Besides its relevance for HLbL studies, a Monte Carlo generator for $e^+e^- \to e^+e^- hadrons$ may also be useful experimentally for modeling the radiative corrections due to two-photon box diagrams in the annihilation channel, thereby helping to obtain cleaner $e^+e^- \to hadrons$ measurements relevant to HVP determinations.
}

On the experimental side, current and future $e^+e^-$ facilities, such as BESIII and Belle~II, can perform detailed measurements of two-photon processes, including kinematic configurations where one or both outgoing leptons are detected (single-tag or double-tag modes). To fully leverage these measurements, a Monte Carlo (MC) event generator that includes the state-of-the-art two-photon amplitudes is required. The existing generator \textsc{Ekhara3.0} \cite{Czyz:2010sp,Czyz:2012nq,Czyz:2017veo,Czyz:2018jpp} has been used for simulating two-photon production of charged pion pairs. However, \textsc{Ekhara3.0} includes only the lowest-order “Born” contribution for $\gamma^*\gamma^* \to \pi^+\pi^-$ and neglects the strong-interaction dynamics that generate resonances in the $\pi^+\pi^-$ invariant mass spectrum. The latest version \textsc{Ekhara3.2} has updated its $\gamma^*\gamma^* \to \pi\pi$ input to include the full dispersive amplitudes for $\pi\pi$ final states, thereby greatly improving the physical accuracy of the simulation. The further development of the \textsc{Ekhara} event generator beyond the version 3.2 will likely not continue \cite{CzyzPrivate}.

Ongoing or foreseen studies of multi-hadron final states other than pion pairs require equivalent activities. Therefore, the generator presented in this work is configured to simulate the production of $\pi^0\eta$, $K\bar{K}$, $\eta\eta$, and $f_1(1285) \to \eta\pi^+\pi^-$. To achieve this, various experimental measurements and theoretical predictions are incorporated.

As mentioned above, beyond their relevance for precision tests, two-photon production processes are of particular importance for hadron spectroscopy. They provide a clean environment for the formation of non-vector states in electron-positron collisions. In the light-meson sector, the density of resonances demands partial-wave analyses (PWAs) to extract the relevant resonance contributions.  Such analyses rely on MC simulations which evenly cover all dimensions of the phase space for fitting and normalization \cite{Shepherd2023-pz, Fritsch2022-ii}. When using a fully phase-space-distributed MC sample to analyze a two-photon process, one must incorporate the QED couplings into the amplitudes. This introduces an additional complication, which can be avoided by using a MC generator that provides a phase-space-distributed sample while already taking into account the QED couplings. However, many available generators capable of producing such MC samples have different kinematic limitations and therefore cannot be used in all cases (see, for example, Refs.~\cite{SCHULER1998279, BaBar:2010jfn, Mustafa2019}).

\subsection{Theoretical Formulation}
To provide such a tool, we begin by formulating the general process
\begin{equation}
e^+ (p_1) + e^- (p_2) \to e^+ (p^\prime_1) + e^- (p^\prime_2) + X (p_x)\,,
\end{equation}
where the quantities in parentheses denote four-momenta. In the overall center-of-mass (c.m.) frame of the initial $e^+e^-$, the initial four-momenta are $p_1=(E, \mathbf{p})$ and $p_2=(E,-\mathbf{p})$ with beam energy $E=\sqrt{s}/2$ and $s=(p_1+p_2)^2$. The outgoing leptons carry momenta $p'_1$ and $p'_2$, and we define the momentum transfers as $q_1=p_1-p'_1$ and $q_2=p_2-p'_2$. Consequently, $q_1$ and $q_2$ are the four-momenta of the photons emitted by the $e^+$ and $e^-$, respectively, that fuse into the $X$ system. The effective squared $\gamma^*\gamma^*$ center-of-mass energy is $W^2=(q_1+q_2)^2=p_X^2$. The photon virtualities are defined as $Q_1^2 \equiv -q_1^2$ and $Q_2^2 \equiv -q_2^2$, with $Q_{1,2}^2 \ge 0$. If $Q^2$ is near zero, the photon is considered as \emph{quasi-real} (such photons are emitted at very small angles and carry very low momentum transfer in the $e^+e^-$ c.m. frame). When $Q^2$ is substantial, the photon is virtual, as is the case when an outgoing lepton is detected at large scattering angle at experiments like BESIII ($\theta\gtrsim\, 20^\circ$, $Q^2\gtrsim 0.1\,$GeV$^2$) or Belle~II ( $\theta\gtrsim\, 17^\circ$, $Q^2\gtrsim3\,$GeV$^2$).

The polarized cross section for the process $e^+ e^- \to e^+ e^- X$, which explicitly depends on the helicities $h_1, h_2$ of the incoming leptons, can be written as
\begin{equation}
d\sigma_{h_1 h_2} = \frac{1}{F} \, d\mathrm{Lips} \sum_{h_1', h_2'} |\mathcal{M}|^2\,,
\end{equation}
where \( \mathcal{M} \) denotes the full scattering amplitude, \( F \) is the flux factor, and \( d\mathrm{Lips} \) is the Lorentz-invariant phase space measure for the final state
\begin{equation}
d\mathrm{Lips} = 
\frac{d^3 \vec p_1^{\, \prime}}{(2\pi)^3 2 E_1'} 
\frac{d^3 \vec p_2^{\, \prime}}{(2\pi)^3 2 E_2'} \, d\Gamma_X\,(2\pi)^4 \delta(p_1 + p_2 - p_1' - p_2' - p_X)\,.\label{eq:full_phase_space}
\end{equation}
{The unpolarized cross section is obtained by averaging over the initial helicities $h_1, h_2$, i.e. $1/4\sum_{h_1,h_2} d\sigma_{h_1 h_2}$.}
The factor $d\Gamma_X$ denotes the phase space element for the intermediate hadronic state.
At tree level, the amplitude takes the form
\begin{equation}
\mathcal{M} = \frac{e^2}{Q_1^2 Q_2^2} 
\left[ \bar{\rm v} (p_1, h_1) \gamma^\mu {\rm v}(p_1', h_1') \right]
\left[ \bar{u}(p_2', h_2') \gamma^\nu u(p_2, h_2) \right] H_{\mu\nu}\,,
\end{equation}
where \( H_{\mu\nu} \) is the hadronic tensor associated with the subprocess \( \gamma^* \gamma^* \to X \). The squared amplitude, summed over the final-state lepton helicities, becomes
\begin{equation}
    \sum_{h_1', h_2'} |\mathcal{M}|^2=\frac{e^4}{Q_1^4 Q_2^4}L_{h_1,\mu\mu'}\,L_{h_2,\nu\nu'}   H^{\mu\nu}\,(H^{\mu'\nu'})^* ,
\label{eq:sqrsum}
\end{equation}
where $L_{h_1,\mu\mu'}$ is the leptonic tensor associated with the positron currents, 
\begin{equation}
L_{h_1,\mu\mu'} = (p_1 + p_1')_\mu (p_1 + p_1')_{\mu'} + (q_1^2 g_{\mu \mu'} - q_{1 \mu} q_{1 \mu'}) + h_1 2 i \varepsilon_{\mu \mu' \kappa \lambda} \, q_1^\kappa \, p_1^\lambda\,, 
\end{equation}
with $\varepsilon_{0123} = +1$, and helicity $h_1=\pm 1$ (in units $\hbar/2$). 
There is an analogous definition for the leptonic tensor $L_{h_2,\nu\nu'}$ associated with the electron currents. 

There are two conceptually distinct ways to compute the $e^+ e^- \to e^+ e^- X$ cross section at this stage. The first proceeds covariantly, using the hadronic tensor $H^{\mu\nu}$ as shown above, and is incorporated in \textsc{Ekhara3.2} for $X=\pi\pi$.  The second approach, which we adopt here, involves decomposing the hadronic tensor in terms of helicity amplitudes for the fusion of two virtual photons into a hadronic system
\begin{equation}
H_{\lambda_1 \lambda_2} \equiv \epsilon^\mu(q_1, \lambda_1) \, \epsilon^\nu(q_2, \lambda_2)\, H_{\mu\nu}\, ,
\label{eq:pipi_matrix_element}
\end{equation}
with virtual photon helicities $\lambda_{1,2}=\pm1,0$. By inserting complete sets of photon polarization states into Eq.~(\ref{eq:sqrsum}) and employing electromagnetic gauge invariance, the squared amplitude can be rewritten as
\begin{equation}
 \sum_{h_1', h_2'} |\mathcal{M}|^2=\frac{e^4}{Q_1^2 Q_2^2} 
\sum_{\lambda_1, \lambda_2, \lambda_1', \lambda_2'} 
\rho_{h_1}^{\lambda_1 \lambda_1'} \, 
\rho_{h_2}^{\lambda_2 \lambda_2'} \, 
H_{\lambda_1 \lambda_2} \, 
H^*_{\lambda_1' \lambda_2'}\,,
\end{equation}
where \( \rho_{h_1}^{\lambda_1 \lambda_1'}  \) and \( \rho_{h_2}^{\lambda_2 \lambda_2'} \)  are the photon density matrices, which encode the polarization information of the virtual photons, defined as:
\begin{eqnarray}
\rho_{h_1}^{\lambda_1 \lambda_1'} \equiv \frac{(-1)^{\lambda_1 + \lambda_1'}}{Q_1^2} \epsilon^{\mu *}(q_1, \lambda_1) \epsilon^{\mu'}(q_1, \lambda_1')  L_{h_1,\mu\mu'}, 
\end{eqnarray}
and an analogous definition for $\rho_{h_2}^{\lambda_2 \lambda_2'} $.

The central object in this representation is the imaginary part of the forward light-by-light scattering amplitude 
$\gamma^\ast(q_1,\lambda_1) \gamma^\ast(q_2,\lambda_2) \to \gamma^\ast(q_1,\lambda'_1) \gamma^\ast(q_2,\lambda'_2) $
\begin{equation}
    \text{Im}\,M_{\lambda'_1 \lambda'_2, \lambda_1 \lambda_2} \equiv W_{\lambda'_1 \lambda'_2, \lambda_1 \lambda_2}\, ,
    \label{eq:forwardLbLampl}
\end{equation}
which defines the helicity-dependent response functions. 
According to unitarity, these response functions can be written as a sum over all possible intermediate hadronic states $X$, yielding
\begin{align}
W_{\lambda'_1 \lambda'_2, \lambda_1 \lambda_2}&= \frac{1}{2} \sum_X \int d \Gamma_X \,(2 \pi)^4\, \delta^4(q_1 + q_2 - p_X) \, H_{\lambda_1 \lambda_2}(q_1,q_2,p_X)\, H^\ast_{\lambda'_1 \lambda'_2}(q_1,q_2,p_X)\nonumber\\
&\equiv \sum_X W^X_{\lambda'_1 \lambda'_2, \lambda_1 \lambda_2}\,.
\label{eq:absW}
\end{align}
The response functions satisfy several symmetry relations due to discrete symmetries and angular momentum conservation. First, angular momentum conservation along the photon-photon axis implies that, in the forward limit, only transitions with equal total helicity \(\Lambda=\Lambda'\) contribute, where $\Lambda=\lambda_1-\lambda_2$  and $\Lambda'=\lambda'_1-\lambda'_2$. Parity invariance imposes the condition
\begin{align}
W_{-\lambda'_1\, -\lambda'_2,\, -\lambda_1\, -\lambda_2}\;=\; W_{\lambda'_1 \lambda'_2,\;\lambda_1 \lambda_2}\,,
\end{align}
and time-reversal symmetry leads to
\begin{align}
W_{\lambda'_1 \lambda'_2,\;\lambda_1 \lambda_2}\;=\; W_{\lambda_1 \lambda_2,\;\lambda'_1 \lambda'_2}\,.
\end{align}
As a result of these constraints, the number of independent response functions is reduced from 81 to 8. This reduced set corresponds to distinct cross sections and interference terms, forming the basis for expressing the full cross section in terms of measurable response functions.

\subsubsection{General cross section for the inclusive $e^+ e^- \to e^+ e^- X$ process}
\label{app:general_xsec}

We begin by considering the inclusive case, in which a sum is taken over all possible hadronic final states $X$ in Eq.~(\ref{eq:absW}). Compared to the exclusive 
$X=\pi\pi$ process, which will be presented in detail in the next subsection, the inclusive cross section can be written compactly as \cite{Budnev:1975poe,Bonneau:1973kg}
\begin{eqnarray}
d \sigma_{h_1,h_2} &=& \frac{\alpha^2}{8 \pi^4 \, Q_1^2 \, Q_2^2} \, \frac{\sqrt{X}}{s (1 - 4 m^2 / s)^{1/2}}  
 \frac{d^3 \vec p_1^{\, \prime}}{E_1^{\prime}} 
\frac{d^3 \vec p_2^{\, \prime}}{E_2^\prime}\left\{ 
4 \, \rho_1^{++} \, \rho_2^{++} \, \frac{1}{2} \left( \sigma_{0} + \sigma_2 \right) 
+  \rho_1^{00} \, \rho_2^{00} \, \sigma_{LL}  \right. 
\nonumber\\
&& \left.
+ 2 \, \rho_1^{++} \, \rho_2^{00} \, \sigma_{TL} 
+ 2 \, \rho_1^{00} \, \rho_2^{++} \, \sigma_{LT} + 2 \, \left( \rho_1^{++} - 1 \right) \, \left( \rho_2^{++} - 1 \right) \, \cos (2 \tilde \phi)  \, \tau_{TT}
\right. \nonumber \\
&& 
+ 8 \, \left[ \left( \rho_1^{00} + 1 \right) \, \left( \rho_2^{00} + 1 \right) \left( \rho_1^{++} - 1 \right) \, \left( \rho_2^{++} - 1 \right)\right]^{1/2} \, \cos \tilde \phi  \, \frac{1}{2} \left( \tau_0 + \tau_1 \right)
\nonumber \\
&&
+ h_1 h_2 \, 4  \left[ \left( \rho_1^{00} + 1 \right) \, \left( \rho_2^{00} + 1 \right) \right]^{1/2} \, \frac{1}{2} \left( \sigma_0 - \sigma_2 \right)\nonumber \\
&&\left.
+ h_1 h_2 \, 8  \left[ \left( \rho_1^{++} - 1 \right) \, \left( \rho_2^{++} - 1 \right) \right]^{1/2} \, \cos \tilde \phi  \, \frac{1}{2} \left( \tau_0 - \tau_1 \right)
\right\}\,,\label{eq:gagacross} 
\end{eqnarray}
where \(X \equiv (q_1 \cdot q_2)^2 - q_1^2 \, q_2^2\),   $\tilde\phi$ is the azimuthal angle  between the two leptonic planes in the \(\gamma^\ast \gamma^\ast\) center-of-mass frame (later called the $\gamma\gamma$ frame), and $m$ is the lepton mass. {This formulation follows a slightly different convention from Refs.~\cite{Budnev:1975poe}, avoiding the use of \(\tau_{TL}\) and \(\tau^{a}_{TL}\) in favor of the \(\tau_{0}\) and \(\tau_{1}\). Also, several typos\footnote{{Compared to Eq.~(A12) of Ref.~\cite{Pascalutsa:2012pr}, we correct three typos:
(i) $(1-4m^2/s) \to (1-4m^2/s)^{1/2}$,
(ii) in the unpolarized $\cos\tilde{\phi}$ term,
$$
\left[\frac{(\rho_1^{00}+1)(\rho_2^{00}+1)}
{(\rho_1^{++}-1)(\rho_2^{++}-1)}\right]^{1/2}\,\tau_{TL}
\to \left[(\rho_1^{00}+1)(\rho_2^{00}+1)(\rho_1^{++}-1)(\rho_2^{++}-1)\right]^{1/2}\,\tau_{TL}^a\,
$$
and (iii) in the $h_1\,h_2$-dependent $\cos\tilde{\phi}$ term,
$\tau_{TL}^a \to \tau_{TL}$.}} have appeared in subsequent works \cite{Pascalutsa:2012pr,Danilkin:2019mhd}, which are fixed here.}
The eight independent response functions that enter the cross section encapsulate the full structure of the $\gamma^*\gamma^* \to X$ sub-processes, and are defined as:
\begin{align}
&\sigma_0  = \frac{1}{2 \sqrt{X}}\, W_{+ +, + +}\,,\quad \sigma_2  = \frac{1}{2 \sqrt{X}}\, W_{+ -, + -}\,,\quad \sigma_{TT}\equiv \frac{1}{2}(\sigma_0+\sigma_2)\,, \nonumber\\
&\sigma_{TL}  = \frac{1}{2 \sqrt{X}}\, W_{+ 0, + 0}\,, \quad\sigma_{LT}  = \frac{1}{2 \sqrt{X}}\, W_{0 +, 0 +}\,,\quad
\sigma_{LL} = \frac{1}{2 \sqrt{X}}\, W_{0 0, 0 0}\,, 
\label{eq:inclresponses} \\
&\tau_{TT}  = \frac{1}{2 \sqrt{X}}\, W_{+ +, - -}\,,\quad \tau_0  = \frac{1}{2 \sqrt{X}}\, W_{+ +, 0 0}\,, \quad \tau_1  = - \frac{1}{2 \sqrt{X}}\, W_{0 +, - 0}\,. \nonumber
\end{align}
These response functions depend on three independent kinematic variables: the two-photon center-of-mass (c.m.) energy $W$ and the virtualities \(Q_1^2\) and \(Q_2^2\). 
{
Equivalently, the c.m. energy $W = \sqrt{s}$ can be traded for the crossing-symmetry variable 
\begin{equation}
\nu \equiv \frac{1}{4}(s - u) = q_1 \cdot q_2, 
\end{equation}
with Mandelstam variables $s = 2 \nu - Q_1^2 - Q_2^2$ and $u = - 2 \nu - Q_1^2 - Q_2^2$.

The above responses are the absorptive parts of the forward light-by-light amplitudes $M_{\lambda'_1 \lambda'_2, \lambda_1 \lambda_2}$ through Eq.~(\ref{eq:forwardLbLampl}). For the latter amplitudes it is useful, for the purpose of writing down forward dispersion relations~\cite{Pascalutsa:2012pr}, to consider the constraint imposed by crossing symmetry, which requires that the amplitudes for the process
\begin{equation}
\gamma^\ast(q_1,\lambda_1) \gamma^\ast(q_2,\lambda_2) \to \gamma^\ast(q_1,\lambda'_1) \gamma^\ast(q_2,\lambda'_2) 
\end{equation}
is equal to the amplitude for the process where the photons with e.g. label 2 are crossed:
\begin{equation}
\gamma^\ast(q_1,\lambda_1) \gamma^\ast(- q_2, -\lambda'_2) \to \gamma^\ast(q_1,\lambda'_1) \gamma^\ast(-q_2, -\lambda_2). 
\end{equation}
Under this photon crossing, which turns $\nu \to - \nu$ one obtains the relation
\begin{eqnarray}
M_{\lambda'_1 \lambda'_2, \lambda_1 \lambda_2}(\nu, Q_1^2, Q_2^2) = 
M_{\lambda'_1 -\lambda_2, \lambda_1 - \lambda'_2}(- \nu, Q_1^2, Q_2^2).   
\end{eqnarray}
The six combinations which are even in $\nu$ are given by:
\begin{eqnarray}
M_{+ +, + +} + M_{+ -, + -},  \quad M_{0 0, 0 0}, 
\quad  M_{0 +, 0 +}, \quad M_{+ 0, + 0}, \quad  M_{+ +, - -}, 
\quad M_{+ +, 0 0} + M_{0 +, - 0}
\end{eqnarray}
whereas the two combinations which are odd in $\nu$ are given by:
\begin{eqnarray}
M_{+ +, + +} - M_{+ -, + -}, \quad M_{+ +, 0 0} - M_{0 +, - 0}.
\end{eqnarray}
}

For specific quantum numbers of the final state $X$, parity conservation imposes the following condition on the helicity amplitudes
\begin{equation}
\label{eq:parity_for_M}
 H_{\lambda_1\lambda_2}=P_X (-1)^{-J_X}H_{-\lambda_1-\lambda_2}\,,
\end{equation}
where $P_X$ is the parity and $J_X$ is the spin of the hadronic system $X$. Using this relation, one finds the following relation between the transverse-transverse interference term and the helicity-0 amplitude for production of states with different $J^{PC}$:
\begin{equation}
\label{eq:extra_relation}
\tau_{TT} = 
\begin{cases}
\;\;\sigma_0 & \text{for } X = 0^{++},\, 2^{++},\, 3^{-+},\, \ldots \\
-\sigma_0 & \text{for } X = 0^{-+},\, 1^{++},\, 2^{-+},\, \ldots \,.
\end{cases}
\end{equation}

The virtual-photon density matrix elements \(\rho_i^{\lambda_i\lambda'_i}\) capture the polarization structure of the virtual photons. The components relevant for Eq.~(\ref{eq:gagacross}) are given by the unpolarized part of $\rho_{h_i}^{\lambda_i\lambda'_i}$, which reads
\begin{align}
\rho_1^{++} &= \frac{1}{2} \left\{ 1 - \frac{4 m^2}{Q_1^2} + \frac{1}{X} \left( 2 \, p_1 \cdot q_2 - q_1 \cdot q_2  \right)^2 \right\}\,, \nonumber \\
\rho_2^{++} &= \frac{1}{2} \left\{ 1 - \frac{4 m^2}{Q_2^2} + \frac{1}{X} \left( 2 \, p_2 \cdot q_1 - q_1 \cdot q_2  \right)^2 \right\}\,,  \nonumber \\
\rho_1^{00} &= \frac{1}{X} \left( 2 \, p_1 \cdot q_2 - q_1 \cdot q_2 \right)^2 - 1\,, \nonumber \\ 
\rho_2^{00} &= \frac{1}{X} \left( 2 \, p_2 \cdot q_1 - q_1 \cdot q_2  \right)^2 - 1\,. 
\label{eq:kincoeff}
\end{align}
From these expressions, one can define the familiar virtual-photon polarization parameters ($0 \leq \varepsilon_i \leq 1$)
\begin{align}
\varepsilon_i \;\equiv\; \frac{\rho_i^{++}-1}{\rho_i^{++}}\,,\quad i=1,2\,,
\end{align}
which allow one to express Eq.~(\ref{eq:gagacross}) in a form that conveniently allows to separate the response functions
\begin{align}
&d \sigma_{h_1,h_2} = \frac{\alpha^2}{8 \pi^4 \, Q_1^2 \, Q_2^2} \, \frac{\sqrt{X}}{s (1 - 4 m^2 / s)^{1/2}}   \frac{d^3 \vec p_1^{\, \prime}}{E_1^{\prime}} 
 \frac{d^3 \vec p_2^{\, \prime}}{E_2^\prime} \frac{4}{(1 - \varepsilon_1)(1 - \varepsilon_2)}\nonumber \\
&\times \Bigg\{ 
\frac{1}{2} \left( \sigma_{0} + \sigma_2 \right) 
+  \left[ \varepsilon_1 + \frac{2 m^2}{Q_1^2} (1 - \varepsilon_1 )\right] \left[ \varepsilon_2 + \frac{2 m^2}{Q_2^2} (1 - \varepsilon_2 )\right] \, \sigma_{LL}   \nonumber \\
&\qquad+  \left[ \varepsilon_2 + \frac{2 m^2}{Q_2^2} (1 - \varepsilon_2 )\right]  \, \sigma_{TL} 
+ \left[ \varepsilon_1 + \frac{2 m^2}{Q_1^2} (1 - \varepsilon_1 )\right] \, \sigma_{LT} 
+ \frac{1}{2}  \varepsilon_1 \varepsilon_2 \, \cos (2 \tilde \phi)  \, \tau_{TT} \nonumber \\
&
\qquad + 2\left[ \varepsilon_1 (1 + \varepsilon_1) + \frac{4 m^2}{Q_1^2} \varepsilon_1 (1 - \varepsilon_1 )\right]^{1/2} \left[ \varepsilon_2 (1 + \varepsilon_2) + \frac{4 m^2}{Q_2^2} \varepsilon_2 (1 - \varepsilon_2 )\right]^{1/2}  \, \cos \tilde \phi  \, \frac{1}{2}\left( \tau_0 + \tau_1 \right)\nonumber\\
&\qquad+ h_1 h_2 \, \left[ 1 - \varepsilon_1^2 + \frac{4 m^2}{Q_1^2} (1 - \varepsilon_1 )^2 \right]^{1/2}  
 \left[ 1 - \varepsilon_2^2 + \frac{4 m^2}{Q_2^2} (1 - \varepsilon_2 )^2 \right]^{1/2}  \, \frac{1}{2} \left( \sigma_0 - \sigma_2 \right) \nonumber \\
& \qquad+ 2\,h_1 h_2 \,  \left[ \varepsilon_1 (1 - \varepsilon_1 \right]^{1/2} \left[ \varepsilon_2 (1 - \varepsilon_2 \right]^{1/2}  \, \cos \tilde \phi  \, \frac{1}{2}\left( \tau_0 - \tau_1 \right)
\Bigg\}\,.
\label{eq:gagacrosseps}
\end{align}

\subsubsection{General cross section for the exclusive $e^+ e^- \to e^+ e^- \pi \pi$ process}

We next consider the exclusive process
\begin{equation} 
e^+(p_1) + e^-(p_2) \to e^+(p'_1) + e^-(p'_2) + \pi_1(p_{\pi_1}) + \pi_2(p_{\pi_2})\,,
\label{eq:gagaexcl}
\end{equation}
where $\pi_1 \pi_2$ stands for the $\pi^+ \pi^-$ or $\pi^0 \pi^0$ state, or in general stands for any two-meson state $M_1 M_2$.  
In contrast to the fully inclusive case, for which the cross section is given by Eq.~(\ref{eq:gagacross}), for the exclusive process of Eq.~(\ref{eq:gagaexcl}), we need to 
retain the full kinematic information on the pion pair, including their polar and azimuthal angular distributions. This allows for a more differential analysis of the reaction dynamics.

\begin{figure}[ht]
\begin{center}
\includegraphics[width=8cm]{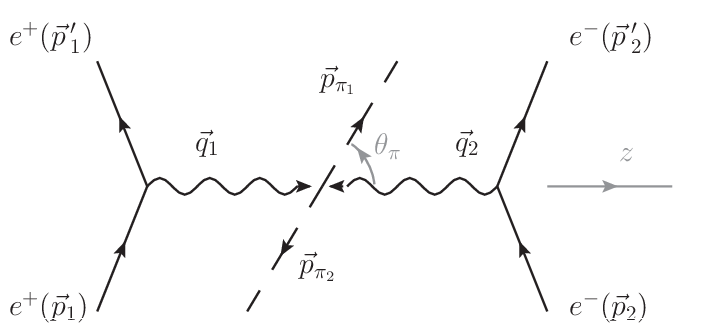}
\end{center}
\caption{{Schematic kinematic configuration of} the $e^+e^- \to e^+e^- \pi_1 \pi_2$ process in the $\gamma^*\gamma^*$ c.m. frame}
\label{fig:process}
\end{figure}

It is convenient to work in the $\gamma\gamma$ frame. By convention, the $z$-axis is chosen along the direction of $\vec{q}_1$ (the momentum of the photon emitted by the $e^+$), as shown in Fig.~\ref{fig:process}. In this frame, the two photons travel back-to-back along $\pm \hat{z}$. 
The pions are produced with momenta $p_{\pi_1}, p_{\pi_2}$ such that $\vec{p}_{\pi_1} + \vec{p}_{\pi_2} = 0$. We define the polar angle $\theta_\pi$ of the $\pi_1$ in the $\gamma\gamma$ frame as the angle between $\vec{p}_{\pi_1}$ and $\vec{q}_1$, so 

\begin{equation}
  \cos\theta_\pi = \hat{\vec{p}}_{\pi_1}\cdot \hat{\vec{q}}_1\,.
\end{equation}

Azimuthal angles are introduced to describe the orientation of the outgoing lepton planes relative to the hadronic plane. First, the “hadron plane” is defined as the plane containing the $\gamma^*\gamma^*$ axis ($\hat{z}$) and the $\pi_1$ momentum; by this definition, the hadron plane is taken to be the $xz$-plane, i.e. $\phi_\pi=0$. The $x$ axis is fixed in the direction of the transverse component of $\vec{p}_{\pi_1}$ (with respect to the direction of $\vec{q}_1$).  

\begin{figure}[t]
\begin{center}
\includegraphics[width=5cm]{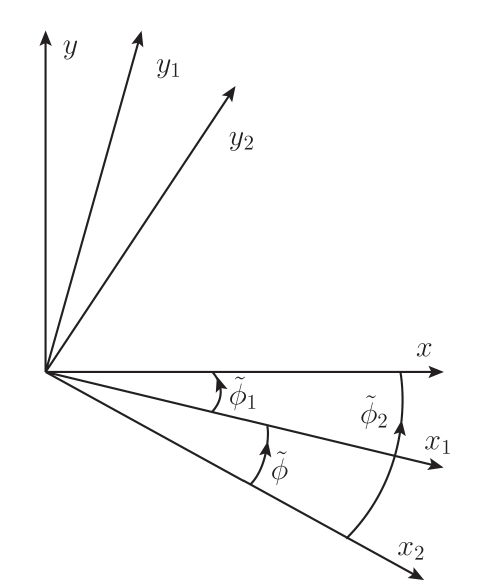}
\end{center}
\caption{The azimuthal angles $\tilde \phi_1$ and $\tilde \phi_2$ in the $\gamma^\ast \gamma^\ast$ c.m. frame between the lepton planes and the hadron plane. The latter is chosen as the $xz$-plane. The momenta $\vec p_1$ and $\vec q_1$ define the $x_1 z$-plane, whereas the momenta $\vec p_2$ and $\vec q_2$ define the $x_2 z$-plane. }
\label{diag}
\end{figure}

Next, the positron plane is defined as the plane containing $\hat{z}$ and the three-momentum of the incoming $e^+$ ($\vec{p}_1$). The angle between the hadron plane and the positron  plane is denoted by $\tilde{\phi}_1$. Similarly, the angle between the hadron plane and the electron plane (containing $\vec{p}_2$ and $\hat{z}$) is $\tilde{\phi}_2$. By construction, $\tilde{\phi}_{1,2}$ lie in $[0,2\pi)$. It is convenient to define the components of lepton momenta $p_i$ transverse to the $\gamma\gamma$ axis in the $\gamma\gamma$ rest frame
\begin{eqnarray}
\vec p_{i \perp} &\equiv& \vec p_i - \frac{\vec p_i \cdot \vec q_1}{|\vec q_1|^2} \vec q_1 \,, \quad i=1,2\,,
\end{eqnarray}
and express $\tilde{\phi}_{1,2}$ in terms of these quantities
\begin{eqnarray}
\hat {\vec p}_{i \perp} \cdot (\hat {\vec q}_1 \times \hat {\vec p}_{\pi^+} ) &=& \sin \theta_\pi \sin \tilde \phi_i\,. 
\end{eqnarray}
The difference $\tilde{\phi}=\tilde{\phi}_2-\tilde{\phi}_1$ is the relative azimuthal angle between the two lepton scattering planes. Equivalently, $\cos\tilde{\phi}$ can also be obtained from the scalar product of the two leptons’ transverse momentum vectors in the $\gamma\gamma$ frame: 
\begin{equation} 
\cos (\tilde \phi_2 - \tilde \phi_1) = \hat {\vec p}_{1 \perp} \cdot \hat {\vec p}_{2 \perp} \,,
\end{equation} 
as shown in Fig.~\ref{diag}.

For the exclusive $e^+ e^- \to e^+ e^- \pi_1 \pi_2$ process, one can introduce the angular integrated hadronic response function, contributing to the inclusive sum in Eq.~(\ref{eq:absW}), as:
\begin{align}
    W^{\pi\pi}_{\lambda'_1 \lambda'_2, \lambda_1 \lambda_2}=\int d \Omega_\pi  \tilde{W}^{\pi\pi}_{\lambda'_1 \lambda'_2, \lambda_1 \lambda_2}\,,
\end{align}
while the non-integrated hadronic response function 
$\tilde{W}^{\pi\pi}_{\lambda'_1 \lambda'_2, \lambda_1 \lambda_2}$ 
can be expressed in terms of the $\gamma^*\gamma^* \to \pi_1 \pi_2$ amplitudes as:
\begin{eqnarray}
\tilde{W}^{\pi\pi}_{\lambda'_1 \lambda'_2, \lambda_1 \lambda_2} \equiv  \frac{1}{2 \pi} \, \frac{\beta_\pi}{32 \pi} H_{\lambda_1 \lambda_2}  H^\ast_{\lambda'_1 \lambda'_2} ,\quad \beta_\pi = \left( 1 - \frac{4 m_\pi^2}{W^2}\right)^{1/2}\,,
\label{eq:Wpipi}
\end{eqnarray}
with $m_\pi$ the pion mass. 

Using parity invariance for the $\gamma^\ast \gamma^\ast \to \pi_1 \pi_2$ process, the hadronic response functions $\tilde{W}^{\pi\pi}_{\lambda'_1 \lambda'_2, \lambda_1 \lambda_2}$  have the property:
\begin{eqnarray}
\tilde{W}^{\pi\pi}_{-\lambda'_1 -\lambda'_2, -\lambda_1 -\lambda_2}  = (-1)^{\Lambda - \Lambda'} \, \tilde{W}^{\pi\pi}_{\lambda'_1 \lambda'_2, \lambda_1 \lambda_2}\,, 
\end{eqnarray}
whereas hermiticity of the hadronic tensor leads to the relation:
\begin{eqnarray}
\tilde{W}^{\pi\pi}_{\lambda'_1 \lambda'_2, \lambda_1 \lambda_2}  = (\tilde{W}^{\pi\pi}_{\lambda_1 \lambda_2, \lambda'_1 \lambda'_2})^\ast\,.
\end{eqnarray}
The process \( e^+ e^- \to e^+ e^- \pi_1 \pi_2 \) allows for contributions from helicity transitions with  
\( \Lambda - \Lambda' = 0, \pm1, \pm2, \pm3, \pm4 \). Accordingly, the fully differential $e^+ e^- \to e^+ e^- \pi_1 \pi_2$ cross section, i.e. also differential in the pion solid angle $d\Omega_\pi$, can be expressed in terms of 25 helicity-dependent cross sections and so-called interference response functions for the subprocess
\( \gamma^\ast \gamma^\ast \to \pi_1 \pi_2 \). The (positive definite) differential cross sections are defined as follows:
\begin{align}\label{Eq:dsiga/dcos_1}
&\frac{d \sigma_0}{d \cos \theta_\pi} \equiv \frac{\beta_\pi}{64 \pi \sqrt{X}} | H_{++} |^2\,, \quad 
\frac{d \sigma_2}{d \cos \theta_\pi} \equiv \frac{\beta_\pi}{64 \pi \sqrt{X}} | H_{+-} |^2\,,  \nonumber \\
&\frac{d \sigma_{TL}}{d \cos \theta_\pi} \equiv \frac{\beta_\pi}{64 \pi \sqrt{X}} | H_{+0} |^2\,, \quad \frac{d \sigma_{LT}}{d \cos \theta_\pi} \equiv \frac{\beta_\pi}{64 \pi \sqrt{X}} | H_{0+} |^2\,, \nonumber\\
&\frac{d \sigma_{LL}}{d \cos \theta_\pi} \equiv \frac{\beta_\pi}{64 \pi \sqrt{X}} | H_{00} |^2\,,
\end{align}
while the interference terms are represented by the real and imaginary parts of products of helicity amplitudes:
\begin{align}\label{Eq:dsiga/dcos_2}
&\frac{d \tau_0}{d \cos \theta_\pi} \equiv \frac{\beta_\pi}{64 \pi \sqrt{X}} 
\Re \left( H^\ast_{++}  H_{00}   \right), \quad \frac{d \tau_1}{d \cos \theta_\pi} \equiv \frac{\beta_\pi}{64 \pi \sqrt{X}} 
\Re \left( H^\ast_{0+}  H_{+0}   \right)\,, \nonumber \\
&\frac{d \tau_{12}}{d \cos \theta_\pi} \equiv \frac{\beta_\pi}{64 \pi \sqrt{X}} 
\Re \left( H^\ast_{+0}  H_{+-}   \right)\,, \quad 
\frac{d \tau_{-12}}{d \cos \theta_\pi} \equiv \frac{\beta_\pi}{64 \pi \sqrt{X}} 
\Re \left(H^\ast_{0+}  H_{+-}   \right)\,, \nonumber\\
&\frac{d \tau_{T2}}{d \cos \theta_\pi} \equiv \frac{\beta_\pi}{64 \pi \sqrt{X}} 
\Re \left( H^\ast_{++}  H_{+-}   \right)\,, \quad
\frac{d \tau_{L2}}{d \cos \theta_\pi} \equiv \frac{\beta_\pi}{64 \pi \sqrt{X}} 
\Re \left( H^\ast_{00}  H_{+-}   \right)\,, \nonumber \\
&\frac{d \tau_{1T}}{d \cos \theta_\pi} \equiv \frac{\beta_\pi}{64 \pi \sqrt{X}} 
\Re \left( H^\ast_{+0}  H_{++}   \right)\,,\quad 
\frac{d \tau_{-1T}}{d \cos \theta_\pi} \equiv \frac{\beta_\pi}{64 \pi \sqrt{X}} 
\Re \left( H^\ast_{0+}  H_{++}   \right)\,,\nonumber\\
&\frac{d \tau_{1L}}{d \cos \theta_\pi} \equiv \frac{\beta_\pi}{64 \pi \sqrt{X}} 
\Re \left( H^\ast_{+0}  H_{00}   \right)\,, \quad
\frac{d \tau_{-1L}}{d \cos \theta_\pi} \equiv \frac{\beta_\pi}{64 \pi \sqrt{X}} 
\Re \left( H^\ast_{0+}  H_{00}   \right)\,,
\end{align}
where the subscripts used in the interference terms 
denote the total $\gamma^* \gamma^*$ helicity of the  
contributing $\gamma^* \gamma^* \to \pi_1 \pi_2$ amplitudes, with $T (L)$ subscripts denote a helicity-zero amplitude arising from transverse (longitudinal) photons respectively.  
The remaining 10 interference response functions are given by the corresponding imaginary parts of the same helicity amplitude combinations. We denote them with a bar. For example, 
\begin{equation}\label{Eq:dsiga/dcos_3}
    \frac{d \bar{\tau}_0}{d \cos \theta_\pi} \equiv \frac{\beta_\pi}{64 \pi \sqrt{X}} 
\Im \left( H^\ast_{++}  H_{00}   \right)\,.
\end{equation}
A summary of for the notation for the two-photon cross sections and interference response functions is given in Tab.~\ref{tab:cross_sections}. {Note that for identical two-meson final states, such as $\pi^0\pi^0$, an additional symmetry factor of $1/2$ has to be included in Eqs.(\ref{Eq:dsiga/dcos_1}-\ref{Eq:dsiga/dcos_3}).}
\begin{table}[h]
\centering
{
\caption{Summary of the notation for the two-photon cross sections and interference response functions used in Eqs.~(\ref{Eq:dsiga/dcos_1}) and (\ref{Eq:dsiga/dcos_2}). The barred quantities $\bar\tau_i$ denote the corresponding imaginary parts.}
\begin{tabular}{lll}
\hline
Quantity & Helicity-ampl. & Meaning \\
\hline
$\sigma_0$   & $|H_{++}|^2$ & cross section, total helicity $0$ \\
$\sigma_2$   & $|H_{+-}|^2$ & cross section, total helicity $2$ \\
$\sigma_{TL}$ & $|H_{+0}|^2$ & cross section, transverse-longitudinal \\
$\sigma_{LT}$ & $|H_{0+}|^2$ & cross section, longitudinal-transverse \\
$\sigma_{LL}$ & $|H_{00}|^2$ & cross section, longitudinal-longitudinal \\
$\tau_0$     & $\mathrm{Re}(H_{++}^\ast H_{00})$ & interference, total helicity $0$ states \\
$\tau_1$     & $\mathrm{Re}(H_{0+}^\ast H_{+0})$ & interference, total helicity $\pm 1$ states \\
$\tau_{12}$  & $\mathrm{Re}(H_{+0}^\ast H_{+-})$ & interference, helicity $1$ and helicity $2$ states \\
$\tau_{-12}$ & $\mathrm{Re}(H_{0+}^\ast H_{+-})$ & interference, helicity $-1$ and helicity $2$ states \\
$\tau_{T2}$  & $\mathrm{Re}(H_{++}^\ast H_{+-})$ & interference, transverse helicity $0$ and helicity $2$ states \\
$\tau_{L2}$  & $\mathrm{Re}(H_{00}^\ast H_{+-})$ & interference, longitudinal helicity $0$ and helicity $2$ states \\
$\tau_{1T}$  & $\mathrm{Re}(H_{+0}^\ast H_{++})$ & interference, transverse helicity $0$ and helicity $1$ states \\
$\tau_{-1T}$ & $\mathrm{Re}(H_{0+}^\ast H_{++})$ & interference, transverse helicity $0$ and helicity $-1$ states \\
$\tau_{1L}$  & $\mathrm{Re}(H_{+0}^\ast H_{00})$ & interference, transverse helicity $0$ and helicity $1$ states \\
$\tau_{-1L}$ & $\mathrm{Re}(H_{0+}^\ast H_{00})$ & interference, transverse helicity $0$ and helicity $-1$ states \\
\hline
\end{tabular}
}

\label{tab:cross_sections}
\end{table}

The fully differential polarized cross section for the exclusive process $e^+ e^- \to e^+ e^- \pi_1 \pi_2$ can be expressed as:
\begin{equation}
    d \sigma_{h_1,h_2}= d \sigma^{(0)}+h_1\, d \sigma^{(1)}+h_2\, d \sigma^{(2)}+h_1\,h_2\, d \sigma^{(12)}\,,
    \label{eq:helcross}
\end{equation}
where $h_{1,2}=\pm 1$ denote the helicities of the incoming leptons (in units $\hbar/2$). 

The lepton helicity averaged (unpolarized) differential cross section $d\sigma^{(0)}$ is then obtained as:
\begin{align}
&d \sigma^{(0)} 
= \frac{\alpha^2}{8 \pi^4 \, Q_1^2 \, Q_2^2} \, \frac{\sqrt{X}}{s (1 - 4 m^2 / s)^{1/2}}  
\frac{d^3 \vec p_1^{\, \prime}}{E_1^{\prime}} 
\frac{d^3 \vec p_2^{\, \prime}}{E_2^\prime} d \cos\theta_\pi
\frac{4}{(1 - \varepsilon_1)(1 - \varepsilon_2)}\nonumber \\
&\times \left\{ 
\frac{1}{2} \left( \frac{d \sigma_{0}}{d \cos \theta_\pi} + \frac{d \sigma_2}{d \cos \theta_\pi} \right) 
+  \left[ \varepsilon_1 + \frac{2 m^2}{Q_1^2} (1 - \varepsilon_1 )\right] 
\left[ \varepsilon_2 + \frac{2 m^2}{Q_2^2} (1 - \varepsilon_2 )\right] \, 
\frac{d \sigma_{LL}}{d \cos \theta_\pi}   \right. \nonumber \\
&+  \left[ \varepsilon_2 + \frac{2 m^2}{Q_2^2} (1 - \varepsilon_2 )\right]  
\left( 1 + \varepsilon_1 \cos (2 \tilde \phi_1) \right) \, \frac{ d \sigma_{TL}}{d \cos \theta_\pi}  
+ \left[ \varepsilon_1 + \frac{2 m^2}{Q_1^2} (1 - \varepsilon_1 )\right] 
\left( 1 + \varepsilon_2 \cos (2 \tilde \phi_2) \right)  \, \frac{d \sigma_{LT}}{d \cos \theta_\pi}  
\nonumber \\
&+ \frac{1}{2}  \varepsilon_1 \varepsilon_2 \, 
 \left[ \cos 2 (\tilde \phi_2 - \tilde \phi_1)  \frac{ d \sigma_{0}}{d \cos \theta_\pi} 
 + \cos 2 (\tilde \phi_1 +  \tilde \phi_2)   \frac{d \sigma_2}{d \cos \theta_\pi} \right]  
-  \left[ \varepsilon_1 \cos (2 \tilde \phi_1)  + \varepsilon_2 \cos (2 \tilde \phi_2)  \right]  
\frac{d \tau_{T2}}{d \cos \theta_\pi}   \nonumber \\
&+ \left[ \varepsilon_1 (1 + \varepsilon_1) + \frac{4 m^2}{Q_1^2} \varepsilon_1 (1 - \varepsilon_1 )\right]^{1/2} 
\left[ \varepsilon_2 (1 + \varepsilon_2) + \frac{4 m^2}{Q_2^2} \varepsilon_2 (1 - \varepsilon_2 )\right]^{1/2}  \, \nonumber \\
&\times \left[ \cos  (\tilde \phi_2 - \tilde \phi_1)  
\left( \frac{ d \tau_{0}}{d \cos \theta_\pi} + \frac{ d \tau_{1}}{d \cos \theta_\pi} \right)  
+ \cos (\tilde \phi_1 +  \tilde \phi_2) 
\left(  \frac{d \tau_1}{d \cos \theta_\pi} - \frac{ d \tau_{L2}}{d \cos \theta_\pi} \right) \right] 
\nonumber \\
&+ \left[ \varepsilon_1 (1 + \varepsilon_1) + \frac{4 m^2}{Q_1^2} \varepsilon_1 (1 - \varepsilon_1 )\right]^{1/2}  
\left[ \cos \tilde \phi_1 \left( \frac{d \tau_{-12}}{d \cos \theta_\pi} - \frac{d \tau_{-1T}}{d \cos \theta_\pi} \right) 
\right. \nonumber \\
&  \hspace{5.cm}
- 2  \left[ \varepsilon_2 + \frac{2 m^2}{Q_2^2} (1 - \varepsilon_2 )\right] 
\cos \tilde \phi_1 \frac{d \tau_{1L}}{d \cos \theta_\pi}  \nonumber \\
&\left.  \hspace{5.cm}+ 
\varepsilon_2 \cos  (\tilde \phi_1 + 2 \tilde \phi_2) \frac{d \tau_{-12}}{d \cos \theta_\pi} 
- \varepsilon_2 \cos  (2 \tilde \phi_2 - \tilde \phi_1) \frac{d \tau_{-1T}}{d \cos \theta_\pi} 
\right] \nonumber \\
&+ \left[ \varepsilon_2 (1 + \varepsilon_2) + \frac{4 m^2}{Q_2^2} \varepsilon_2 (1 - \varepsilon_2 )\right]^{1/2}  
\left[ \cos \tilde \phi_2 \left( \frac{d \tau_{12}}{d \cos \theta_\pi} - \frac{d \tau_{1T}}{d \cos \theta_\pi} \right) 
\right. \nonumber \\
&\left.  \hspace{5.cm}
- 2  \left[ \varepsilon_1 + \frac{2 m^2}{Q_1^2} (1 - \varepsilon_1 )\right] 
\cos \tilde \phi_2 \frac{d \tau_{-1L}}{d \cos \theta_\pi} \right. \nonumber \\
&\left. \left. \hspace{5.cm}+ 
\varepsilon_1 \cos  (2 \tilde \phi_1 + \tilde \phi_2) \frac{d \tau_{12}}{d \cos \theta_\pi} 
- \varepsilon_1 \cos  (2 \tilde \phi_1 - \tilde \phi_2) \frac{d \tau_{1T}}{d \cos \theta_\pi} 
\right] 
\right\}\,. \nonumber \\
\label{eq:gagapipiunpolcross}
\end{align}

The helicity dependent cross section for the exclusive \( e^+ e^- \to e^+ e^- \pi_1 \pi_2 \) process in the case of polarized lepton beams, proportional to  
\( h_1 = \pm1 \) and \( h_2 = \pm1 \) in Eq.~(\ref{eq:helcross}), cannot be exploited at present due to missing beam polarization at existing experimental facilities. For future reference, we present these expressions for completeness in the Appendix. 

{
Note that the above expressions for the observables are fully symmetric under the interchange of leptons 1 and 2, which also implies the interchange of the virtual photons 1 and 2. For the kinematic variables, this interchange amounts to:
\begin{align}
& Q_1^2 \leftrightarrow Q_2^2, \quad \varepsilon_1 \leftrightarrow \varepsilon_2, \quad \tilde \phi_1 \leftrightarrow - \tilde \phi_2, \quad \cos \theta_\pi \leftrightarrow  -\cos \theta_\pi.
\end{align}
For the cross sections and interferences of Eqs.~(\ref{Eq:dsiga/dcos_1}) and (\ref{Eq:dsiga/dcos_2}) entering in the unpolarized cross section in Eq.~(\ref{eq:gagapipiunpolcross}), this interchange amounts to the symmetry:
\begin{align}
& \frac{d \sigma_0 }{d \cos \theta_\pi}(\nu, Q_1^2, Q_2^2, \cos \theta_\pi) = \frac{d \sigma_0 }{d \cos \theta_\pi}(\nu, Q_2^2, Q_1^2, - \cos \theta_\pi),
\label{eq:symmrel1}
\end{align}
and similarly for $\sigma_2,\sigma_{LL},\tau_{0},\tau_{1},\tau_{T2},\tau_{L2}$. For the response functions whose labels are also interchanged, it holds
\begin{align}
& \frac{d \sigma_{LT}}{d \cos \theta_\pi} (\nu, Q_1^2, Q_2^2, \cos \theta_\pi) = \frac{d \sigma_{TL}}{d \cos \theta_\pi}(\nu, Q_2^2, Q_1^2, -\cos \theta_\pi)\,,\nonumber \\
& \frac{d \tau_{-1 T}}{d \cos \theta_\pi} (\nu, Q_1^2, Q_2^2, \cos \theta_\pi) = \frac{d \tau_{1 T}}{d \cos \theta_\pi}(\nu, Q_2^2, Q_1^2, -\cos \theta_\pi)\,, \nonumber \\
&\frac{d \tau_{-1 L}}{d \cos \theta_\pi} (\nu, Q_1^2, Q_2^2, \cos \theta_\pi) = \frac{d \tau_{1 L}}{d \cos \theta_\pi}(\nu, Q_2^2, Q_1^2, -\cos \theta_\pi)\,, \nonumber \\
& \frac{d \tau_{-1 2}}{d \cos \theta_\pi} (\nu, Q_1^2, Q_2^2, \cos \theta_\pi) = \frac{d \tau_{1 2}}{d \cos \theta_\pi}(\nu, Q_2^2, Q_1^2, -\cos \theta_\pi)\,.
\label{eq:symmrel2}
\end{align}
For the cross sections and interferences which are given by the imaginary parts, as in Eq.~(\ref{Eq:dsiga/dcos_3}), and which enter the cross sections for one lepton beam polarization, as given in the Appendix, similar relations hold as Eqs.~(\ref{eq:symmrel1},\ref{eq:symmrel2}), except for one interference structure where the imaginary part induces a sign change:
\begin{align}
& \frac{d \bar \tau_1}{d \cos \theta_\pi} (\nu, Q_1^2, Q_2^2, \cos \theta_\pi) = -  \frac{d \bar \tau_1}{d \cos \theta_\pi}(\nu, Q_2^2, Q_1^2, -\cos \theta_\pi),
\end{align}
ensuring the symmetry under $1 \leftrightarrow 2$ between Eqs.~(\ref{eq:gagapipipolcross1}) and (\ref{eq:gagapipipolcross2}).
}

Furthermore, note that upon integrating Eq.~(\ref{eq:gagapipiunpolcross}) over \( d\Omega_\pi \) \footnote{To restore the trivial $\phi_\pi$ dependence in the kinematic factors in Eq.~(\ref{eq:gagapipiunpolcross}), which is shown for the choice $\phi_\pi = 0$, one needs to replace $\tilde{\phi}_1 \to \tilde{\phi}_1-\phi_\pi$ and $\tilde{\phi}_2 \to \tilde{\phi}_2-\phi_\pi$.} yields the \( \pi\pi \) contribution to the fully inclusive cross section given in Eq.~(\ref{eq:gagacrosseps}). It is important to keep in mind that for the exclusive case \( X = \pi\pi \), only final states with quantum numbers \( J^P = 0^{++},2^{++},4^{++},.. \) are allowed. Consequently, the transverse-transverse interference term satisfies \( \tau_{TT} = \sigma_0 \) as indicated in Eq.~(\ref{eq:extra_relation}).

The cross section for the single-tagged exclusive $e^+ e^- \to e^+ e^- \pi \pi$ process, where photon 1 has finite virtuality is obtained by integrating Eq.~(\ref{eq:gagapipiunpolcross})  over $\tilde \phi_2$ and taking the limit $Q^2_2 \to 0$. This yields:
\begin{eqnarray}\label{eq:gagapipiunpolcross_singlevirtual}
d \sigma^{(0)}  |_{Q^2_2 \to 0}
&=& \frac{\alpha^2}{8 \pi^4 \, Q_1^2 \, Q_2^2} \, \frac{\sqrt{X}}{s (1 - 4 m^2 / s)^{1/2}}  
\frac{d^3 \vec p_1^{\, \prime}}{E_1^{\prime}} 
 \frac{d^3 \vec p_2^{\, \prime}}{E_2^\prime} d\cos\theta_\pi
\frac{4}{(1 - \varepsilon_1)(1 - \varepsilon_2)}\nonumber \\
&\times& \left\{ 
\frac{1}{2} \left( \frac{d \sigma_{0}}{d \cos \theta_\pi} + \frac{d \sigma_2}{d \cos \theta_\pi} \right) 
+ \left[ \varepsilon_1 + \frac{2 m^2}{Q_1^2} (1 - \varepsilon_1 )\right] 
 \, \frac{d \sigma_{LT}}{d \cos \theta_\pi}  
- \varepsilon_1 \cos (2 \tilde \phi_1)  \frac{d \tau_{T2}}{d \cos \theta_\pi}
\right. \nonumber \\
&&
\left.+ \left[ \varepsilon_1 (1 + \varepsilon_1) + \frac{4 m^2}{Q_1^2} \varepsilon_1 (1 - \varepsilon_1 )\right]^{1/2}  \cos \tilde \phi_1 \left( \frac{d \tau_{-12}}{d \cos \theta_\pi} - \frac{d \tau_{-1T}}{d \cos \theta_\pi} \right) \right\}\,,  
\end{eqnarray}
which is fully consistent with the cross section equation for the unpolarized electron scattering off an unpolarized target~\cite{Drechsel:1992pn}. An important feature of the expression above is its explicit dependence on the azimuthal correlations between the lepton and hadronic planes. In particular, the terms proportional to $\cos (\tilde \phi_1)$ and $\cos (2\tilde \phi_1)$ reflect characteristic angular modulations of the cross section. These contributions arise from interference response functions such as $\tau_{T2}$, $\tau_{-12}$, and $\tau_{1L}$. While $\tau_{T2}$ describes interference between two different transverse photon amplitudes, $\tau_{-12}$ and $\tau_{1L}$ result from interferences between transverse and longitudinal photons. The presence of these azimuthal modulations therefore provides a novel opportunity to experimentally access the differential distributions of these response functions, thereby offering additional information.

The equations above constitute a convenient starting point for experimental investigations aimed at isolating and interpreting different response functions.

\subsubsection{Two-Photon Luminosity Functions}
\label{sec:lumi_function}
The experimentally measurable quantities, namely the two-photon cross sections and response functions, depend only on a limited set of variables: the two-photon center-of-mass energy $W$, the photon virtualities $Q_1^2$ and $Q_2^2$, and, in the case of multi-hadron final states, certain scattering angles of the decay products. Consequently, the differential cross sections given in Eqs.~(\ref{eq:gagacross})~and~(\ref{eq:gagapipiunpolcross}) are typically studied after suitable change of variables and integration over the remaining lepton variables, see Ref.~\cite{Bonneau:1973kg}. In this way, the inclusive cross section of Eq.~(\ref{eq:gagacross}) can be written, for unpolarized beams, as
\begin{align}
    \frac{\diff^3 \sigma}{\diff W \diff Q_1^2 \diff Q_{2}^2} &= \frac{\diff^3 L_\mathrm{TT}}{\diff W \diff Q_1^2 \diff Q_2^2}\sigma_\mathrm{TT}+\frac{\diff^3 L_\mathrm{LL}}{\diff W \diff Q_1^2 \diff Q_2^2}\sigma_\mathrm{LL}+
    \frac{\diff^3 L_\mathrm{TL}}{\diff W \diff Q_1^2 \diff Q_2^2}\sigma_\mathrm{TL} \nonumber\\
    &+\frac{\diff^3 L_\mathrm{LT}}{\diff W \diff Q_1^2 \diff Q_2^2}\sigma_\mathrm{LT}
    +\frac{\diff^3 \tilde L_\mathrm{TT}}{\diff W \diff Q_1^2 \diff Q_2^2}\tau_\mathrm{TT}
    +\frac{\diff^3 \tilde L_\mathrm{TL}}{\diff W \diff Q_1^2 \diff Q_2^2} \frac{1}{2}\left( \tau_0  + \tau_1  \right)\,,\label{sq:lumi_functions}
\end{align}
where $L_\mathrm{TT}$, $L_\mathrm{LL}$, $L_\mathrm{TL}$, $L_\mathrm{LT}$, $\tilde L_\mathrm{TT}$, and $\tilde L_\mathrm{TL}$ are the two-photon luminosity functions, which collect all lepton-side kinematics. These functions provide the link between the measurable cross section of the process $e^+e^-\to e^+e^-X$ and the underlying photon-photon subprocess. For the inclusive case, these functions have been obtained analytically in Ref.~\cite{Bonneau:1973kg}. In exclusive processes, however, the same approach is only applicable for the parts without explicit $\tilde\phi_{1,2}$ dependence. The terms involving $\tilde\phi_1$ and $\tilde\phi_2$ generally require numerical evaluation.

\subsection{Related Monte Carlo Generators}
Having established the theory framework, we briefly review existing two-photon MC tools to motivate our choices in \HTP. Over the years, several generators have been developed for studies of two-photon physics, each optimized for specific processes and subject to certain limitations:

\begin{itemize}

\item The KLOE collaboration used the code by A. Courau \cite{Courau:1984ia} for untagged measurements of $e^+e^-\to e^+e^-\pi^0\pi^0$ \cite{Bellucci:1994id}. The code is based on the double equivalent-photon approximation. \item A code written by F. Nguyen, F. Piccinini, and A.D. Polosa for two-photon physics at DA$\Phi$NE \cite{Nguyen:2006sr}. It has been used, for example, in the measurement of the $\eta$ radiative width at KLOE-2 \cite{KLOE-2:2012lws}. 

\item The Belle and Belle~II collaborations use the code by S. Uehara \cite{Uehara:1996bgt}. It has been applied in several studies, including those presented in Refs.~\cite{Belle:2003xlt, Belle:2010ckn, Belle:2013eck}. It uses an equivalent photon approximation in which the photon virtualities are taken into account. 

\item The code GALUGA2.0 by S. Schuler \cite{SCHULER1998279} is intended for LEP-type physics and uses appropriate models for high energies and a highly efficient code to generate the $e^+e^-\to e^+e^-X$ phase space, taking into account the strong dependence of the cross section on the momentum transfers. 

{ 
\item There are Monte Carlo generators (i.e. LPAIR \cite{Baranov:1991yq}) for the two-photon production of different final states at very high energies which are based on the work of J. A. M. Vermaseren and H. Krasemann \cite{Krasemann:1980ck,Vermaseren:1982cz}.

}

\item The BaBar collaboration used the \textsc{GGResRC} Monte Carlo generator \cite{DRUZHININ2014236} developed by the Novosibirsk group in their studies of transition form factors (TFFs) of pseudoscalars \cite{BaBar:2009rrj, BaBar:2018zpn}. The generator includes radiative corrections as described in Ref.~\cite{Ong:1988kg}. 

\item The \textsc{Ekhara} code by H. Czy\.z \cite{Czyz:2010sp, Czyz:2018jpp} and collaborators implements the same highly efficient phase-space generation algorithm as \textsc{GALUGA2.0} and pairs it with the determination of the cross section using the actual matrix element for $e^+e^-\to e^+e^-(\pi^0,\,\eta,\,\eta^\prime,\,\eta_c)$, including NLO radiative corrections. An updated version of the generator (“\textsc{Ekhara3.2}”) incorporates the double-virtual $\gamma^\ast\gamma^\ast\to\pi\pi$ amplitudes from Ref.~\cite{Danilkin:2019opj} and, unlike other generators, also simulates $e^+e^-\to e^+e^-\pi^+\pi^-$ via a virtual photon radiated in Bhabha scattering in addition to two-photon scattering. It ensures a complete description of the $\pi^+\pi^-$ channel.

\end{itemize}

\section{Phase Space Generation}

\subsection{Generation of the $e^+e^-X$ Phase Space}

The generation of the $e^+e^-X$ phase space is largely based on the highly efficient algorithm developed by Schuler, which is implemented in the \textsc{GALUGA} generator and described in detail in Ref.~\cite{SCHULER1998279}. The same method was later incorporated into the \textsc{Ekhara} event generator \cite{Czyz:2010sp, Czyz:2018jpp}, from which parts of the code have been adapted for this work. {When this phase space generation algorithm is combined with a proper generation of the lepton momentum transfers using a logarithmic mapping, the \((Q_1^2\,Q_2^2)^{-1}\) dependencies in Eqs.~(\ref{eq:gagacross}), (\ref{eq:gagacrosseps}), and (\ref{eq:gagapipiunpolcross}) cancel improving the numerical stability of the code. The logarithmic mapping also preferably generates events at small momentum transfers, roughly mimicking the behavior of the cross section, thereby increasing generation efficiency.} This section provides a brief overview of the phase-space generation, with a focus on the differences from the implementations in \textsc{GALUGA} and \textsc{Ekhara}. A full description of the algorithm can be found in Refs.~\cite{SCHULER1998279, Czyz:2010sp}.

The algorithm produces the final-state vectors by using five Lorentz invariants
\begin{align}  W^2&=p_X^2\,,   \\ 
    t_1 &=-Q_1^2\,, &\quad  t_2&=-Q_2^2\,,  \\
    s_1 &= (p_1^\prime + p_X)^2\,, &\quad s_2 &= (p_2^\prime + p_X)^2\,,
\end{align}
together with the constant center-of-mass energy $\sqrt{s}$.

In the first step, the $e^+e^-X$ four vectors are generated from six numbers $x_1 \dots x_6$ randomly distributed in $[0,1)$. This maps the Lorentz invariant phase space element onto a six-dimensional unit-1 hypercube:
\begin{align}
d\text{Lips}_{eeX} &= \frac{\diff ^3\vec{p}^\prime_1}{(2\pi)^3 2E_1'} \frac{\diff ^3\vec{p}^\prime_2}{(2\pi)^3 2E_2'} \frac{\diff ^3\vec{p}_X}{(2\pi)^3 2E_X} (2\pi)^4\delta^{(4)}\left( p_1+p_2-p^\prime_1-p^\prime_2-p_X \right)\nonumber\\
                    &= \frac{2W}{(2\pi)^5}\, \frac{\pi^2}{4 \beta s} \, \delta \,\delta_1 \, \delta_2 \,\delta_W \, \diff x_1 ... \diff x_6 \,,\label{eq:lips}
\end{align}
where $\beta = \sqrt{1-4m^2/s}$ is the electron velocity in the $e^+e^-$ c.m. frame. The remaining terms in this equation, which stem partially from the calculations in Refs.~\cite{SCHULER1998279,Czyz:2010sp} and depend on the use of different mappings, will be explained later in this section. The factor $2W$ arises from transforming the differential cross section from $\mathrm{d}W^2$ (as used in Ref.~\cite{SCHULER1998279}) to $\mathrm{d}W$.
The Jacobian factors $\delta$, $\delta_1$, $\delta_2$, and $\delta_W$ are detailed below. Using random numbers, the three-body final state is generated in the following steps:

\begin{enumerate}
	\item \textit{Generation of the hadronic mass $W$:} \\
	If $W$ is not fixed by the user to a constant value, it is restricted to the range $[W_\text{min},W_\text{max}]$ by kinematic and user-defined conditions. For cross-checks, two different mapping have been implemented:
	\begin{enumerate}
		\item Flat mapping: \\
		In this case, every mass between $W_\text{min}$ and $W_\text{max}$ is generated with equal probability:
		\begin{equation}
			W = W_\text{min} + \left( W_\text{max}-W_\text{min} \right)\, x_6 \,,
		\end{equation}
		which leads to a contribution to the phase space element of
		\begin{equation}
			\delta_W \equiv \frac{\diff W}{\diff x_6} = W_\text{max}-W_\text{min} \,.
		\end{equation}
		
		\item Logarithmic Mapping: \\
		For better description of the kinematic factors in Eqs.~(\ref{eq:gagacross}) and (\ref{eq:gagapipiunpolcross}), which typically drop rapidly with increasing $W$, a Logarithmic mapping is implemented. This enhances the generation of events at smaller masses:
		\begin{equation}
			W = W_\text{min} \, \text{exp}\left( x_6 \, \text{log}\left(\frac{W_\text{max}}{W_\text{min}} \right)\right)\,.
		\end{equation}
		The corresponding contribution to the phase space element is
	\begin{equation}
			\delta_W =  W\, \text{log}\left(\frac{W_\text{max}}{W_\text{min}} \right)  \, .
		\end{equation}
	\end{enumerate}
	In general, the choice of mapping does not affect the final physics result but has a significant influence the efficiency of the code. For the generation of arbitrary final states used in PWAs, both mappings perform equally well. For channels that use hadronic input to describe a physical process (see Sec.~\ref{sec:tp_xs_and_resp}), the flat mapping is recommended.\\
	
	\item \textit{Generation of the lepton momentum transfers ${t_1}$ and ${t_2}$:} \\
	The electron momentum transfer $t_2$ is limited by kinematic and user-defined constraints to the range $[t_{2,\text{min}},t_{2,\text{max}}]$. For this variable, both flat and logarithmic mappings are implemented, following the same structure as for the hadronic mass.
	\begin{enumerate}
		\item Flat Mapping: \\
		The momentum transfer is generated as
    \begin{equation}
			t_2 = t_{2,\text{min}} + \left(t_{2,\text{max}} - t_{2,\text{min}}\right)\, x_1 \, ,
	\end{equation}
		which leads to a contribution to the phase-space element of
	\begin{equation}
			\delta_2 = \frac{\diff t_2}{\diff x_1} = \left(t_{2,\text{max}} - t_{2,\text{min}}\right) \, .
	\end{equation}
		
		\item Logarithmic Mapping: \\
	\begin{equation}
			t_2 = t_{2,\text{min}} \, \text{exp}\left( x_1 \, \text{log}\left(\frac{t_{2,\text{max}}}{t_{2,\text{min}}} \right)\right)\,,
	\end{equation}
		\begin{equation}
		\delta_2 = t_2\, \text{log}\left(\frac{t_{2,\text{max}}}{t_{2,\text{min}}} \right)\,.
		\end{equation}
	\end{enumerate}
	After the generation of the electron momentum transfer $t_2$, the positron momentum transfer $t_1$ (and its contribution to the phase-space element $\delta_1$) is obtained in the same manner using the random variable $x_2$. In principle, both mappings yield the same physical result, but the logarithmic mapping is significantly more efficient, since it better describes the rapidly dropping behavior of the kinematic factor in Eqs.~(\ref{eq:gagacross}) and (\ref{eq:gagapipiunpolcross}). Moreover, the logarithmic mapping cancels the pole in Eqs.~(\ref{eq:gagacross}) and (\ref{eq:gagapipiunpolcross}) at $t_1t_2=0$, which may otherwise cause numeric instabilities at small values of $t_1$ and $t_2$.\\

    \item \textit{Generation of the subsystem center-of-mass energies ${s_1}$ and ${s_2}$:}\\
    The generation of $s_1$ and $s_2$ follow the procedure described in Ref.~\cite{SCHULER1998279}. Since these invariants are typically not used in studies of two-photon interactions, user cuts on these variables are not implemented. The possibility to directly restrict the energy of the final state leptons during the generation of the sub-system center-of-mass energies, which would act on the limits $s_1$ and $s_2$ are generated in, is not implemented for the same reason. Thus, $s_1$ and $s_2$ are restricted only by kinematics. User-defined cuts on the final lepton's energies are possible once the final state vectors are generated. The less efficient option is provided for completeness, but is assumed to be unnecessary for normal use cases. The invariants are generated using the random numbers $x_3$ and $x_4$. Their contribution to the phase space element is given by \cite{SCHULER1998279}
    \begin{equation}
        \delta = \frac{s\,(1+\beta)^2}{( \nu + \sqrt{X} )\left(1+y_1\right)\left(1+y_2\right)}\,,\quad y_{1/2}\equiv\sqrt{1-4\,m^2/t_{1/2}} \,.
    \end{equation}\\
with
\begin{equation}\nonumber
\nu \equiv (q_1 \cdot q_2)=\frac{1}{2}(W^2-t_1-t_2)\,\quad  X=\nu^2-t_1 t_2\,.
\end{equation}
    \item \textit{Calculation of the final state four-vectors:}\\
    When all six Lorentz invariants are known, the four-vectors of the final state can be calculated, except for some rotation around the azimuthal angle. The numerically stable form of the vectors is taken from Ref.~\cite{SCHULER1998279}. As the vectors are determined only relative to one another, they need to be rotated around the azimuthal angle by a random number between 0 and $2\pi$, which is determined using $x_5$. 
\end{enumerate}
At this point, the final state kinematics of the process $e^+e^-\to e^+e^-X$ is generated, and the Lorentz invariant phase space element of this process is calculated. 

\subsection{Generation of the Final State Hadronic Decay}

After generating the overall four-momentum of the hadronic system $X$ with invariant mass $W$, its decay into $N$ final-state particles with masses $m_1\dots m_N$ and four-momenta $\ell_1\dots \ell_N$ needs to be simulated. This is achieved using a recursive method (similar to the Genbod algorithm \cite{James:1968gu}), where the $N$-body decay is constructed as a sequence of two-body decays. The sequence is realized by introducing a set of intermediate invariant masses:
\begin{align}
    M^2_1 &= m^2_1\,,\nonumber\\
    M_i^2 &= \left(\sum_{j=1}^i\ell_j\right)^2\,,\quad i=2,...,N-1\,,\\
    M_{N}^2& =p_X^2 \equiv W^2\,,\nonumber 
\end{align}
where each $M_i$ corresponds to the invariant mass of a subsystem composed of the first $i$ particles. The key idea is that at each step an intermediate state of mass $M_{i+1}$ decay into two bodies: one real particle of mass $m_{i+1}$ and a residual system of invariant mass $M_{i}$. To generate the actual values of the intermediate masses $i=2,...,N-1$, they are randomly sampled within their kinematic limits
\begin{align} M_{i,\text{min}}&=\sum_{j=1}^{i}m_j\,,\quad 
    M_{i,\text{max}}=M_{i+1} - m_{i+1}\,,\nonumber\\
    M_i^2 &= M^2_{i,\text{min}} + \left(M^2_{i,\text{max}} - M^2_{i,\text{min}}\right) \tilde x_i\,.
\end{align}
In each two-body step, the magnitude of the three-momentum of the decay products in the rest frame of the parent $(M_{i+1})$ is given by
\begin{equation}
    p^\ast_i \left( M_{i+1}, M_{i}, m_{i+1} \right) = \frac{\sqrt{\lambda\left(  M^2_{i+1}, M^2_{i}, m^2_{i+1} \right)}}{2M_{i+1}}\,,
\end{equation}
where
\begin{equation}
\lambda(x,y,z)=x^2+y^2+z^2-2xy-2xz-2yz\,,
\end{equation}
is the Käll\'en triangle function. By using the additional random variables $\hat x_{i,1},\hat x_{i,2} \in [0,1]$, the  azimuthal and polar angles of the particle in the parent rest frame can be generated as
\begin{align}
    \cos\theta_i^\ast &= 2\,\hat x_{i,1} -1,\quad
    \phi_i^\ast = 2\pi\, \hat x_{i,2} \,,\\
    d\Omega^\ast_i&=d \cos\theta_i^\ast\,d\phi_i^\ast = 4\pi\,d\hat x_{i,1}\,d\hat x_{i,2}\,.
    \nonumber 
\end{align}

Using the generated decay angles, two-body momentum, and the particle masses, the four-vectors of the decay products are first constructed in the rest frame of their parent particle. At each stage of the recursive decay chain, these four-vectors are then boosted from the parent rest frame to the overall $e^+e^-$ center-of-mass frame by a Lorentz transformation. 

The Lorentz invariant phase space element for the decay of a state $X$ into $N$ particles is given by
\begin{equation}
    d\text{Lips}_\text{N}(M_N^2) =\left(\prod_{i=1}^N\frac{\diff ^3\vec{\ell}_i}{(2\pi)^3 2E_i}\right)(2\pi)^4\delta^{(4)}\left(p_X-\sum_{i=1}^{N}\ell_i\right)\,.
\end{equation}
This expression can be written recursively in terms of $\text{dLips}_\text{N-1}$ 
\begin{align}
    d\text{Lips}_\text{N}(M_N^2) &=  \frac{\diff M_{N-1}^2}{2\pi}\,d\text{Lips}_\text{2}(p_N;p_{N-1},l_N)\,d\text{Lips}_\text{N-1}(M_{N-1}^2)\,,
    \label{eq:recursion_1}
\end{align}
where $d\text{Lips}_\text{2}(p_N;p_{N-1},l_N)$ is the two-body phase space for the decay
\begin{equation}
    p_N \to p_{N-1}+\ell_N\quad p_i\equiv \sum_{j=1}^{i}\ell_j\,,
\end{equation}
and $d\text{Lips}_\text{N-1}(M_{N-1}^2)$ is the $(N-1)$ body phase space of the subsystem with total four-momentum $p_{N-1}$ (satisfying $p_{N-1}^2=M_{N-1}^2$). Inserting the two-body element in the spherical coordinates, one obtains
\begin{equation}
    d\text{Lips}_\text{N}(M_N^2) =  \frac{\diff M^2_{N-1}}{2\pi}\diff \Omega^\ast_{N-1}\frac{\sqrt{\lambda\left(  M^2_N, M^2_{N-1}, m^2_N \right)}}{8\pi\, M_N^2}\,\text{dLips}_\text{N-1}(M_{N-1}^2)\,.
\end{equation}
Iterating (\ref{eq:recursion_1}) yields the standard product form
\begin{equation}
    d\text{Lips}_\text{N}(M_N^2) = \left(\prod_{i=2}^{N-1}\frac{\diff M^2_{i}}{2\pi}\right)
    \left(\prod_{i=1}^{N-1}
    \frac{\sqrt{\lambda\left(  M^2_{i+1}, M^2_{i}, m^2_{i+1}\right)}}{8\pi\, M_{i+1}^2}\diff \Omega^\ast_{i}\right)\,.
\end{equation}
Using the random variables introduced earlier ($\tilde{ x}_{i}$ for the invariant masses, and $\hat x_{i,1}, \hat x_{i,2}$ for the angles), the full recursive expression for the $N$-body phase space can be written as
\begin{align}
    d\text{Lips}_N(W^2) &=\left( \prod_{i=2}^{N-1}\frac{\left(M^2_{i,\text{max}} - M^2_{i,\text{min}}\right)\diff \tilde{x}_i}{2\pi}  \right) \,\left(\prod_{i=1}^{N-1}\frac{\sqrt{\lambda\left(  M^2_{i+1}, M^2_{i}, m^2_{i+1} \right)}}{2\,M_{i+1}^2}\diff \hat x_{i,1}\,\diff \hat x_{i,2}\right)\,.
\end{align}
The Lorentz invariant phase space of the full final state (which is given in Eq.~(\ref{eq:full_phase_space})) can then be obtained by
\begin{equation}
    d\mathrm{Lips} = d\mathrm{Lips}_{eeX} \times d\mathrm{Lips}_N\,.
\end{equation}

\subsection{Application of User Cuts and Consistency Checks}
After all final state four-vectors have been generated, they are numerically checked for energy and momentum conservation with a tolerance of 1\,eV, which is negligible for typical particle physics experiments. { Additionally, we verify the numerical stability of the calculated sines for the final-state lepton and the combined hadronic system, ensuring their polar and azimuthal angles remain within the physical range of -1 to +1. Because direct calculation of the cosines for these angles is numerically less stable, we retain only the sign of the cosine. Its precise magnitude is then derived using the trigonometric identity \(\cos^2(x)=1-\sin^2x\).} When the code is compiled using quadruple precision floating point numbers, the final state vectors always fulfill this requirement, and no unstable events have been observed. With double precision, {around 6 out of 10$^7$ events show numerical instablities. The precise number is dependent on the final state and the restrictions to the phase space, but generally remains much smaller than 0.001\,\%.} These events are rejected and regenerated.

Following the stability check, several user-defined cuts are applied. These include cuts on the polar angles of the hadronic decay products as well as on the total transverse momentum of the hadronic final state. The latter is a commonly used observable in studies involving two quasi-real photons.

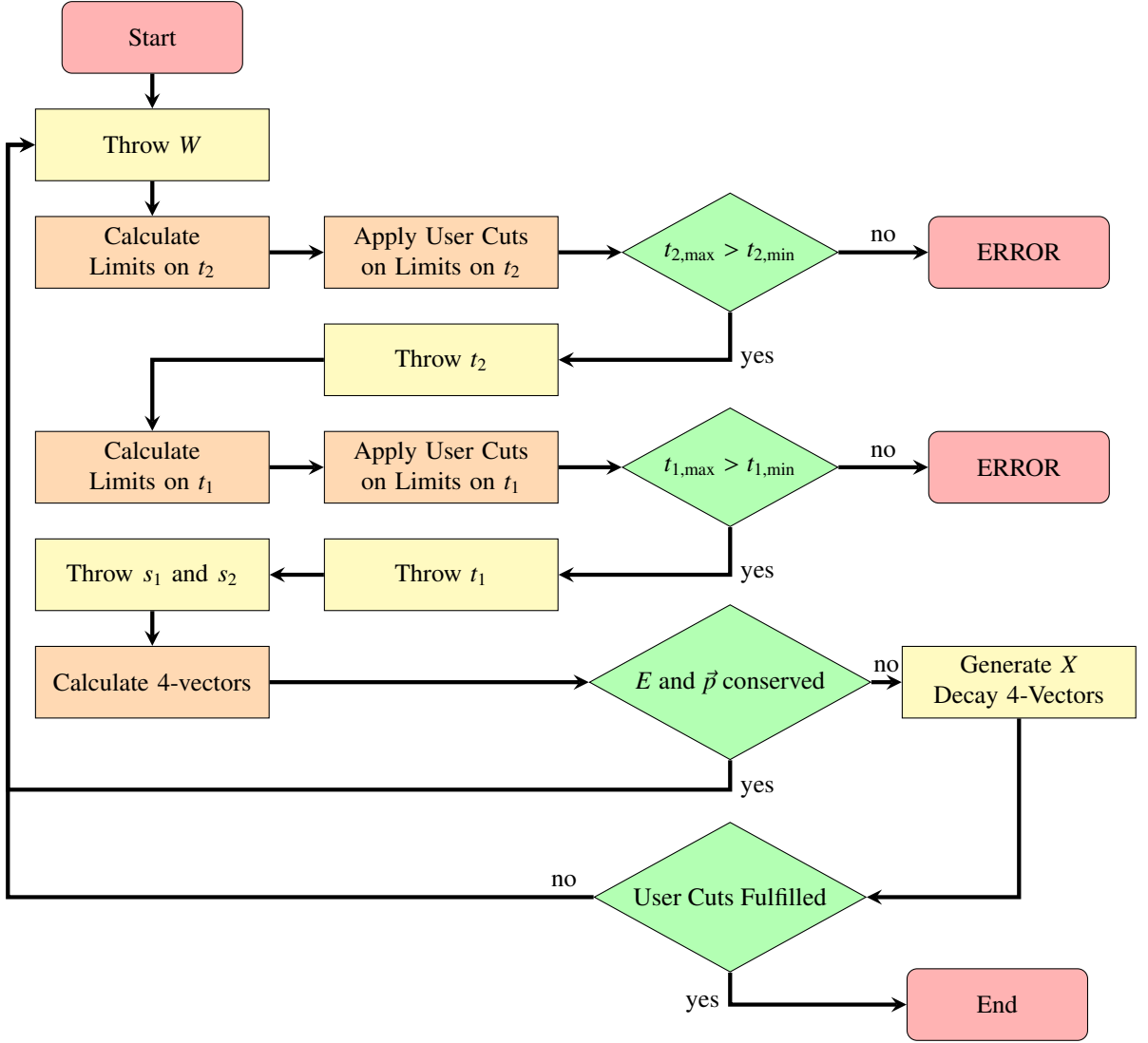
\begin{figure}[t!b]
	\centering
	\begin{tikzpicture}[node distance=1.5cm]
		\centering
		
		\node (start) [startstop] {Start};	
		\node (genW) [generate, below of=start] {Throw $W$};
		\node (limitT2) [process, below of=genW] { Calculate Limits on $t_2$ };
		\node (userCutsT2) [process, right of=limitT2, xshift=2.5cm] { Apply User Cuts on Limits on $t_2$ };
		\node (checkPHSP2) [decision, right of=userCutsT2, xshift=2.5cm] {$ t_{2,\text{max}}>t_{2,\text{min}} $};
		\node (errT2) [startstop, right of=checkPHSP2, xshift=2.5cm] {ERROR};
		\node (genT2) [generate, below of=userCutsT2] { Throw $t_2$ };
		
		\draw [arrow] (start) -- (genW);
		\draw [arrow] (genW) -- (limitT2);
		\draw [arrow] (limitT2) -- (userCutsT2);
		\draw [arrow] (userCutsT2) -- (checkPHSP2);		
		\draw [arrow] (checkPHSP2) -- node[anchor=south ] {no} (errT2);
		\draw [arrow] (checkPHSP2) |- node[anchor=west] {yes} (genT2);
		
		\node (limitT1) [process, below of=limitT2, yshift = -1.5cm] { Calculate Limits on $t_1$ };
		\node (userCutsT1) [process, right of=limitT1, xshift=2.5cm] { Apply User Cuts on Limits on $t_1$ };
		\node (checkPHSP1) [decision, right of=userCutsT1, xshift=2.5cm] {$ t_{1,\text{max}}>t_{1,\text{min}} $};
		\node (errT1) [startstop, right of=checkPHSP1, xshift=2.5cm] {ERROR};
		\node (genT1) [generate, below of=userCutsT1] { Throw $t_1$ };
		
		\draw [arrow] (genT2) -| (limitT1);
		\draw [arrow] (limitT1) -- (userCutsT1);
		\draw [arrow] (userCutsT1) -- (checkPHSP1);		
		\draw [arrow] (checkPHSP1) -- node[anchor=south ] {no} (errT1);
		\draw [arrow] (checkPHSP1) |- node[anchor=west] {yes} (genT1);
		
		\node (genS) [generate, below of=limitT1] {Throw $s_1$ and $s_2$};
		\node (calc4)[process, below of=genS] {Calculate 4-vectors};
		\node (acc) [decision, below of=checkPHSP1, yshift = -1.5cm] { $E$ \& $\vec p$ conserved};
		
		\draw [arrow] (genT1) -- (genS);
		\draw [arrow] (genS) -- (calc4);
		\draw [arrow] (calc4) -- (acc);
		\draw [arrow] (acc) |- node[anchor=west] {no} (-2,-10.5) |- (genW);
		
		\node (usercuts) [decision, below of=acc, yshift = -1.5cm] {User Cuts Fulfilled};
		\node (4body) [generate, right of=acc, xshift=2.5cm] {Generate $X$ Decay 4-Vectors};
		\draw [arrow] (acc) -- node[anchor=south] {yes} (4body);
		\draw [arrow] (4body) |- (usercuts);
		\draw [arrow] (usercuts) -| node[anchor=south, xshift=7.7cm] {no} (-2,-7.5) |- (genW);
		
		\node (end) [startstop, below of=usercuts, xshift = 3.7cm] {End};
		\draw [arrow] (usercuts) |- node[anchor = east] {yes} (end);

	\end{tikzpicture}
	\caption{Flowchart of the four-vector generation of the $e^+e^-X$ final state}
	\label{fig:sorrow_4_vectors}
\end{figure}

Once these steps are completed, the generation of the final state four vectors and the calculation of the phase-space element are finished. Figure~\ref{fig:sorrow_4_vectors} illustrates the four-vector generation procedure as a flow chart.

\section{Two-Photon cross sections and Responses}
\label{sec:tp_xs_and_resp}
The two-photon cross sections and response functions entering Eqs.~(\ref{eq:gagacross}) and (\ref{eq:gagapipiunpolcross}) must be specified in order to obtain physically meaningful results. Several modes are implemented to describe different classes of final states, which are briefly discussed in this section. A key feature of the generator is its flexibility: users can selectively switch between different cross sections and response functions, enabling the study of each contribution either independently or in combination.

\subsection{$e^+e^-\to e^+e^- \pi \pi$ Modes}
\label{sec:pipi_mode}

To simulate the process \( e^+e^- \to e^+e^- \pi\pi \), one must accurately model the differential response functions of the subprocess \( \gamma^\ast \gamma^\ast \to \pi\pi \). In \HTP, these are implemented using the dispersive formalism developed in Refs.~\cite{Danilkin:2019opj,Danilkin:2018qfn,Danilkin:2019mhd}, which are summarized in this section. In the $\gamma\gamma$ frame, the helicity amplitudes \( H_{\lambda_1 \lambda_2} \) are decomposed into partial waves according to
\begin{eqnarray}
\label{p.w.expansion}
H_{\lambda_1 \lambda_2}=e^{i\phi_\pi(\lambda_1-\lambda_2)} \sum_{J\,\text{even}}(2J+1)\,h^{(J)}_{\lambda_1\lambda_2}(W)\,d_{\Lambda,0}^{(J)}(\theta_\pi)\,, 
\end{eqnarray}
where \( \Lambda = \lambda_1 - \lambda_2 \) and \( d^{(J)}_{\Lambda,0}(\theta_\pi) \) are Wigner rotation functions. The hadronic plane is fixed by choosing \( \phi_\pi = 0 \), so that all azimuthal dependence arises through 
$\tilde \phi_1$ and $\tilde \phi_2$ on the lepton side as in Eq.~(\ref{eq:gagapipiunpolcross}). The main idea of the dispersive approach is to impose analyticity and unitarity on the partial-wave amplitudes. This is achieved using a modified Muskhelishvili--Omnès representation~\cite{Garcia-Martin:2010kyn}, applied to specific linear combinations of partial-wave amplitudes that are free from kinematic singularities \cite{Danilkin:2020pak}. For the scalar channel (\( J=0,I=0 \)), a coupled-channel dispersive analysis is employed to account for both \( \pi\pi \) and \( K\bar{K} \) intermediate states. This allows a unified treatment of the broad \( f_0(500) \) and narrow \( f_0(980) \) resonances. For the tensor channel (\( J=2,I=0 \)), dominated by the \( f_2(1270) \) resonance, a single-channel Omnès representation based on \( \pi\pi \) rescattering is sufficient. Contributions from isospin-2 channels are also treated within a single-channel approach. Re-scattering effects for higher partial waves (\( J \geq 4 \)) are expected to be suppressed at low to moderate energies and are approximated by their Born contributions. The S-wave left-hand cut is dominated by the Born terms, while for the D-wave, additional contributions from vector-meson exchange (e.g., \( \rho \), \( \omega \)) are included. As input for the pion and kaon vector form factors, standard VMD parameterizations are used, following the treatment in Ref.~\cite{Danilkin:2021icn}. The \( \pi\rho \) and \( \pi\omega \) TFFs are modeled using a phenomenological extension of VMD~\cite{Danilkin:2019mhd}, designed to reproduce the correct asymptotic behavior at large photon virtualities. The parameters of these TFFs are determined by matching to low \( Q^2 \) input from VMD for \( \pi\rho \) and dispersive results for \( \pi\omega \) \cite{Niecknig:2012sj,Schneider:2012ez,Danilkin:2014cra}, and by matching to high \( Q^2= 4.5~\mathrm{GeV}^2 \) experimental data from the Belle collaboration~\cite{Belle:2015oin}. Unsubtracted dispersion relations are employed to ensure predictive power for virtual photons with \( Q_i^2 \neq 0 \), while respecting low-energy theorems and matching smoothly to real-photon data. The dispersively constructed amplitude for the scalar-isoscalar channel (\( J=0, I=0 \)), which includes the \( f_0(500) \) and \( f_0(980) \) resonances, is currently used as input for computing their contributions to hadronic light-by-light scattering contribution of the muon anomalous magnetic moment \( (g-2)_\mu \)~\cite{Aliberti:2025beg}.

\begin{figure*}[!t]
\centering
\includegraphics[width=0.96\linewidth]{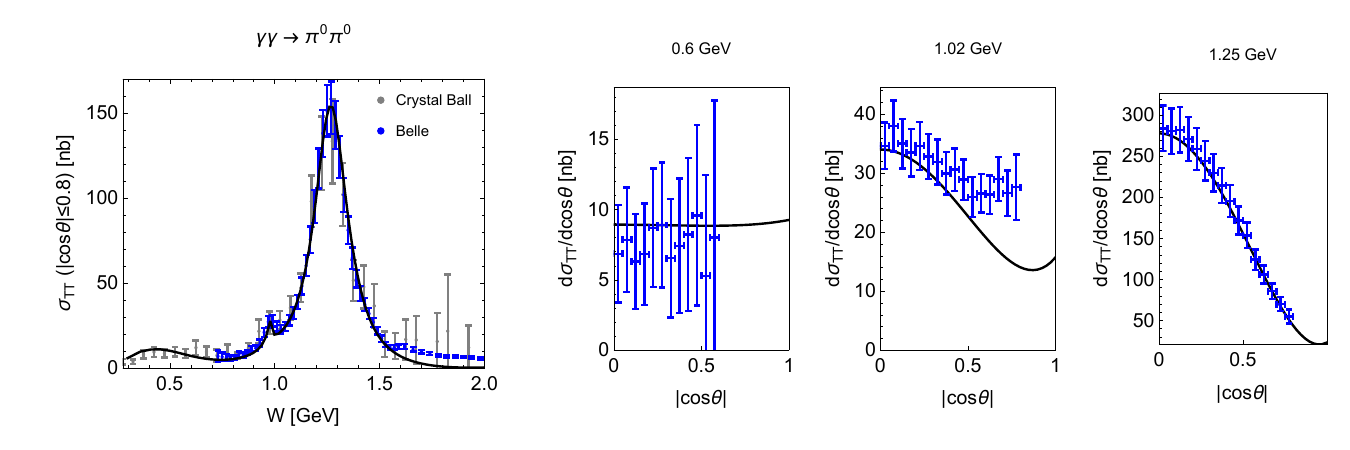}\\
\includegraphics[width=0.96\linewidth]{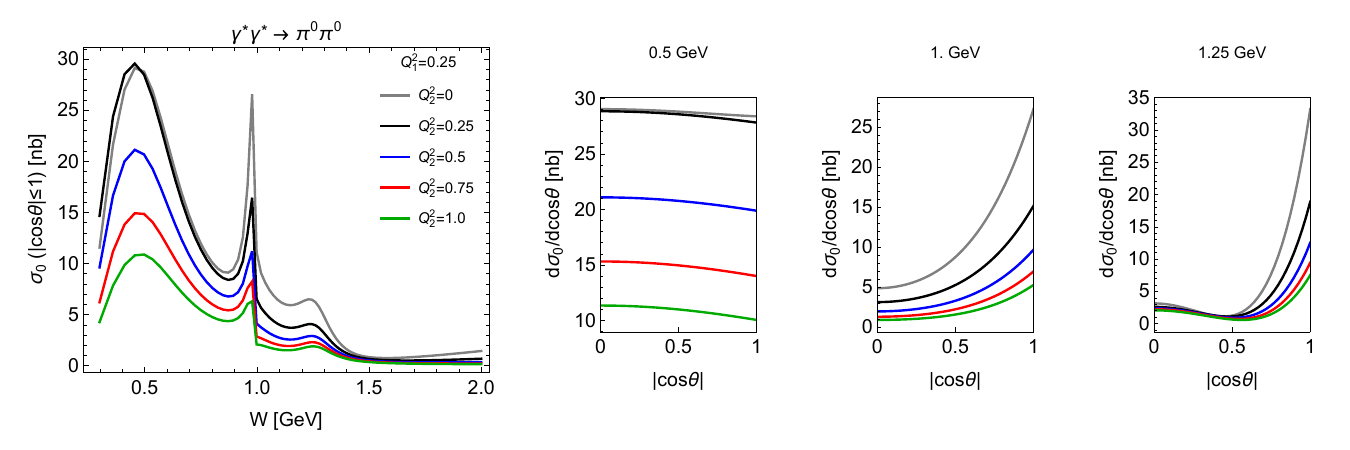}\\
\includegraphics[width=0.96\linewidth]{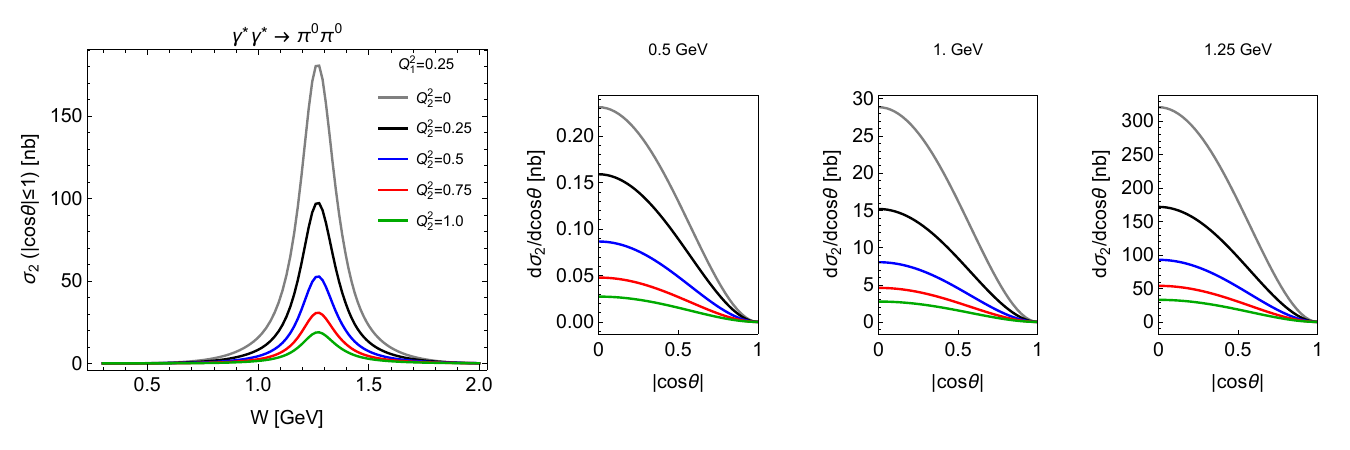}\\
\includegraphics[width=0.96\linewidth]{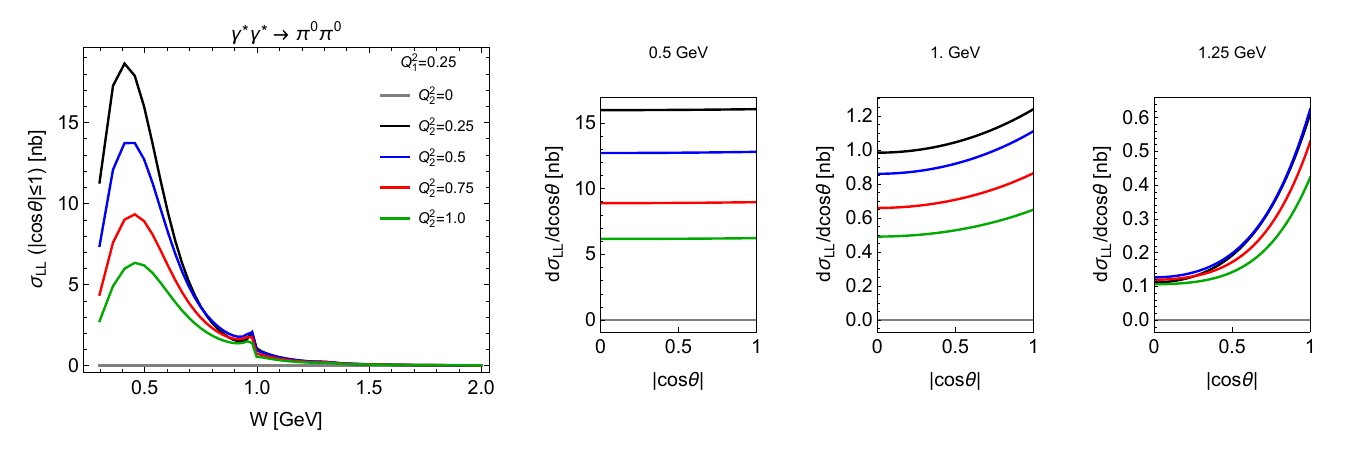}
\caption{Top row of panels: prediction for $\sigma_{TT}$ for  $\gamma \gamma \to \pi^0 \pi^0$ compared with data. 
Lower panels: predictions for $\sigma_{0}$, $\sigma_{2}$, and $\sigma_{LL}$ for $\gamma^* \gamma^* \to \pi^0 \pi^0$ with $Q_1^2 = 0.25 \,\text{GeV}^2$ and $Q_2^2 = 0, 0.25, 0.5, 0.75, 1.0 \,\text{GeV}^2$, shown for full angular coverage ($|\cos\theta| \leq 1$). {The experimental data are taken from \cite{Belle:2009ylx,CrystalBall:1990oiv}.}
\label{fig:pipiQ1Q2_1}}
\end{figure*}

\begin{figure*}[!t]
\centering
\includegraphics[width=0.96\linewidth]{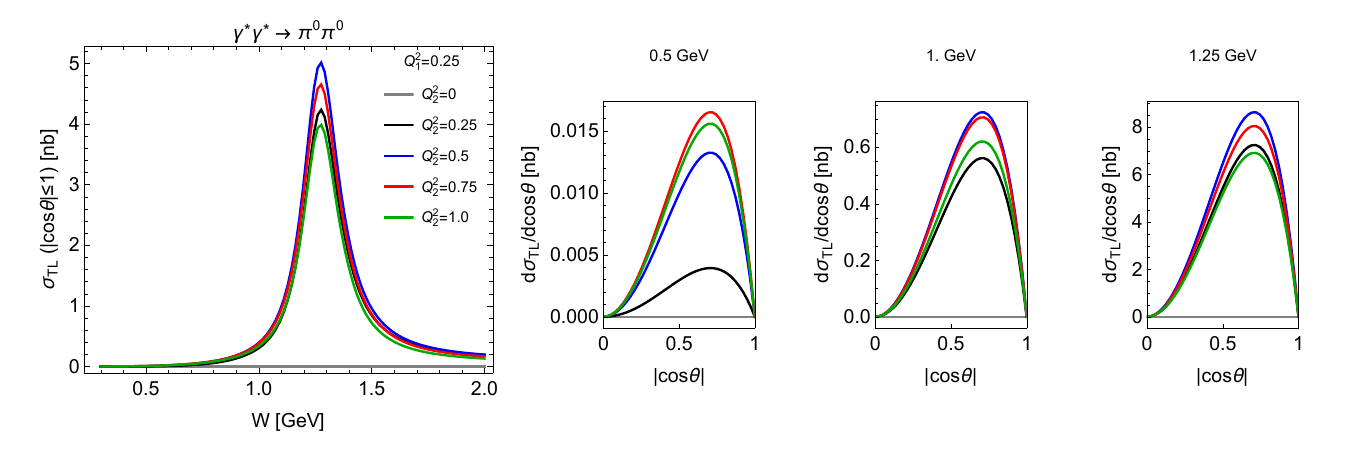}\\
\includegraphics[width=0.96\linewidth]{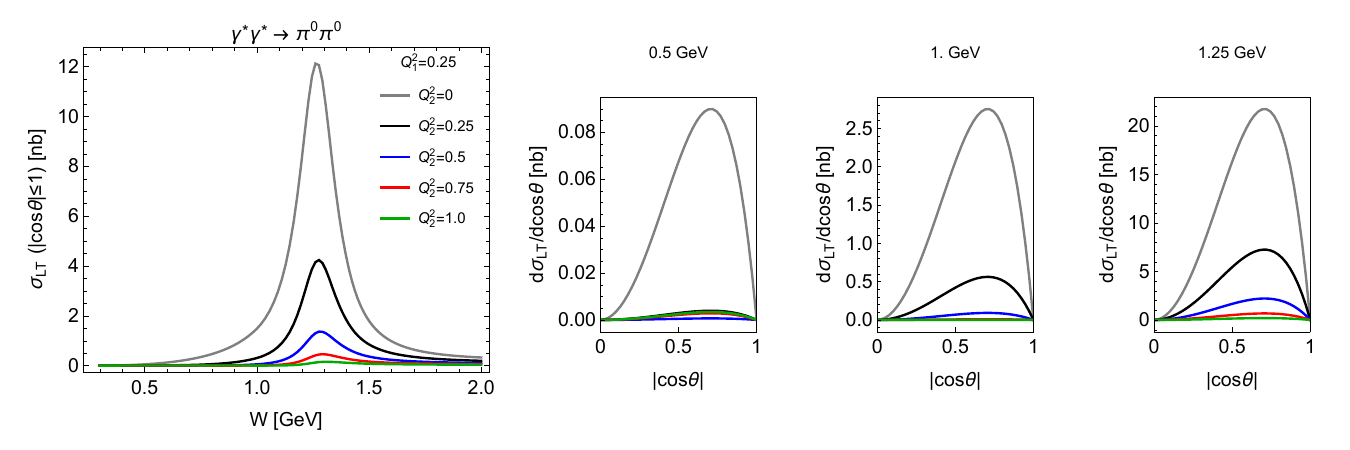}\\
\includegraphics[width=0.96\linewidth]{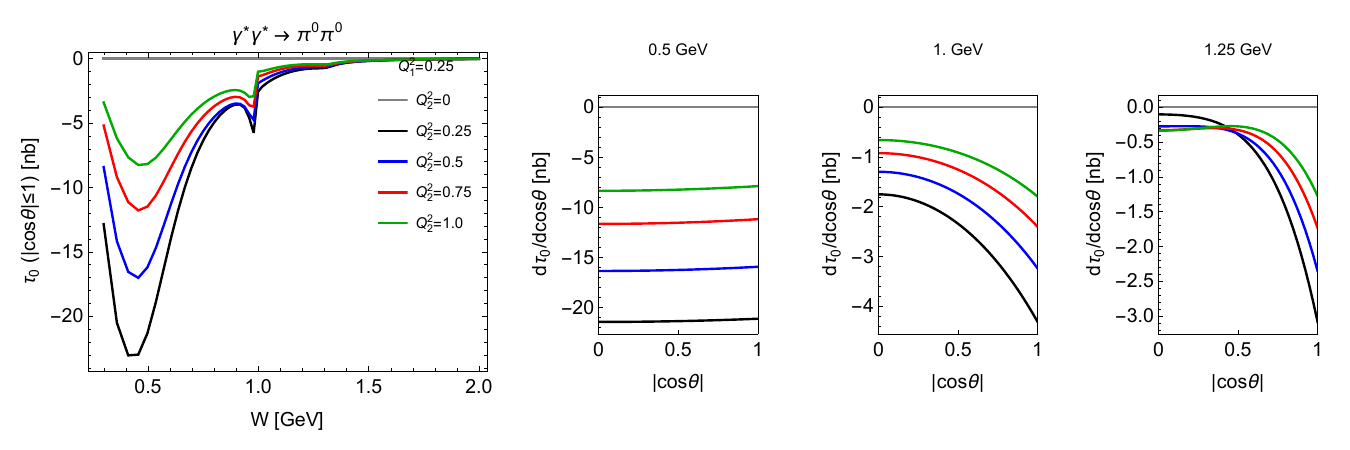}\\
\includegraphics[width=0.96\linewidth]{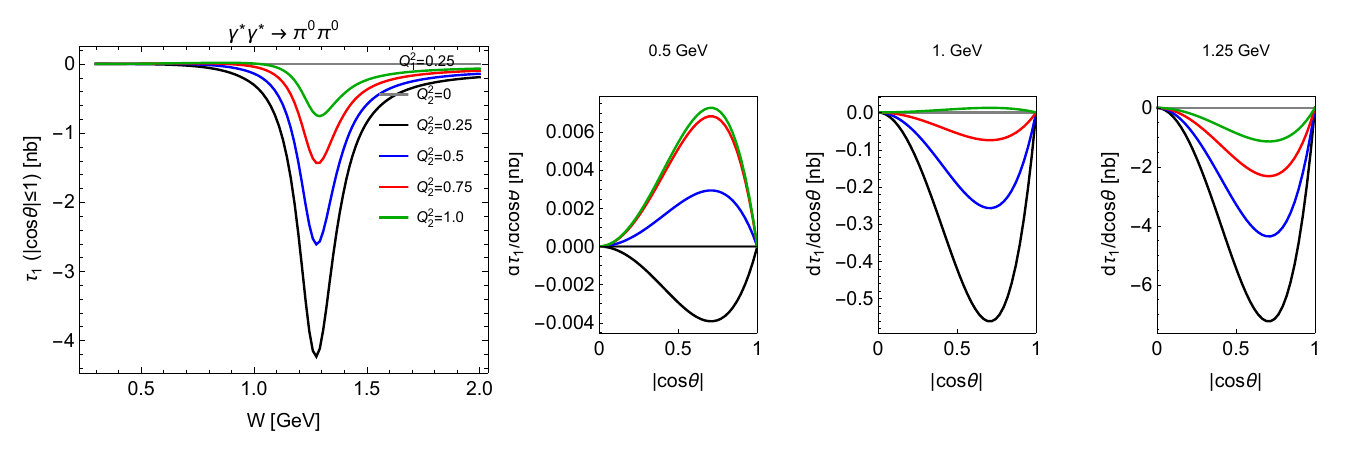}
\caption{Predictions for $\sigma_{TL}$, $\sigma_{LT}$, $\tau_{0}$, and $\tau_{1}$ for $\gamma^* \gamma^* \to \pi^0 \pi^0$ with $Q_1^2 = 0.25 \,\text{GeV}^2$ and $Q_2^2 = 0, 0.25, 0.5, 0.75, 1.0 \,\text{GeV}^2$, shown for full angular coverage ($|\cos\theta| \leq 1$).
\label{fig:pipiQ1Q2_2}}
\end{figure*}

\begin{figure*}[!t]
\centering
\includegraphics[width=0.96\linewidth]{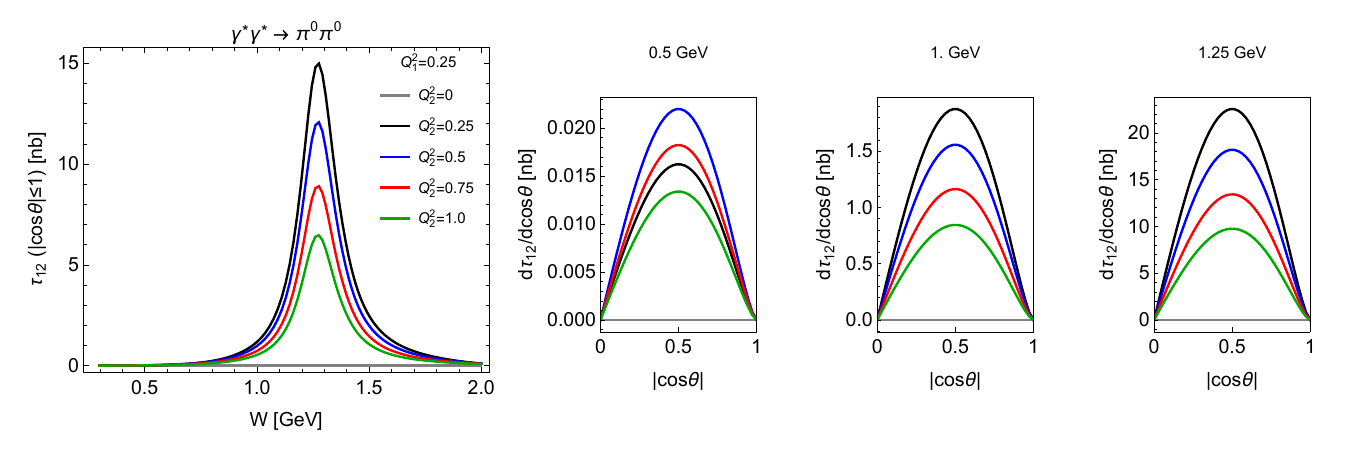}\\
\includegraphics[width=0.96\linewidth]{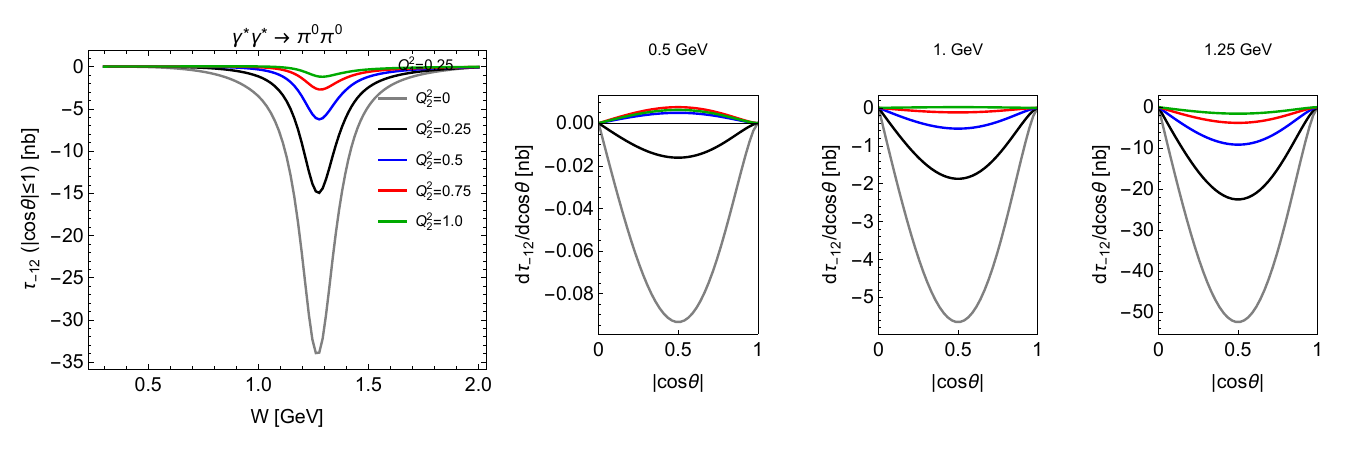}\\
\includegraphics[width=0.96\linewidth]{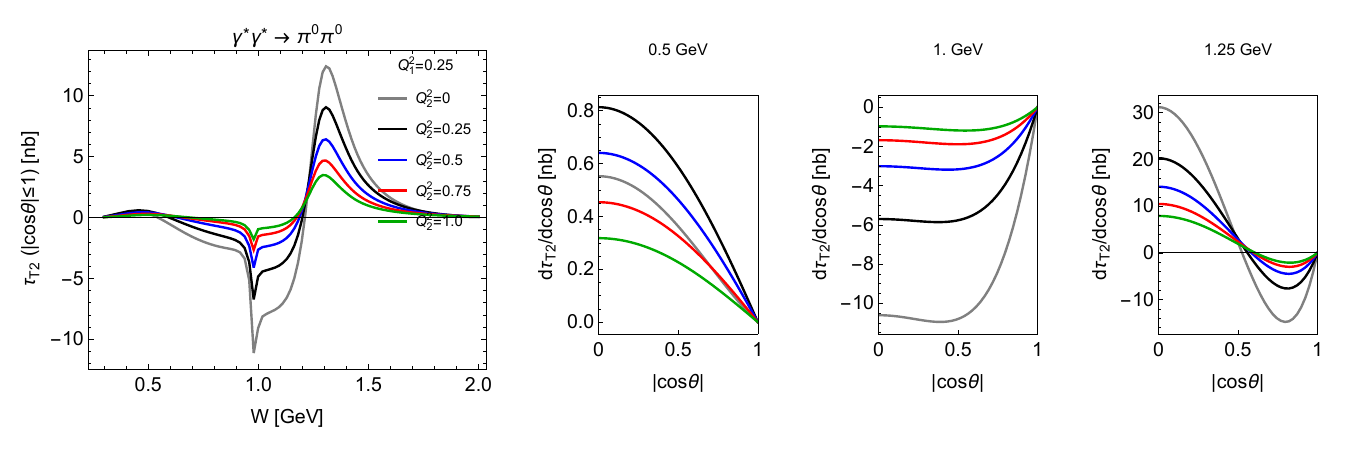}\\
\includegraphics[width=0.96\linewidth]{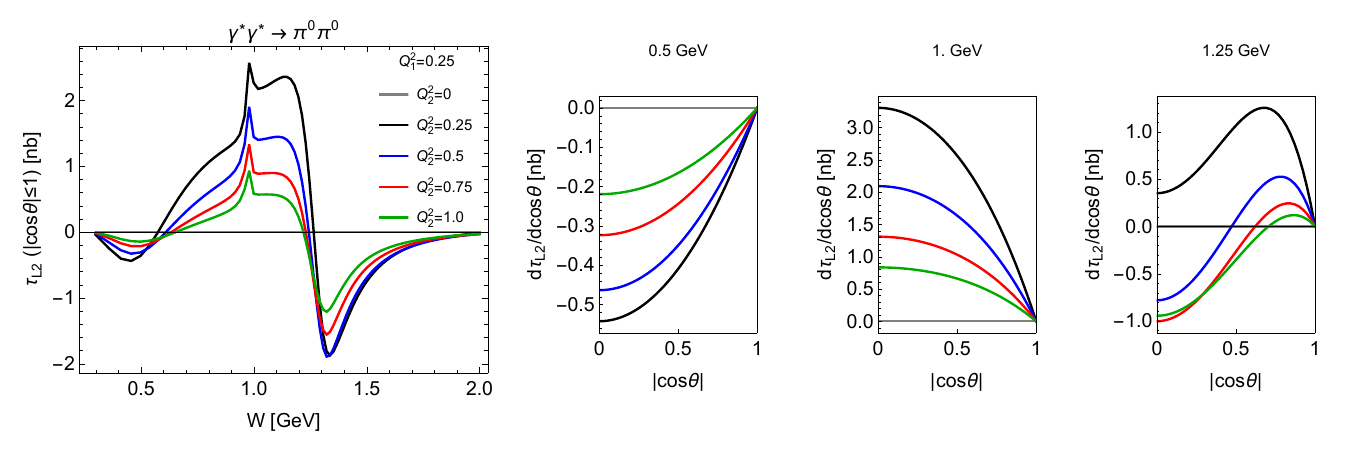}
\caption{Predictions for $\tau_{12}$, $\tau_{-12}$, $\tau_{T2}$, and $\tau_{L2}$ for $\gamma^* \gamma^* \to \pi^0 \pi^0$ with $Q_1^2 = 0.25 \,\text{GeV}^2$ and $Q_2^2 = 0, 0.25, 0.5, 0.75, 1.0 \,\text{GeV}^2$, shown for full angular coverage ($|\cos\theta| \leq 1$).
\label{fig:pipiQ1Q2_3}}
\end{figure*}

\begin{figure*}[!t]
\centering
\includegraphics[width=0.96\linewidth]{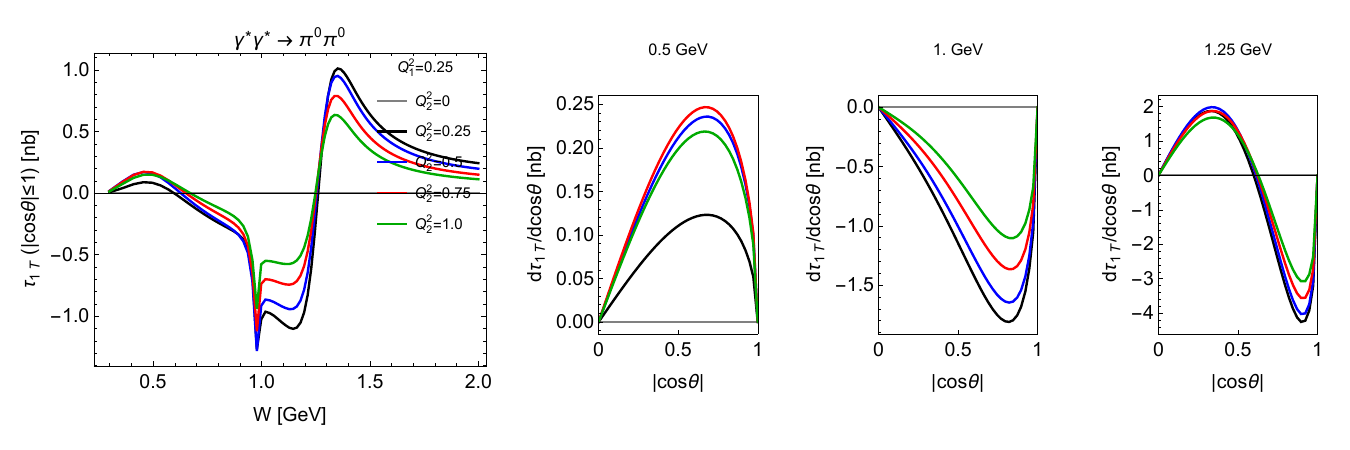}\\
\includegraphics[width=0.96\linewidth]{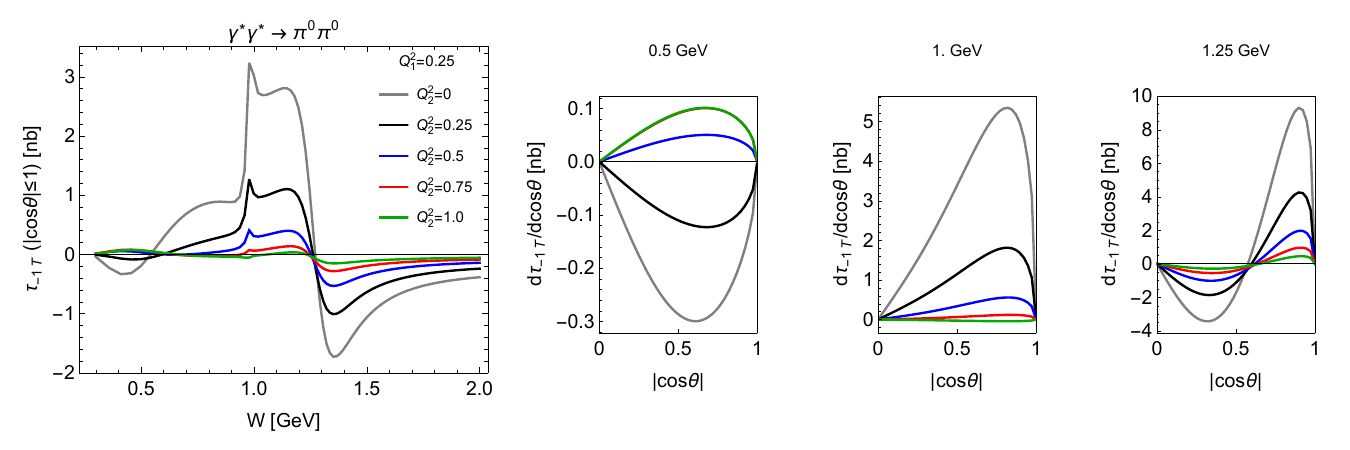}\\
\includegraphics[width=0.96\linewidth]{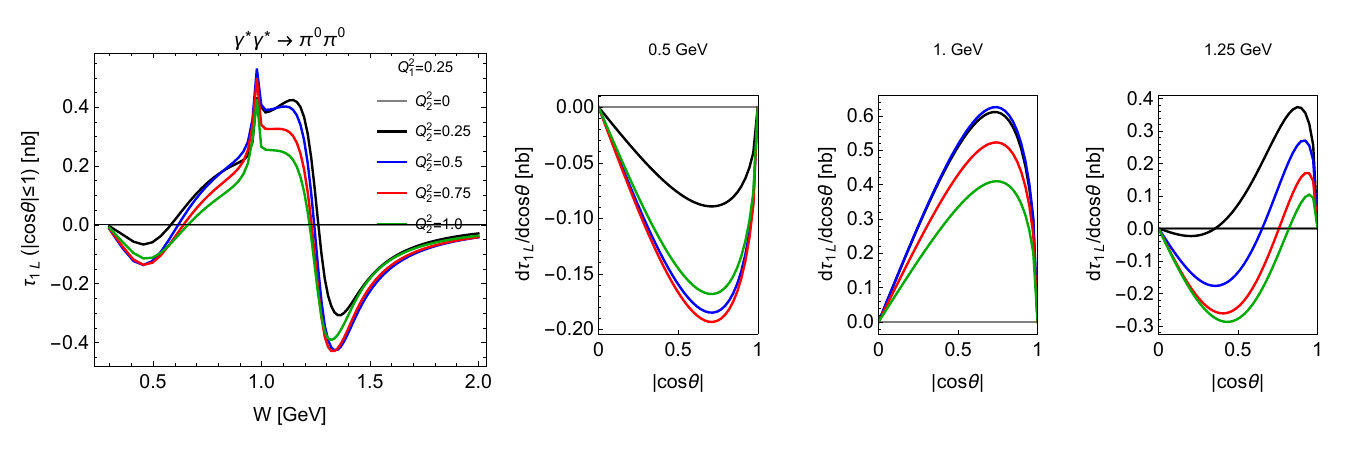}\\
\includegraphics[width=0.96\linewidth]{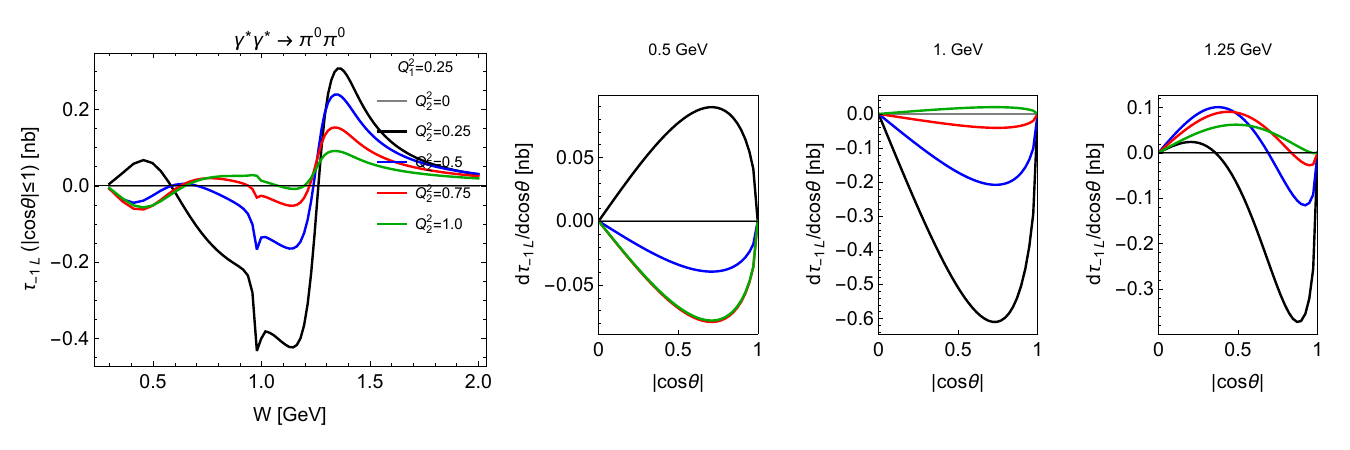}
\caption{Predictions for $\tau_{1T}$, $\tau_{-1T}$, $\tau_{1L}$, and $\tau_{-1L}$ for $\gamma^* \gamma^* \to \pi^0 \pi^0$ with $Q_1^2 = 0.25 \,\text{GeV}^2$ and $Q_2^2 = 0, 0.25, 0.5, 0.75, 1.0 \,\text{GeV}^2$, shown for full angular coverage ($|\cos\theta| \leq 1$).
\label{fig:pipiQ1Q2_4}}
\end{figure*}

The cross sections are provided in numerical form as functions of $W$, $Q_1^2$, $Q_2^2$, and the scattering angle $|\cos\theta_\pi|$, within the kinematic limits $2m_\pi < W < 2\,\text{GeV}$ and $0 \leq Q_{1,2}^2 \leq 4\,\text{GeV}^2$. The results for the differential cross sections and response functions used in Section 1.1 for $\gamma^* \gamma^* \to \pi^0 \pi^0$ are illustrated in Figs.~\ref{fig:pipiQ1Q2_1}, \ref{fig:pipiQ1Q2_2}, \ref{fig:pipiQ1Q2_3}, and \ref{fig:pipiQ1Q2_4}. As expected, the dominant contributions are $\sigma_{0}$ and $\sigma_{2}$, which survive in the real-photon limit. In single-virtual kinematics, $\sigma_{TL}$ also becomes sizable. Owing to the new exclusive treatment of the $\pi\pi$ channel presented in Eq.~(\ref{eq:gagapipiunpolcross_singlevirtual}), it will be possible to resolve the dependence on the azimuthal angle through terms proportional to $\cos\tilde{\phi}_1$ and $\cos 2\tilde{\phi}_1$, where the interference responses $\tau_{T2}$ and $\tau_{-12}, \tau_{12}$ emerge as dominant structures. For differential responses that involve helicity flip, the differential distributions vanish in the forward and backward directions and, in some cases, also at $\cos\theta_\pi = 0$. The latter is a direct consequence of the odd helicity-flip nature of the entering amplitudes.

Cross sections at any given kinematic point are obtained by interpolating the theoretical input using three methods: linear spline interpolation, a nearest-neighbour approach, and an inverse-distance weighted average. The primary method is a linear spline interpolation of the nearest neighbors across all four dimensions ($W$, $Q_1^2$, $Q_2^2$, $\cos\theta_\pi$), which is validated against a simple nearest-neighbor algorithm and an inverse distance–weighted average, where the nearest points in all dimensions are averaged, weighted by the inverse of their Euclidean distance raised to a user-defined power. While all three approaches yield comparable results, linear spline interpolation consistently demonstrates superior stability with the available input and is therefore recommended.


To generate events beyond the theory range, the user can extrapolate the theory curves. The cross section is assumed to decrease with a $1/W^2$ behavior, while the large $Q^2$ behavior is modeled using the form factor frameworks discussed in the subsection on the production through a single resonance (Sec.~\ref{sec::custom_mode}). Both the choice of form factor model and its parameters can be adjusted by the user. We emphasize that the theoretical input is valid only within its specified kinematic range and any extrapolation should therefore be regarded as a rough estimate rather than an accurate prediction.


For the $\pi^+\pi^-$ channel, it should be noted that the simulation currently includes only the two-photon production of pions in the reaction $e^+e^-\to e^+e^-\pi^+\pi^-$. The production of charged pion pairs via the radiation of a virtual photon in Bhabha scattering events is not simulated, although it is expected to play an important role in experimental studies of this channel \cite{Guo:2019gjf}. An updated version of the \textsc{Ekhara} event generator is expected to incorporate both production mechanisms as well as their interference \cite{CzyzPrivate}.

\subsection{$e^+e^-\to e^+e^- \pi^0 \eta$ Mode}
\label{sec:pieta_mode}

For the $S$-wave contribution to $\gamma^*\gamma^* \to \pi^0\eta$, the double-virtual partial waves from the dispersive analysis of~\cite{Deineka:2024mzt} are used, which enforces analyticity, coupled-channel $\pi\eta/K\bar{K}_{I=1}$ unitarity, and the necessary kinematic constraints. This treatment is analogous to the two-pion mode discussed in the previous section, with the only difference being that real-photon data are used to constrain the coupled-channel Omnès representation. The dispersively constructed $S$-wave amplitude is currently employed as input for computing the $a_0(980)$ contribution to hadronic light-by-light (HLbL) scattering in $(g-2)_\mu$~\cite{Aliberti:2025beg}.

The $D$-wave is modeled with a relativistic Breit–Wigner representation of the $a_{2}(1320)$, including an energy-dependent width and $L=2$ Blatt–Weisskopf barrier factors. To extend the result for the real-photon production of a tensor meson ($T$) of Ref.~\cite{Deineka:2024mzt} to the double-virtual case, the $\gamma^*\gamma^*  \to T$ Lorentz decomposition is taken from Ref.~\cite{Hoferichter:2020lap}, with the TFFs provided by the quark-model calculation of Ref.~\cite{Schuler:1997yw}, where only $F^T_1(Q_1^2,Q_2^2)$ is nonzero. The normalization $|F^T_1(0,0)|$ is fixed by the on-shell decay width $\Gamma_{a_2\to\gamma\gamma}$, and the sign is chosen by comparing to the dispersive description of $\gamma^*\gamma^*  \to \pi^0\pi^0$ in the vicinity of the $f_2(1270)$. This construction ensures helicity-2 dominance in the real-photon limit and provides a controlled continuation to the double-virtual case. The mass scale is taken as $m_T \equiv m_{a_2}$, as in the original quark-model calculation. This differs from~\cite{Hoferichter:2024bae}, which adopted a VMD scale for the $(g-2)_\mu$ calculation. Currently, tensor-meson contributions to the hadronic light-by-light in $(g-2)_\mu$ are attributed an uncertainty of about 400\% \cite{Aliberti:2025beg}, making the reduction of model dependence in tensor form factors particularly timely.

\begin{figure*}[!t]
\centering
\includegraphics[width=0.96\linewidth]{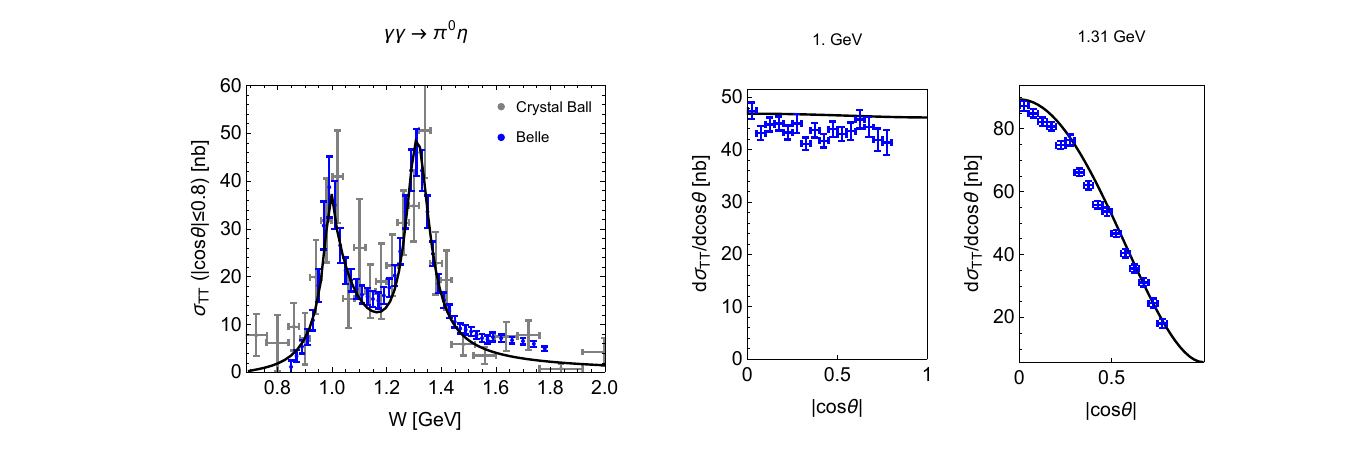}\\
\includegraphics[width=0.96\linewidth]{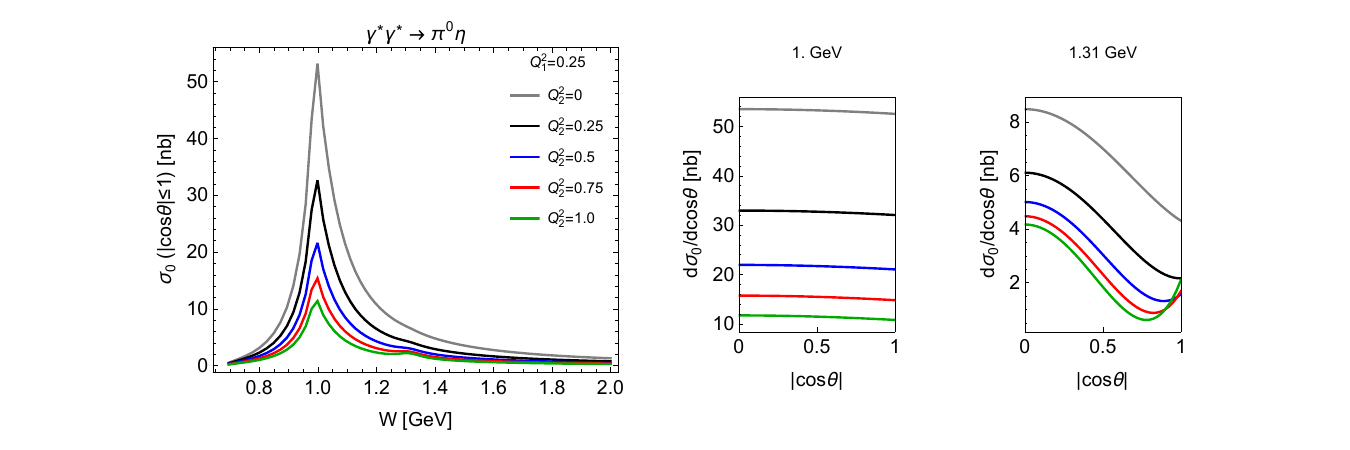}\\
\includegraphics[width=0.96\linewidth]{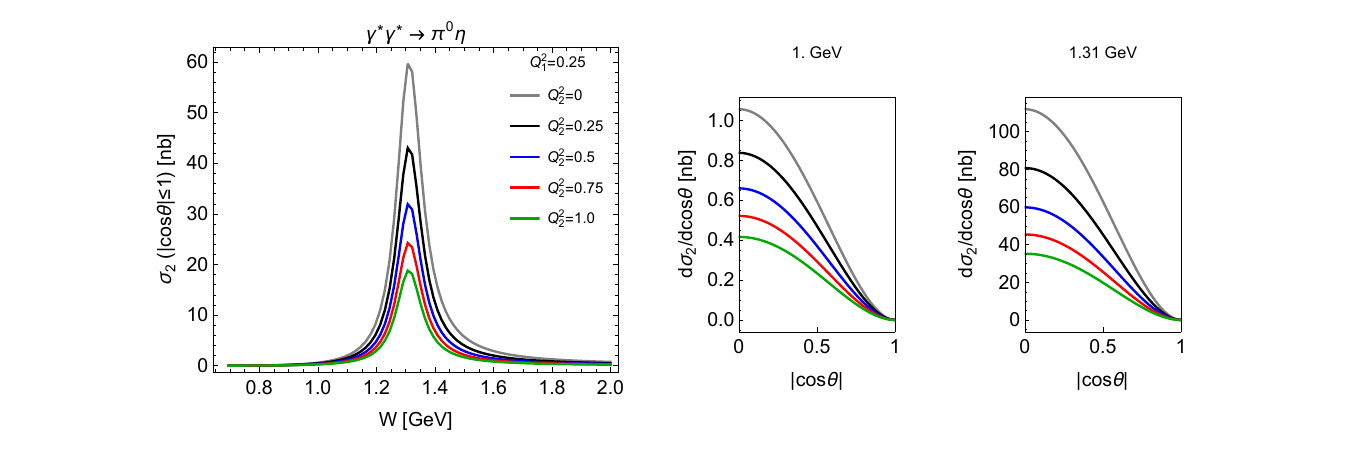}\\
\includegraphics[width=0.96\linewidth]{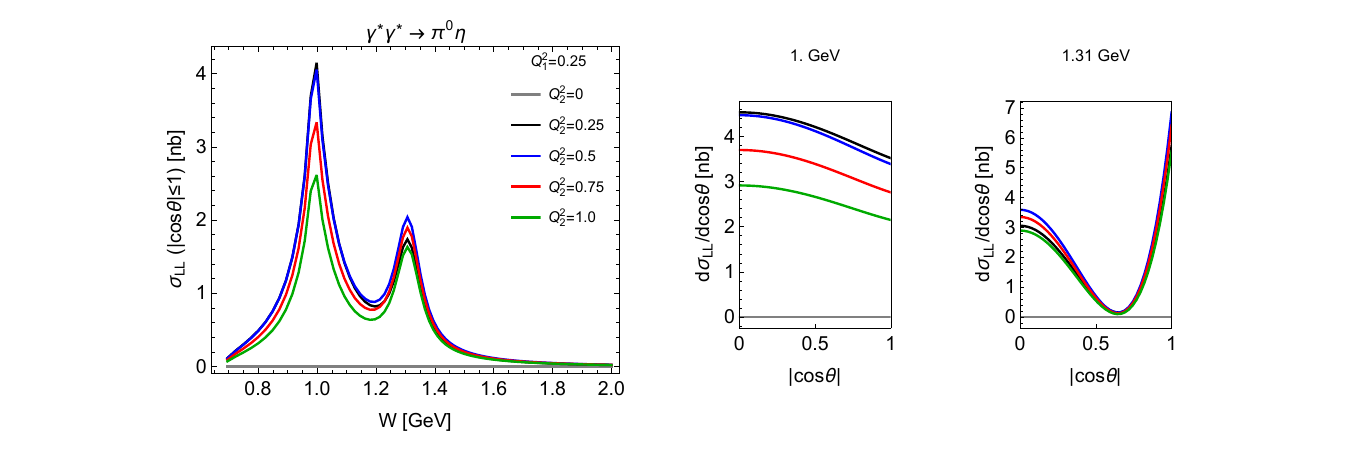}
\caption{Top row of panels: predictions for $\sigma_{TT}$ for  $\gamma \gamma \to \pi^0 \eta$ compared with data. 
Lower panels: predictions for $\sigma_{0}$, $\sigma_{2}$ and $\sigma_{LL}$ for $\gamma^* \gamma^* \to \pi^0 \eta$ with $Q_1^2 = 0.25 \,\text{GeV}^2$ and $Q_2^2 = 0, 0.25, 0.5, 0.75, 1.0 \,\text{GeV}^2$, shown for full angular coverage ($|\cos\theta| \leq 1$). {The experimental data are taken from \cite{Belle:2009xpa,CrystalBall:1985mzc}}
\label{fig:pietaQ1Q2_1}}
\end{figure*}

\begin{figure*}[!t]
\centering
\includegraphics[width=0.96\linewidth]{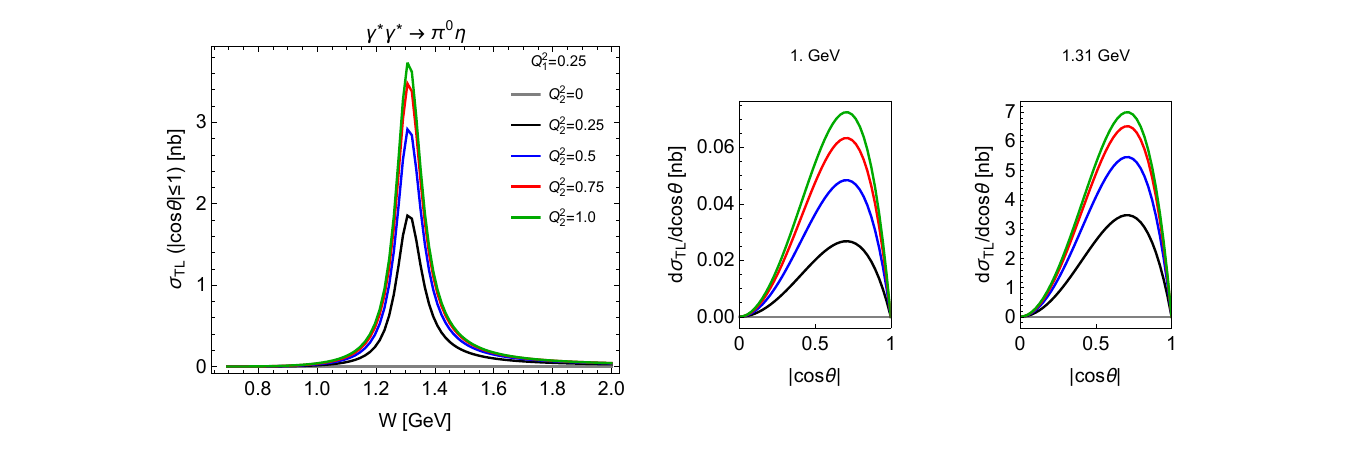}\\
\includegraphics[width=0.96\linewidth]{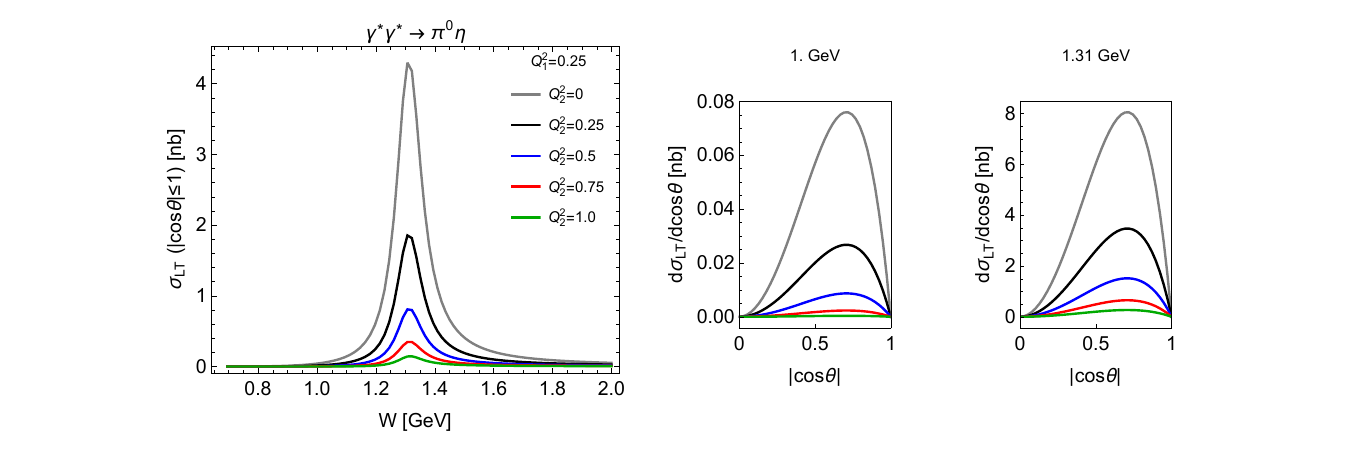}\\
\includegraphics[width=0.96\linewidth]{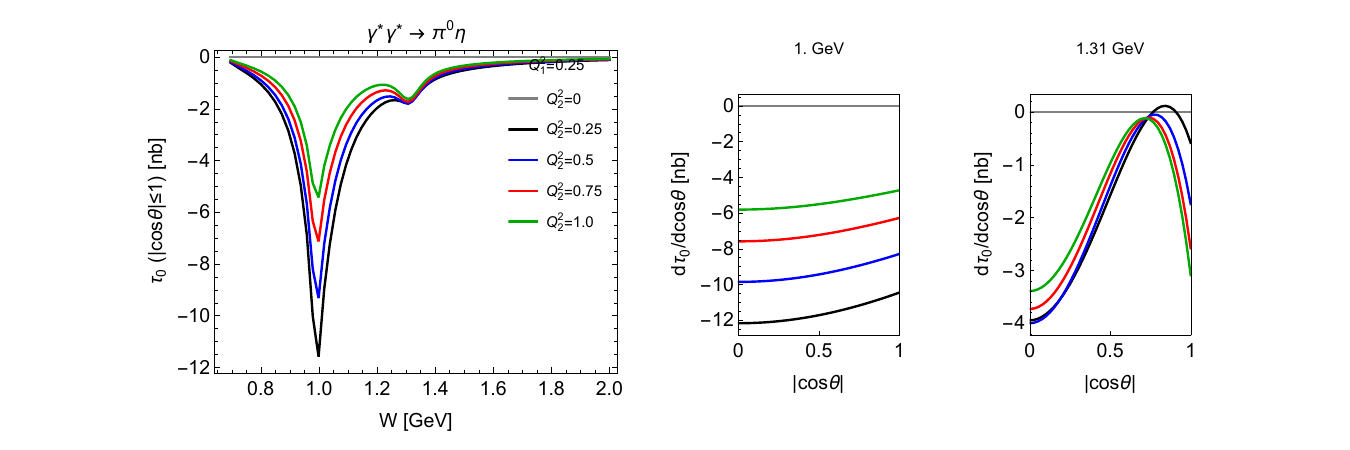}\\
\includegraphics[width=0.96\linewidth]{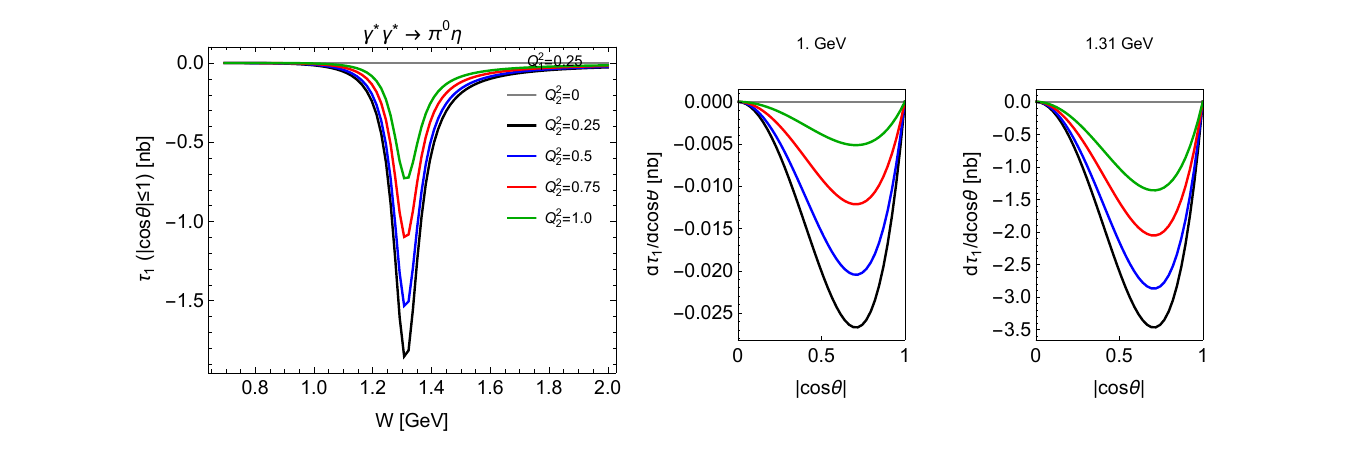}
\caption{Predictions for $\sigma_{TL}$, $\sigma_{LT}$, $\tau_{0}$, and $\tau_{1}$ for $\gamma^* \gamma^* \to \pi^0 \eta$ with $Q_1^2 = 0.25 \,\text{GeV}^2$ and $Q_2^2 = 0, 0.25, 0.5, 0.75, 1.0 \,\text{GeV}^2$, shown for full angular coverage ($|\cos\theta| \leq 1$).
\label{fig:pietaQ1Q2_2}}
\end{figure*}

\begin{figure*}[!t]
\centering
\includegraphics[width=0.96\linewidth]{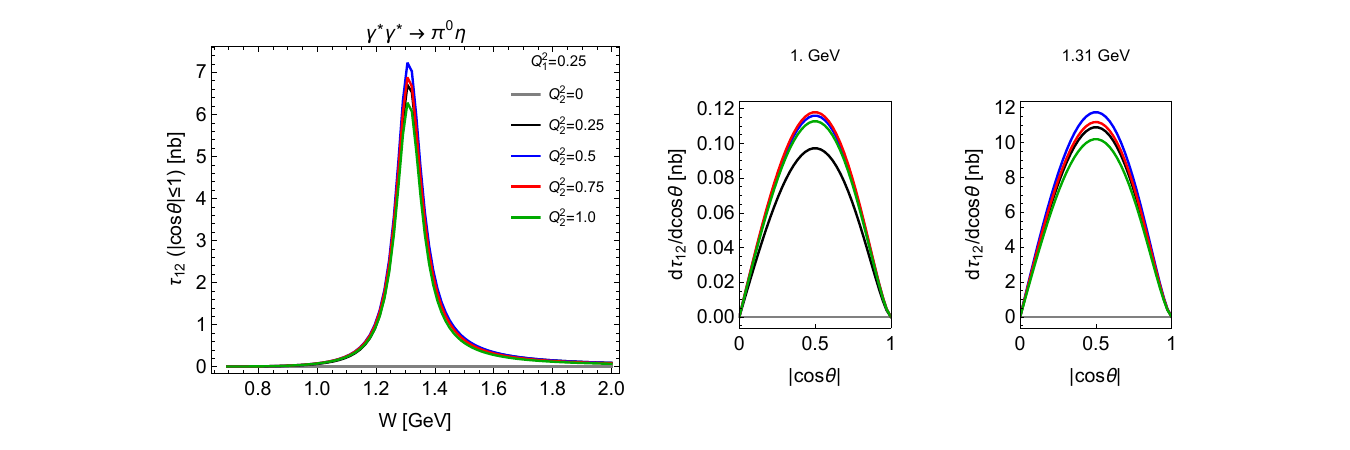}\\
\includegraphics[width=0.96\linewidth]{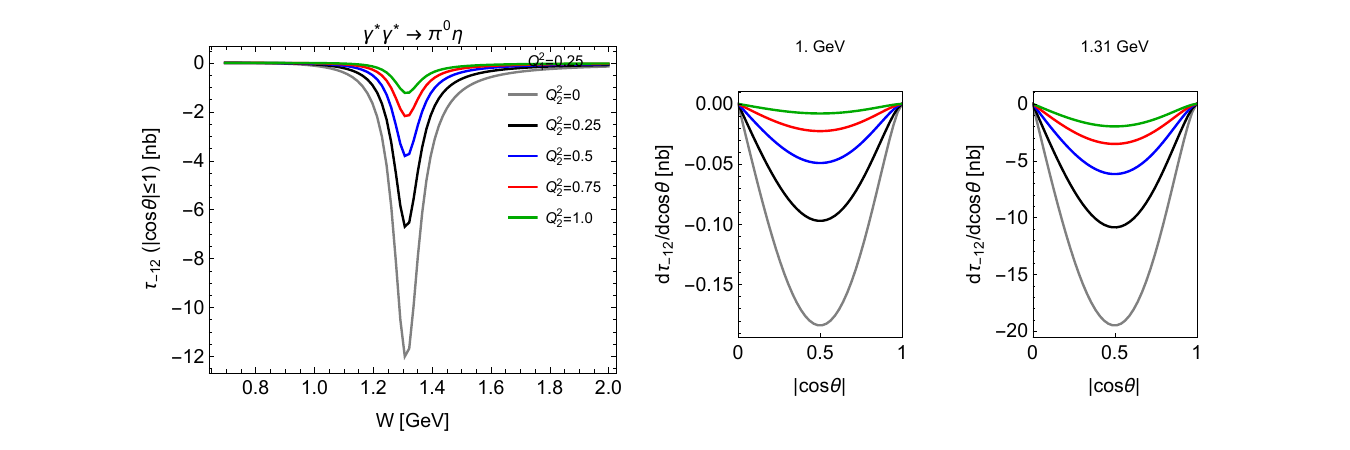}\\
\includegraphics[width=0.96\linewidth]{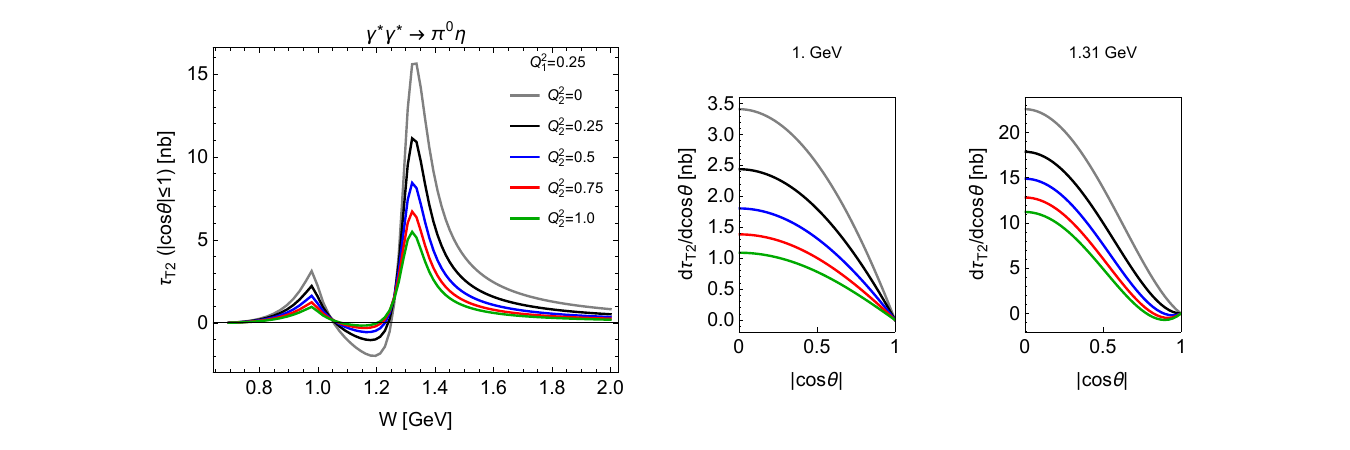}\\
\includegraphics[width=0.96\linewidth]{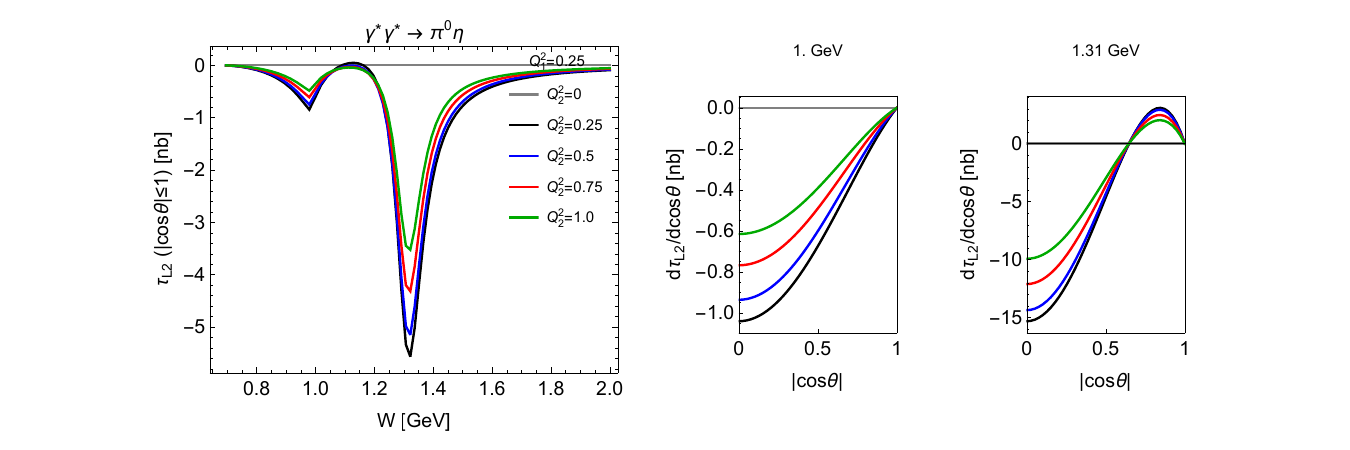}
\caption{Predictions for $\tau_{12}$, $\tau_{-12}$, $\tau_{T2}$, and $\tau_{L2}$ for $\gamma^* \gamma^* \to \pi^0 \eta$ with $Q_1^2 = 0.25 \,\text{GeV}^2$ and $Q_2^2 = 0, 0.25, 0.5, 0.75, 1.0 \,\text{GeV}^2$, shown for full angular coverage ($|\cos\theta| \leq 1$).
\label{fig:pietaQ1Q2_3}}
\end{figure*}

\begin{figure*}[!t]
\centering
\includegraphics[width=0.96\linewidth]{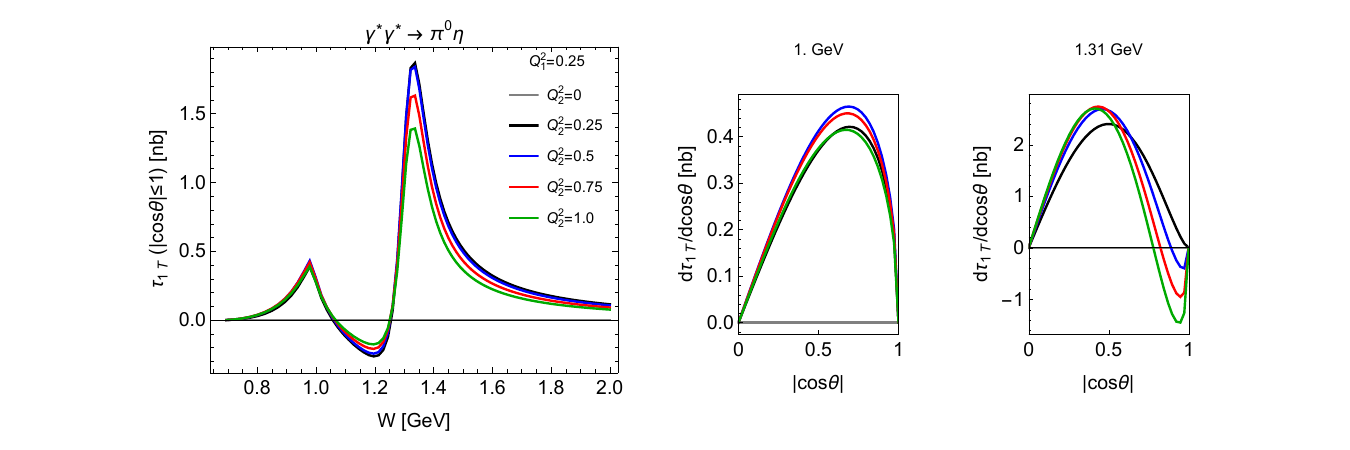}\\
\includegraphics[width=0.96\linewidth]{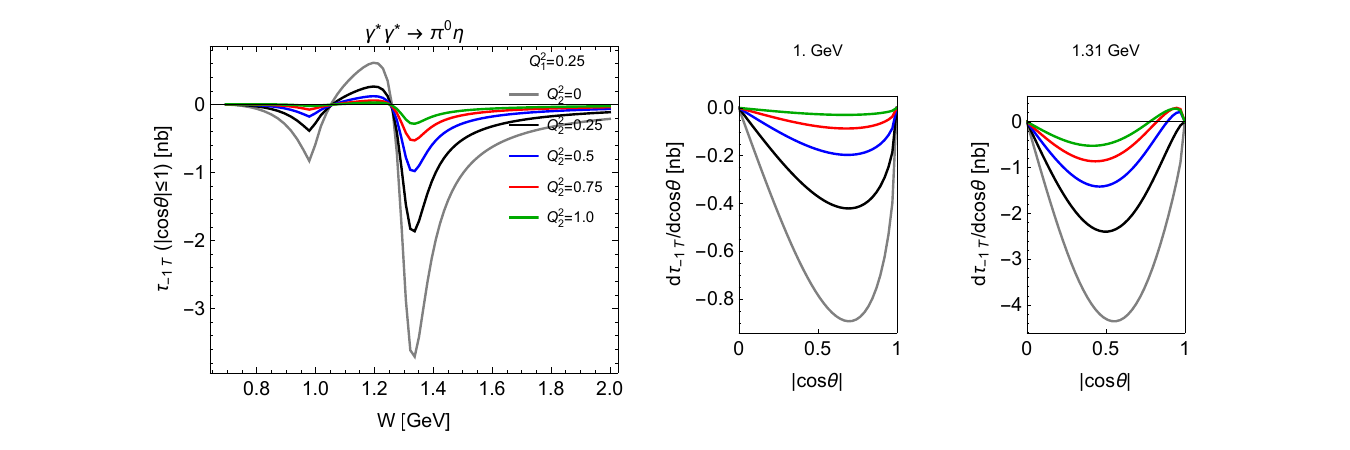}\\
\includegraphics[width=0.96\linewidth]{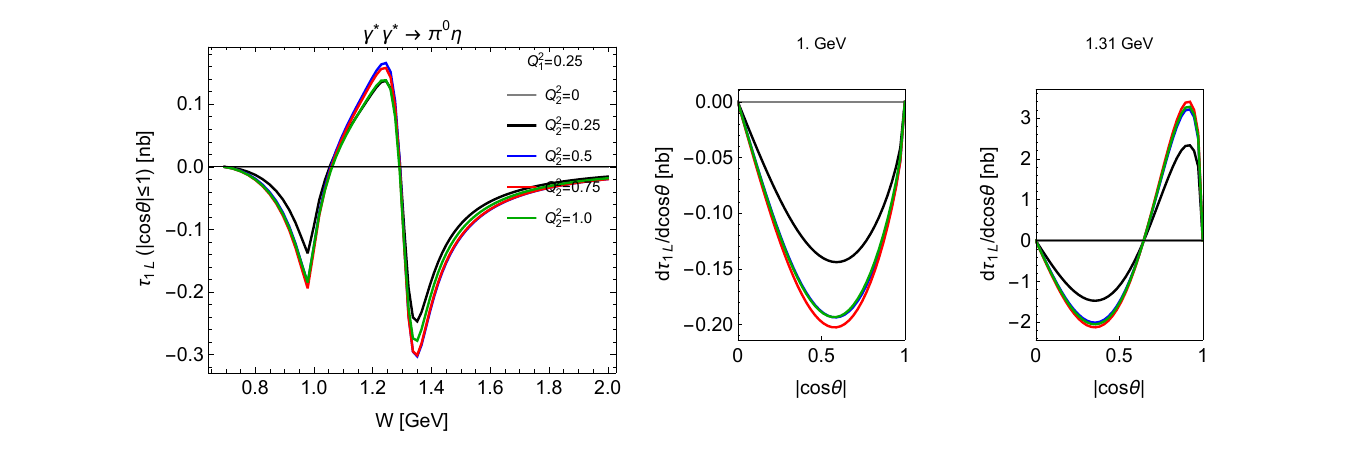}\\
\includegraphics[width=0.96\linewidth]{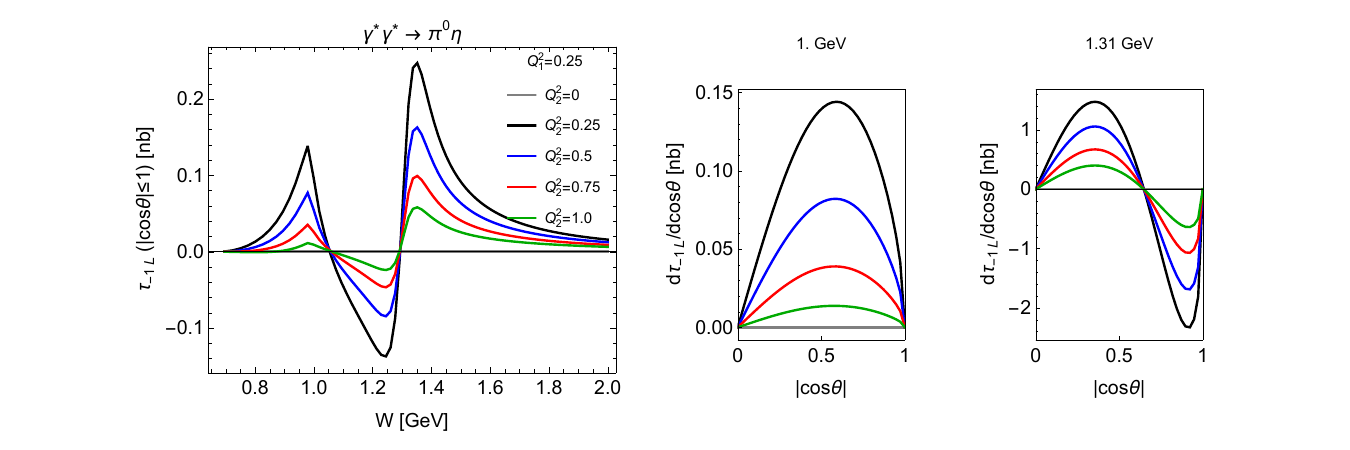}
\caption{Predictions for $\tau_{1T}$, $\tau_{-1T}$, $\tau_{1L}$, and $\tau_{-1L}$ response functions for $\gamma^*\gamma^* \to \pi^0\eta$ for $Q_1^2=0.25$ GeV$^2$ and $Q_2^2=0, 0.25,0.5,0.75,1.0$ GeV$^2$ and for full angular coverage $|\cos\theta| \leq 1$.
\label{fig:pietaQ1Q2_4}}
\end{figure*}

This $\gamma^*\gamma^* \to \pi^0\eta$ input yields a reliable description of the $a_{0}(980)$ and $a_2(1320)$ regions and can be directly interpolated on a $(W,Q_1^2,Q_2^2)$ grid using the same interpolation and extrapolation algorithms as discussed for the $\gamma^\ast\gamma^\ast\to \pi^0\pi^0$ case. The $\gamma^*\gamma^* \to \pi^0\eta$ response functions are shown in Figs.~\ref{fig:pietaQ1Q2_1}, \ref{fig:pietaQ1Q2_2}, \ref{fig:pietaQ1Q2_3}, and \ref{fig:pietaQ1Q2_4} as functions of the momentum transfers $Q_{1,2}^2$ in the range $0-2$ GeV$^2$, and invariant masses from threshold up to $2$ GeV.

\subsection{$e^+e^-\to e^+e^-K^+K^-$, $e^+e^-\to e^+e^-K^0_SK^0_S$, and $e^+e^-\to e^+e^-\eta\eta$ Modes}
\label{sec::kk_etaeta_mode}
Combined dispersive analyses discussed in Secs.~\ref{sec:pipi_mode} and ~\ref{sec:pieta_mode}, can provide $Q^2$-dependent estimates for the $K\bar{K}$ channel, but are presently restricted to invariant masses below 1.4\,GeV \cite{Deineka:2024mzt}.
To extend the generator coverage, it is also possible to employ experimental data as direct input for the generator. 

The modes $e^{+}e^{-}\to e^{+}e^{-}K^{+}K^{-}$, $e^{+}e^{-}\to e^{+}e^{-}K_{S}^{0}K_{S}^{0}$, and $e^{+}e^{-}\to e^{+}e^{-}\eta\eta$ are thus modeled using available experimental results from Belle and preliminary data from BESIII. Since both collaborations have performed untagged measurements with $Q_1^2 = Q_2^2 \approx 0$, only the TT-polarized cross sections are accessible, and information on the momentum-transfer dependence is not available. Nonetheless, such estimates are experimentally valuable: for example, the ongoing BESIII initial state radiation determination of the kaon form factor suffers from background due to two-photon production of charged kaon pairs in $e^+e^-\to e^+e^-K^+K^-$ \cite{BESIII:ISRKK}. Providing cross section inputs for $e^+e^-\to e^+e^-K^+K^-$ and the related $K_S^0K_S^0$ and $\eta\eta$ channels based on available untagged data thus aids background modeling. 


The Belle collaboration investigated the two-photon production of $K^+K^-$ \cite{Belle:2003xlt,Belle:2004bpk}, $K^0_SK^0_S$ \cite{Belle:2013eck}, and $\eta\eta$ \cite{Belle:2010ckn}, covering invariant-mass ranges of $1.4 - 4.1$\,GeV, $1.5 - 4.0$\,GeV, and  $1.1 - 3.8$\,GeV, respectively. The measurements cover different angular regions: $|\cos\theta_K| < 0.6$ for $K^+K^-$, $|\cos\theta_K| < 0.8$ or $|\cos\theta_K| < 0.6$ for $K_SK_S$, and $|\cos\theta_\eta| < 0.9$ or with full angular coverage for $\eta\eta$. Furthermore, the BESIII collaboration has released preliminary results of a coupled-channel PWA of the $\pi^0\pi^0$, $K^+K^-$, and $\pi^0\eta$ final states for masses below 2\,GeV and within the angular range $|\cos\theta_{\pi,\eta,K}|<0.8$. These results can be extrapolated to full angular coverage \cite{Kussner:2024ryb,Kuessner2022}. 

For the $K^ +K^-$ channel, the preliminary BESIII results are used up to masses of 2\,GeV as they cover the threshold region. At higher masses, the Belle results are employed. For the $K_SK_S$ and $\eta\eta$ modes, the available Belle data are used exclusively.

Since the Belle measurements do not cover the full angular range and the collaboration has not supplied results extended to full coverage, an extrapolation must be implemented. To achieve this, the differential cross section is described by
\begin{equation} 
    \frac{\diff\sigma}{\diff |\cos\theta_{h}|} \propto \left| S(W)  Y^0_0(\theta_{h},0) + D_0(W)  Y^0_2(\theta_{h},0) e^{i\phi(W)} \right|^2 + \left| D_2(W)Y^2_2(\theta_{h},0) \right|^2\,,\quad h=K,\eta\,,
    \label{eq:xs_fit_curve} 
\end{equation} 
where $Y^m_n(\theta_h,\phi)$ are spherical harmonics, $S$, $D_0$, and $D_2$ denote the relative strengths of the corresponding amplitudes, and $\phi(W)$ denotes the relative phase between the $S$ and $D_0$ amplitudes at a given invariant mass $W$. Since HadroTOPS does not require knowledge of the actual intermediate states, a quantitative fit at each energy point is sufficient to determine the numerical input.

\begin{figure}[b!]
    \centering
    \includegraphics[width=\linewidth]{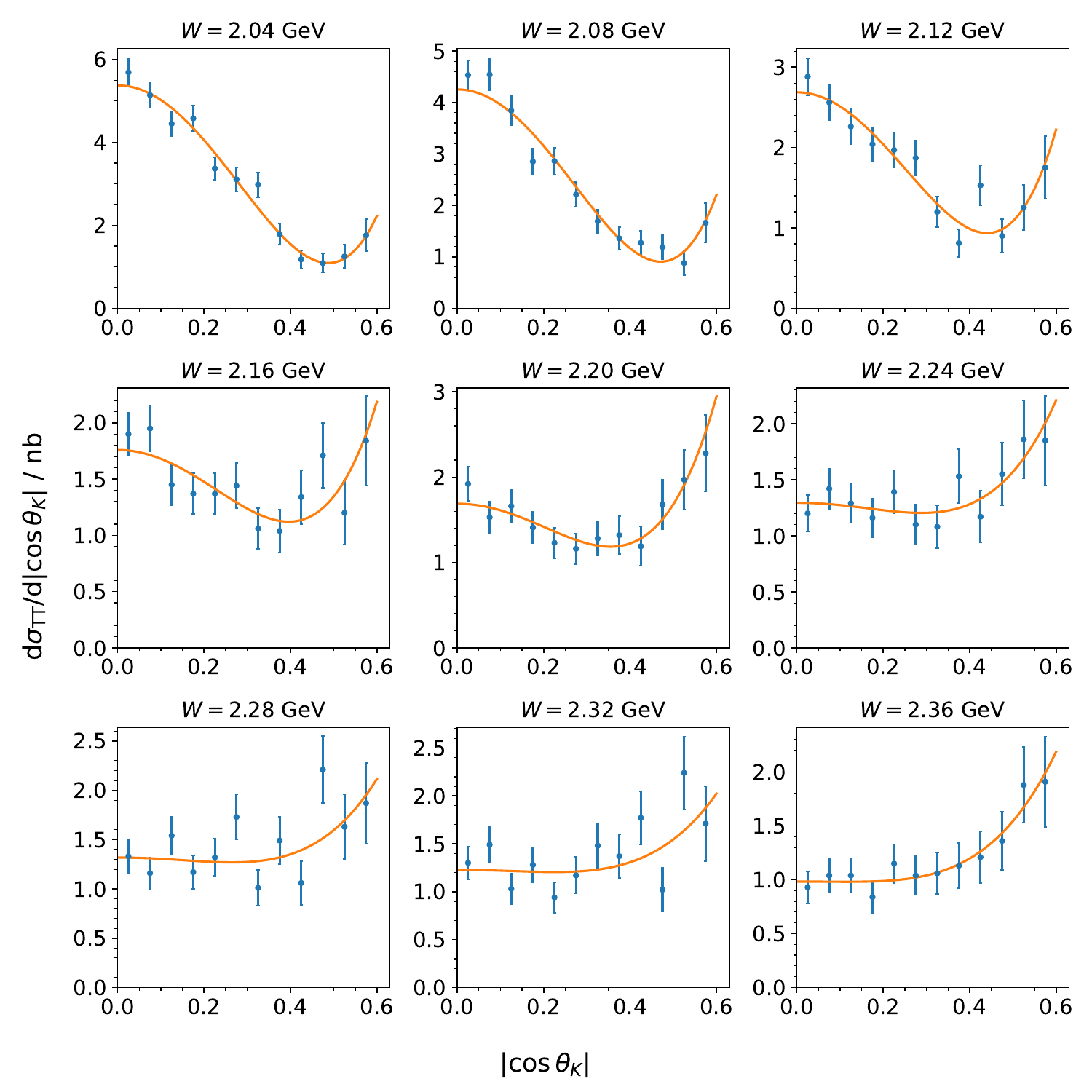}
    \caption{Fit of the cross section model (Eq.~\ref{eq:xs_fit_curve}) to the Belle $\gamma\gamma\to K^+K^-$ cross sections as a function of the helicity angle $|\cos\theta_K|$. The blue points show the Belle data from Ref.~\cite{Belle:2003xlt}, while the orange curve denotes the fit result. The different panels correspond to different $K^+K^-$ invariant masses. }
    \label{fig:KpKm_fit}
\end{figure}

Figure~\ref{fig:KpKm_fit} presents the fit results for the Belle measurement of the $K^+K^-$ cross section below $W < 2.4$\,GeV. Equation~(\ref{eq:xs_fit_curve}) reproduces the data within the quoted experimental uncertainties. The same fitting procedure is applied to the higher-mass $K^+K^-$ data from Belle \cite{Belle:2004bpk}, as well as to the $K^0_SK^0_S$ and $\eta\eta$ cross sections. 

\begin{figure}[p]
    \centering
    \includegraphics[width=0.9\linewidth]{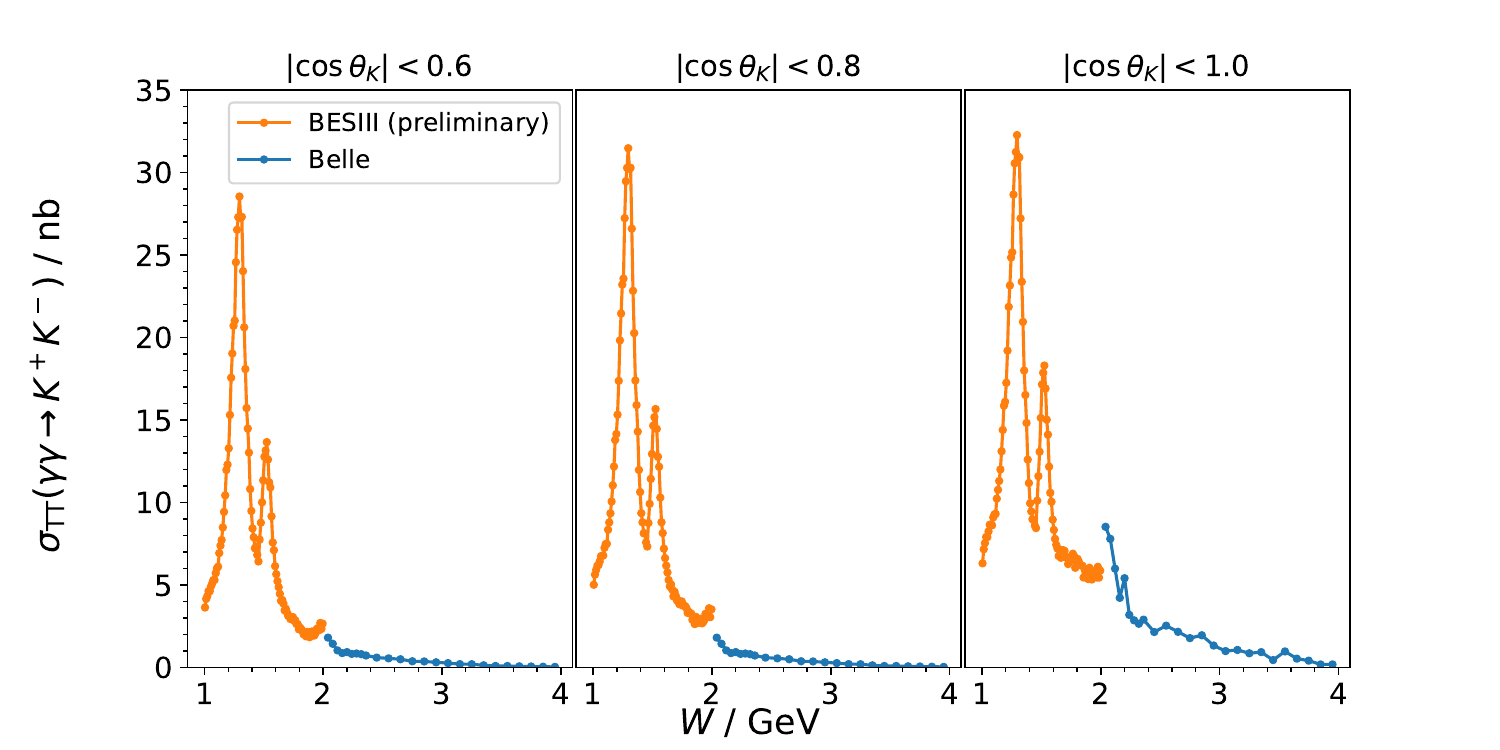}
    \includegraphics[width=0.9\linewidth]{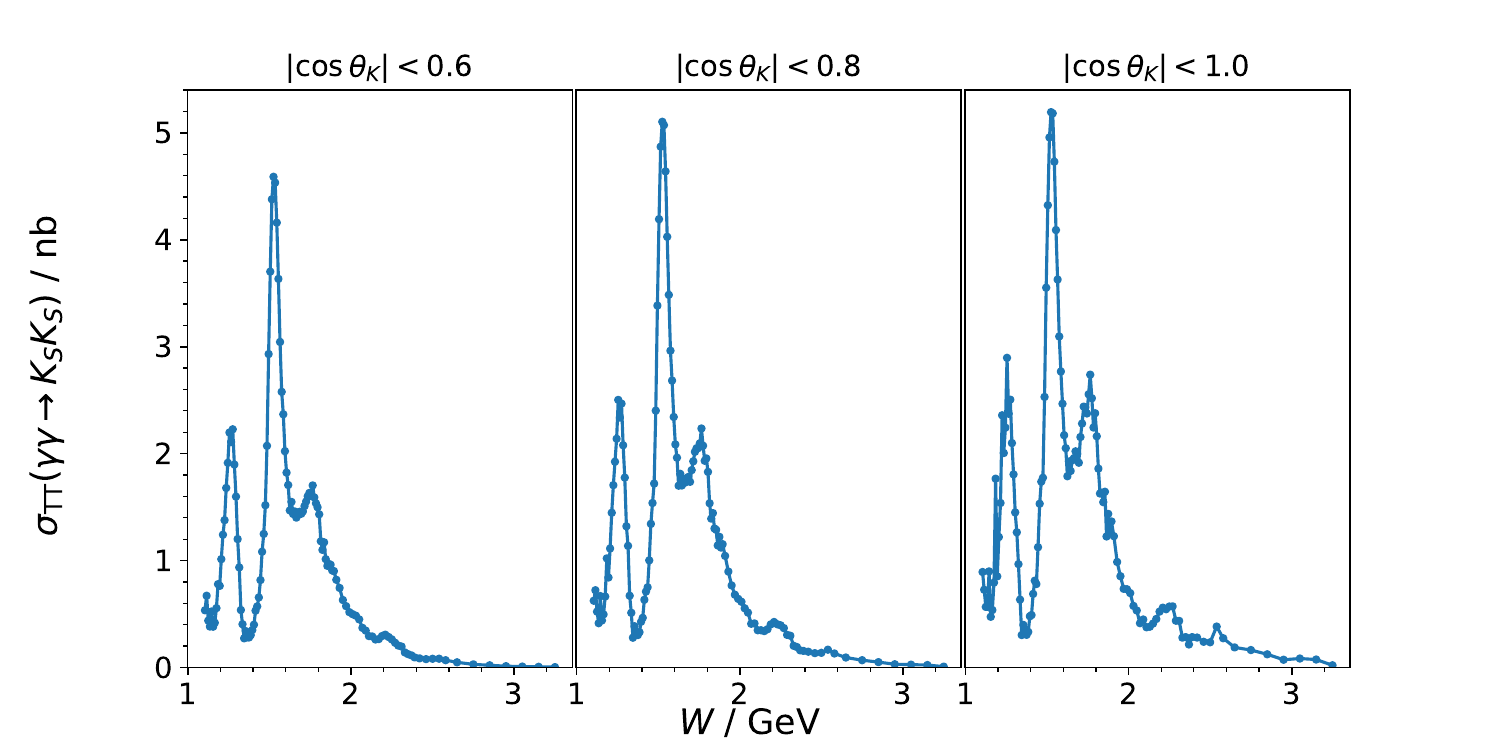}
    \includegraphics[width=0.9\linewidth]{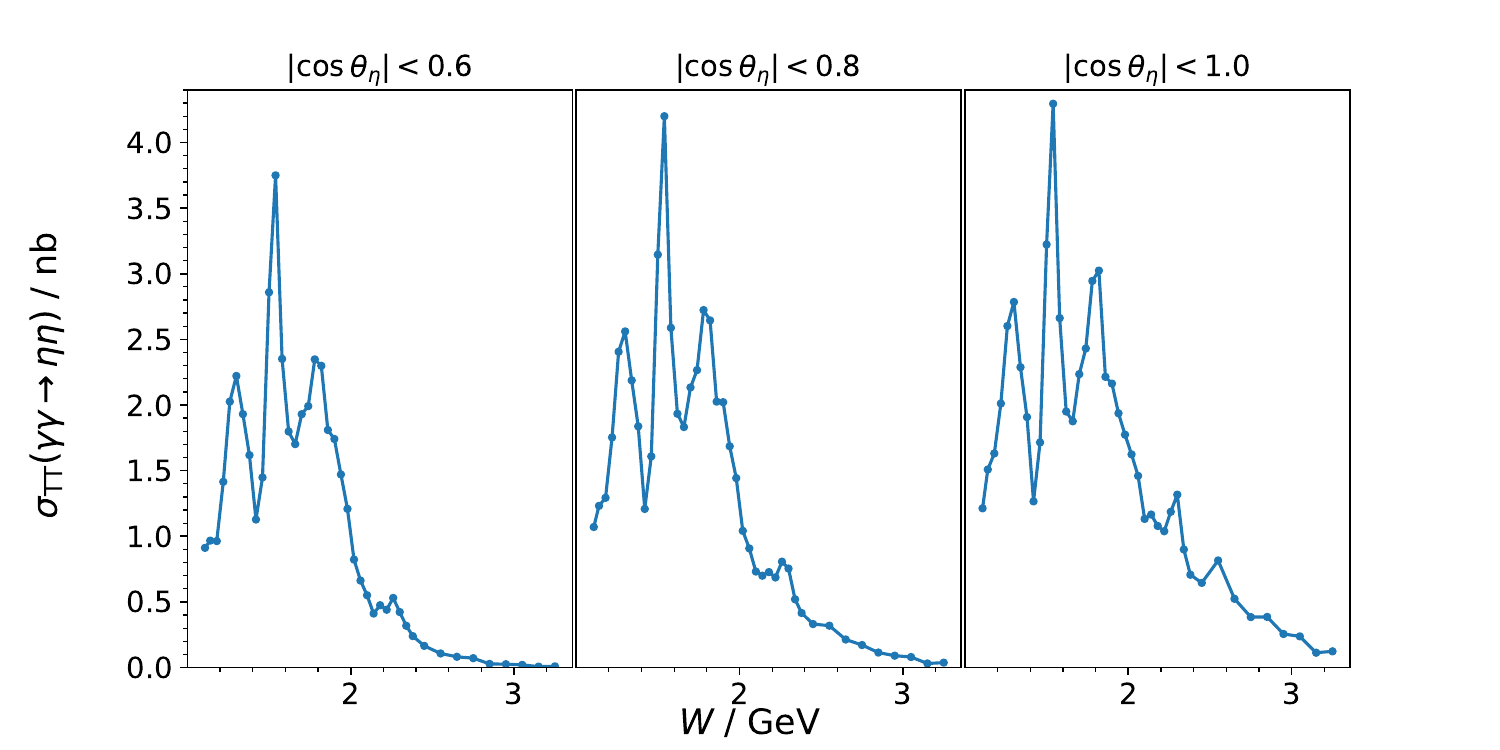}
    \caption{Integrated BESIII PWA (orange) and fits to the $\gamma\gamma\to K^+K^-$ (top), $\gamma\gamma\to K^0_SK^0_S$ (center) and $\gamma\gamma\to \eta\eta$ (bottom) cross sections (blue) over different helicity angle ranges. The points (crosses) represent the actual mass points, the lines are linear interpolation between the results.}
    \label{fig:xs_extrapolations}
\end{figure}

Figure~\ref{fig:xs_extrapolations} shows the numerical integration of the differential cross section over various helicity angle ranges. The integration of the fits to the Belle data shows larger scattering when extended beyond the range of the experimental input. This is particularly evident in the intermediate mass region ($W < 2.4$\,GeV) and $|\cos\theta_K| < 1$ for the $K^+K^-$ channel, which is constrained by limited statistics and helicity angle coverage.

\newpage
\subsection{$e^+e^-\to e^+e^-f_1(1285)\to e^+e^-\eta\pi^+\pi^-$ Mode}
\begin{figure}[h!]
    \centering
    \includegraphics[width=0.5\linewidth]{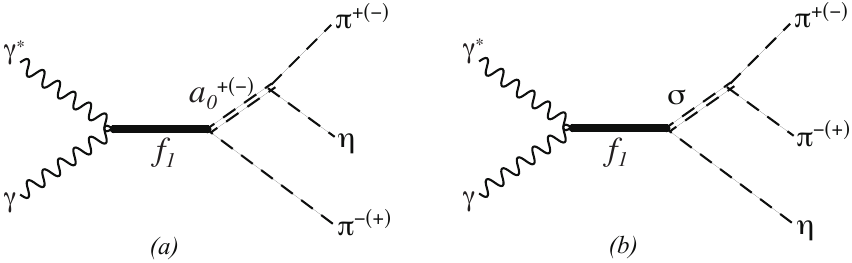}
    \caption{Feynman diagrams of the $\gamma^*\gamma \to \eta\pi^+\pi^-$ process via the $f_1(1285) \to a_0(980)^\pm \pi^\mp$ and $f_1(1285)\to f_0(500)\eta$ decays.}
    \label{fig:f1Feyn}
\end{figure}

In addition to the two-body modes, \HTP can simulate the two-photon production of the axial-vector state $f_1(1285)$ and its subsequent decay into $\eta \pi^+ \pi^-$ via the intermediate states $a_0^\pm(980) \pi^\mp$ and $f_0(500)\eta$. The theoretical basis for this process follows Ref.~\cite{Ren:2024uui}, where the single-virtual process $\gamma^*\gamma \to \eta \pi^+\pi^-$ is studied within an effective Lagrangian framework. As illustrated in Fig.~\ref{fig:f1Feyn}, the two dominant channels $\gamma^*\gamma\to f_1(1285)\to a_0(980)^\pm \pi^\mp$ and $\gamma^*\gamma\to f_1(1285)\to f_0(500)\eta$ are included. The total two-photon amplitude is given by
\begin{equation} 
\mathcal{M} = \mathcal{M}_{a_0^+(980)\pi^-} + \mathcal{M}_{a_0^-(980)\pi^+} + \mathcal{M}_{f_0(500)\eta} \,e^{i\phi}\,,
\end{equation}
where the relative phase $\phi$ controls the interference between the channels. 
Motivated by the L3 measurement~\cite{L3:2001cyf}, Ref.~\cite{Ren:2024uui} assumes fully destructive interference with $\phi=180^\circ$. {The couplings  $g_{f_1a_0\pi}$, $g_{a_0\eta\pi}$, and $g_{f_1 f_0(500)\eta}$, are assumed to be real and are fixed from the corresponding decay widths. The $f_{1}(1285)$ and $a_0(980)^\pm$ propagators are modeled by Breit-Wigner forms with energy-dependent widths, and the p-wave vertices are supplemented by the Blatt-Weisskopf factor. For the decay $f_0(500)\to \pi^+\pi^-$, the $I=0$ $s$-wave Omn\`{e}s function function is used to account for $\pi\pi$ rescattering effects.
}

The model predicts invariant-mass spectra, angular distributions, and total cross sections. In the current implementation, the user has a full flexibility to specify coupling constants, branching fractions, scaling factors, and phase $\phi$. Default values correspond to those used in Ref.~\cite{Ren:2024uui}.

In the current implementation, the framework is extended to account for a small but nonzero second virtuality $Q_2^2$ (which is assumed to be smaller than $Q_1^2$), thereby generalizing the original single-tagged study to more realistic experimental conditions. Specifically, the matrix elements in Eqs.~(22) and (31) of Ref.~\cite{Ren:2024uui} are modified to incorporate a symmetric $Q_2^2$-dependent term. For the dominant LT contribution, the corresponding TFFs $F_{LT}(Q_1^2, Q_2^2)$ amounts to the dipole parameterization
\begin{equation}
    F_{LT}(Q_1^2,Q_2^2) = \frac{F_{LT}(0,0)}{\left(1+(Q_1^2+Q_2^2)/\Lambda_{f_1}^2\right)^2}\,,
\end{equation}
where $F_{LT}(0,0)$ and the scale parameter $\Lambda_{f_1}$ are defined as in Ref.~\cite{Ren:2024uui}. 
For the subdominant TT contribution, the TFF $F_{TT}(Q_1^2, Q_2^2)$ is estimated via the quark model relation:
\begin{equation}
F_{TT}(Q_1^2, Q_2^2) = - \frac{X}{\nu (\nu + Q_2^2)} F_{LT}(Q_1^2, Q_2^2)\,,
\end{equation}
with $\nu \equiv q_1 \cdot q_2 = (W^2 + Q_1^2 + Q_2^2)/2$, and $X = \nu^2 - Q_1^2 Q_2^2$.
Finally, the TL contribution is obtained using the symmetry relation
\begin{equation}
\frac{d\sigma_{TL}}{d|\cos\theta^*|}(Q_2^2,Q_1^2) = \frac{d\sigma_{LT}}{d|\cos\theta^*|}(Q_1^2,Q_2^2)\,,
\end{equation}
while the LL contribution is not included. The model is optimized for use in single-tagged studies. The Omn\`es function, which is used to describe the $f_0(500)$ amplitude in the matrix element of the $f_0(500)\eta$ intermediate state, is linearly interpolated and set to zero outside of its defined range.

\subsection{Production through a Single Resonance}
\label{sec::custom_mode}

So far, only  the simulation of specific final states has been discussed. \HTP also offers the possibility to simulate the production of arbitrary hadronic final states through a single resonance using a custom final state mode. This feature enables experimental collaborations to estimate reconstruction efficiencies for arbitrary final states and to use the generated events in standard PWA tools. Since no universal theoretical model exists for the production and decay of generic final states in two-photon collisions, simplified phenomenological models are implemented for such processes.

{In the custom single-resonance mode,} the two-photon cross sections are {modeled by the phenomenological} factorized {form, which includes} a mass-dependent part, $f(W)$, and a TFF, $|\mathcal{F}_{AB}|^2$,
\begin{equation}
\sigma_{AB}(W, Q_1^2, Q_2^2) \approx f(W) \, \big| \mathcal{F}_{AB}(Q_1^2, Q_2^2) \big|^2\quad A, B \in \{T, L\}\,.
\end{equation}  
The default functional forms for $f(W)$, $h_T$ and $h_L$ are given below. When both are set to unity, the generated {events} can be directly used in PWA frameworks such as AmpTools \cite{Shepherd2023-pz} or Pawian \cite{Fritsch2022-ii}, without explicitly incorporating the two-photon luminosity functions into the amplitudes.

For the TFF, a factorized dependence on the photon virtualities is assumed, with correct real-photon limits
\begin{equation}
|\mathcal{F}_{AB}(Q_1^2,Q_2^2)|^2 \approx h_A(Q_1^2)\, h_B(Q_2^2)\quad h_T(0)=1\quad h_L(0)=0 \,.
\end{equation}
All other cross sections and response functions are implemented analogously to $\sigma_\text{TT}$. 
Since experimental studies of these cross sections and responses involve studies analyzing the azimuthal dependence, normalizing them to unity allows Monte Carlo samples to include the full lepton-side kinematics. This, in turn, enables the extraction of hadronic structure information from fits of the Monte Carlo samples to experimental data.

For the mass-dependent part, three options are available. The first one excludes any explicit mass dependence and simulates the mass distribution solely from the kinematic factors in Eqs.~(\ref{eq:gagacross}) and (\ref{eq:gagapipiunpolcross}), i.e. $f(W)=1\,$GeV$^{-2}$. A simulation at a fixed mass, $f(W)=\delta(W^2-W^2_\text{user})$, is also possible. Alternatively, a Breit-Wigner mass distribution can be used,
\begin{equation}
    f(W)=\frac{1}{\left(W^2 - M^2\right)^2 + \Gamma^2 M^2}\,\frac{M\,\Gamma}{\pi}\,,
\end{equation}
where the resonance mass $M$ and width $\Gamma$ are specified by the user.

Similarly to the mass dependency, a constant TFF model has also been implemented, $|\mathcal{F}_{AB}|=1$. For a more realistic description of the TFF, a vector-pole model following the approach of the \textsc{GALUGA2.0} code \cite{SCHULER1998279} is implemented
\begin{equation}
    h_T(Q^2) = \left(\frac{1}{1+Q^2/M_V^2}\right)^2\,, \quad  h_L(Q^2)= \xi \frac{Q^2}{M_V^2}\left(\frac{1}{1+Q^2/M_V^2}\right)^2\,,
\end{equation}
where the pole mass $M_V$ and the scaling parameter $\xi$ can be freely chosen. Alternatively, a simplified vector meson dominance (VMD) model with multiple poles is available \cite{Sakurai:1972wk}. In this case, the functions are
\begin{equation}
    h_\text{T}(Q^2) = \sum_{i=0}^3 r_i\left( \frac{1}{1 + Q^2/M_{V_i}^2} \right)^2\,,\quad 
    h_\text{L}(Q^2) = \sum_{i=1}^3 \xi \frac{Q^2}{M_{V_i}^2} r_i\left( \frac{1}{1 + Q^2/M_{V_i}^2} \right)^2\,,
\end{equation}
where the masses $M_{V_0}\dots M_{V_3}$ and the parameters $\xi$, $r_1\dots r_3$ are free inputs, constrained by $r_0=1-r_1-r_2-r_3$ to ensure normalization. The resonance with index 0 normalizes the transverse term and does not contribute to the longitudinal one. A generalized VMD model \cite{Bezurkov:1980wu} is also included:
\begin{align}
    h_\text{T}&= r\, \left( \frac{1}{1+Q^2/M_1^2} \right)^2 + (1-r)\,\frac{1}{1+Q^2/M_2^2}\,, \nonumber \\
    h_\text{L}&= \xi r\, \left( \frac{1}{1 + Q^2/M_1^2} \right)^2 + \xi (1-r)\, \left[ \frac{M_2^2}{Q^2}\text{log}\left( 1+\frac{Q^2}{M_2^2} \right) - \frac{1}{1 + Q^2/M_2^2} \right]\,.
\end{align}
This implementation reproduces the results of \textsc{GALUGA2.0} without relying on high-energy approximations. All parameters are configurable by the user.

Helicity-angle dependencies of the decay products in the final state have not been implemented. Instead, all final states decay according to a completely flat phase-space distribution. Since no reliable theoretical information exists on the relative normalization of the different cross sections for arbitrary hadronic states, a common absolute scale is used for all. Therefore, simulating combinations of cross sections simultaneously is not meaningful. Ratios of generated Monte Carlo samples can instead be used to study effects arising from lepton-side kinematics.

The previously introduced TFF models can be employed to extrapolate the numerical input for the $\pi\pi$, $\pi\eta$, $K\bar{K}$, and $\eta\eta$ channels beyond their defined virtuality range. The two-photon cross sections and response functions are assumed to continue proportional to $h_T(Q_1^2)\,h_T(Q_2^2)$ outside of their defined range. The mass dependence is modeled as $\sim 1/W^2$. This extrapolation should, however, be regarded only as a rough estimate rather than a reliable prediction of the cross sections outside the defined input region. In particular, for the $K\bar{K}$ and $\eta\eta$ channels, only the $\sigma_{TT}$ cross section is available, meaning that essential components of the required information are missing for simulations at finite virtuality.

\subsection{Luminosity Function Mode}
Besides generating $e^+e^-\to e^+e^-X$ events, the code can also be used to calculate the luminosity functions introduced in Sec.~\ref{sec:lumi_function}. For this purpose, every two-photon cross section or response is set to unity thereby removing hadronic dynamics, and only the $e^+e^-\to e^+e^-X$ phase-space element is evaluated. In other words, the $d\text{Lips}_{eeX}$ is used and the hadronic decay factor is set to unity to exclude any dynamics from the decay of $X$. The same type of calculation can be performed for the corresponding functions associated with the two-photon cross sections in Eq.~(\ref{eq:gagapipiunpolcross}).


\section{Program Structure}
\HTP is distributed as source code organized into a specific directory structure. The core of the generator is implemented as a class, enabling straightforward integration into detector software frameworks. The main header file for \HTP, \texttt{HadroTOPS.hh}, is located in the \texttt{include/} folder. Alongside it, the \texttt{common.hh} header file includes essential quadruple precision functions, while the header files for the Lorentz and three-vector objects (\texttt{LorentzVector} and \texttt{Vector3}) are also located in this directory. The source code for the vector objects resides in the \texttt{src/vector} directory, while the \HTP generator's source code is located in \texttt{src/hadrotops}. The main function, which reads the job options and instantiates the \HTP object accordingly, is implemented in \texttt{src/main.cc}.

The \HTP source code is split into several files:
\begin{itemize} 
    \item \texttt{HadroTOPS.cc}: Contains the constructor of the \HTP object and handles the initialization and finalization of the event generation procedure. All constants are defined here. 
    
    \item \texttt{HadroTOPS.aux.cc}: Contains auxiliary functions such as consistency checks of user input, a convenience function for the random number generator, and the display of status bars. 
    
    \item \texttt{HadroTOPS.crosssection.cc}: Implements all functions that calculate the event weight, which is the full cross section element, for the current kinematic setting. 
    
    \item \texttt{HadroTOPS.generate.cc}: Includes all necessary functions for the generation of the final-state Lorentz vectors and the calculation of the associated phase space elements and other kinematic factors. 
    
    \item \texttt{HadroTOPS.getters.cc}: Contains all getters. 
    
    \item \texttt{HadroTOPS.mc.cc}: Contains all functions necessary to handle the Monte Carlo procedure. The weighted and unweighted event generation loops, as well as the Monte Carlo integration procedure, are implemented here. 
    
    \item \texttt{HadroTOPS.out.cc}: Includes all functions necessary for the output of the program. 
    
    \item \texttt{HadroTOPS.setters.cc}: Includes all setters. 
    
    \item \texttt{HadroTOPS.theory.cc}: Contains all functions necessary to read and interpolate the theory input for the two-photon cross sections. 
\end{itemize}

\subsection{Double and Quadruple Precision Floating Point Numbers}
The code can be compiled to be executed either in double or quadruple precision. The latter offers a slight improvement in stability, but results in significantly longer execution times. Nevertheless, using quadruple precision is generally recommended.

For quadruple precision, the \texttt{libquadmath} library and its quadruple precision floating-point type \texttt{\_\_float128} are used. For convenience, this is set as the custom data type \texttt{quad}. When only double precision is required, the precompiler flag \texttt{-DDOUBLE\_PREC} can be set during compilation, making the \texttt{quad} data type equivalent to \texttt{long double} and replacing the quadruple precision functions with their standard, double-precision counterparts.

\subsection{Input Scheme}
The executable \texttt{HadroTOPS.exe} is produced in the \texttt{bin/} directory. When executed, it reads the configuration settings from the file \texttt{jobOptions.txt}, unless a different file is specified by the user. The input file defines all necessary settings for input, output, and the generation process itself. The available settings are discussed in the \HTP manual.

It should be noted that the generator can also operate without a dedicated input file. If integrated into a larger framework, all settings can be adjusted using the appropriate setter functions. The numerical theory input is placed in the folder \texttt{data/}.

\subsection{Output Scheme}
The program is capable of providing event-by-event output as well as histograms filled with the (differential) cross section output. The histogram functionality requires a \texttt{ROOT} installation~\cite{ROOT} with properly set \texttt{\$ROOTSYS} variables.

Weighted and unweighted events are stored in two different ASCII files {using the HepMC3 format \cite{Buckley:2019xhk}}. 

Histograms are stored in a \texttt{ROOT} file, the name and path of which is set in the \texttt{jobOptions.txt} file. 

In addition to the output files, information about the current run is displayed in the console, which can also serve for logging purposes.

\subsection{Implementation of Vector3 and LorentzVector classes}
Since the code may use quadruple precision floating-point calculations, the standard double precision classes implementing three- and four-vectors are not suitable. Therefore, a simplified and very incomplete re-implementation of the most important functions of the \texttt{TVector3} and \texttt{TLorentzVector} classes of the \texttt{ROOT} framework~\cite{ROOT}, using quadruple precision floats and functions, is utilized. The code for these re-implementations can be found in the respective header files and \texttt{src/vector/Vector3.cc} and \texttt{src/vector/LorentzVector.cc}, respectively.

\subsection{Important Variables}
The most important kinematic variables required for by many functions are implemented as member variables. All of them are stored in units of GeV for energies, masses, widths, and momenta, and in radians for angles. These variables include the scalar (type \texttt{quad}) variables for the center-of-mass energy (\texttt{Ecm} and its square \texttt{s}), the two-photon center-of-mass energy (\texttt{W}), the lepton momentum transfers (\texttt{t1}, \texttt{t2}, \texttt{Q1s}, and \texttt{Q2s}), the subsystem center-of-mass energies (\texttt{s1} and \texttt{s2}), the modulation angles $\tilde\phi$, $\tilde\phi_1$, and $\tilde\phi_2$ (\texttt{phiTilde}, \texttt{phiTilde1}, \texttt{phiTilde2}), and the polar and azimuthal angles of the final state hadrons in the two-photon center-of-mass system (\texttt{cosThetaStar} and \texttt{phiStar}).

The four-vectors of the initial and final state positron ($p_1$ denoted in the code as \texttt{k1} and $p_1^\prime$ denoted as \texttt{p1}) and electron ($p_2$ denoted as \texttt{k2} and $p_2^\prime$ denoted as \texttt{p2}), the hadronic system four-vector $p_X$ (\texttt{px}), and the final state hadrons (\texttt{pdec}) are stored as \texttt{LorentzVector} (or vector thereof) objects. 

Objects, such as the current event weight (\texttt{WEIGHT}), the maximum event weight observed during the weighted event loop (\texttt{maxWEIGHT}), and the phase space elements $\mathrm{dLips}{eeX}$, $\frac{\mathrm{dLips}{eeX}}{(t_1 \cdot t_2)}$, and $\mathrm{dLips}_N$ (\texttt{dLips}, \texttt{dLipsDiv}, \texttt{dLipsDec}) are also member variables of type \texttt{quad}.

\subsection{Important Functions}
This subsection provides an overview of a selected set of key functions implemented in the \HTP source code.

\begin{itemize} 

    \item Functions for the generation of final state four-vectors 
    
    \begin{itemize} 
    
        \item \texttt{void HadroTOPS::generateFixW()}\\ Generates the $e^+e^-X$ final state four-vectors and calculates $\mathrm{dLips}_{eeX}$, along with several kinematic variables, at given (globally defined) values of $s$ and $W$. The function takes into account user-defined and kinematic limits of the final-state vectors. 
    
        \item \texttt{void HadroTOPS::decayFinalState()}\\ Handles the decay of the hadronic system $X$ into its final-state particles and calculates the corresponding phase-space element. 
    
        \item \texttt{bool HadroTOPS::generateEvent()}\\ Generates a numerically stable weighted event. If necessary, $W$ is generated and decay products of the hadronic final state $X$ are generated. The function determines final state Lorentz vectors in the two-photon center-of-mass system and calculates remaining necessary kinematic variables. It returns a Boolean value indicating whether the generated event satisfies the user-defined cuts and calls \texttt{calcXS()} to determine event weight. 
    
        \item \texttt{bool HadroTOPS::acceptEvent()}\\ Checks user-defined constraints and returns \texttt{false} if the event shall be rejected. Additional user-specific constraints can be implemented within this function.
    
        \item \texttt{void HadroTOPS::generateWeightedEvent()}\\ Generates a weighted event under the specified user conditions. This function handles Monte Carlo integration and event output, and internally calls \texttt{generateEvent()} and \texttt{acceptEvent()}. 
    
        \item \texttt{void HadroTOPS::runWeighted()}\\ Implements the weighted events loop. It generates as many weighted events as requested by the user within user constraints and determines the maximum event weight and stores it in \texttt{maxWEIGHT}. 
    
        \item \texttt{void HadroTOPS::generateUnweightedEvent()}\\ Generates an unweighted event within user-specific conditions using the rejection sampling method \cite{vonNeumann}. This function performs Monte Carlo integration and produces unweighted event output. 
        The maximum event weight (\texttt{maxWEIGHT}) must be predefined by running \texttt{runWeighted()}. It internally calls \texttt{generateWeightedEvent()}.
    
        \item \texttt{void HadroTOPS::runUnweighted()}\\ Generates as many unweighted events within user-specific constraints as requested. 
    
    \end{itemize}
    
    \item Functions for the calculation of the cross section
    
    \begin{itemize}
        \item \texttt{void HadroTOPS::interpolateCrossSectionInput()}\\Wrapper function for the interpolation of the cross section input. It calls the necessary functions to interpolate the theory input to the current kinematic variables. The interpolated two-photon cross sections are stored in the member variable \texttt{quad itXS[18]}.
        
        \item \texttt{quad HadroTOPS::getFFTT()}\\ Calculates the TFF for two transversely polarized virtual photons, according to the user-selected model. Corresponding functions for transverse–longitudinal, longitudinal–transverse, and longitudinal–longitudinal photon polarizations are provided as \texttt{getFFTL()}, \texttt{getFFLT()}, and \texttt{getFFLL()}, respectively.
        
        \item \texttt{void HadroTOPS::calcXS()}\\ Calculates the cross section using the current kinematics and the user-defined theoretical model.
    \end{itemize}

\end{itemize}

\section{Running the Simulation}
\HTP operates in a four-step procedure: 
\begin{enumerate} 
    \item Initialization 
    \item Weighted-event generation and upper bound estimation loop
    \item Unweighting loop (optional) 
    \item Finalization \quad . 
\end{enumerate} 

Detailed instructions for running the generator both as a standalone application and as an embedded component in a C++ program are provided in the \HTP manual. During the initialization process, a new \HTP object is created, and the simulation options are set using, the setter functions, which are summarized in detail in the manual. Afterwards, the function \texttt{void HadroTOPS::init()} must be called to validate the settings and initialize the output functionalities, the random-number generator, and the Monte Carlo integration variables.

Next, the function \texttt{void HadroTOPS::runWeighted()} is called to generate a predefined number of weighted events within the user-defined cuts. In this way, the maximum weight within user cuts is determined, which is necessary for the unweighting loop. Alternatively, the user may call the \texttt{void HadroTOPS::generateWeightedEvent()} function to generate a single weighted event that satisfies the user constraints. The function \texttt{double HadroTOPS::getWeight()} returns the current event weight in units of nanobarns (or dimensionless in luminosity function mode, with the correct dimension of the luminosity function of GeV$^{-5}$ arising from bin normalization). Note that the single-event function cannot be used to determine the maximum event weight. For all functions, the events are filled into histograms and/or written to the output files, as specified by the user.

The weighted-event loop may optionally be followed by the unweighting loop. Similar to the weighted-events procedure, the function \texttt{void HadroTOPS::runUnweighted()} generates a predefined number of unweighted events, while \texttt{void HadroTOPS::generateUnweightedEvent()} produces a single unweighted event within the user-defined cuts. The results are also reported to the histogram and output files as specified by the user. All events must be scaled to the total cross section using the same event weight. {The upper bound used in the unweighing procedure is constrained by generating a user defined number of weighted events and checking for the largest event weight. The maximum weight can be increased by an enlargement factor to use a more conservative upper bound estimate.}

The last step is the program finalization, in which the total cross section of the simulated physics process is calculated and reported. The cross section and its corresponding uncertainty for the weighted events (which also takes into account all trials needed in the unweighting procedure) are given by
\begin{align}
    \sigma &= \frac{1}{N_w}\sum_{i=1}^{N_w} \diff \sigma_i  \nonumber  \\
    (\Delta\sigma)^2 &= \frac{1}{N_w-1} \left( \frac{1}{N_w}\sum_{i=1}^{N_w}(\diff \sigma_i)^2 - \sigma^2 \right)\,,
\end{align}
where $\diff \sigma_i$ is the weight (cross section) of event $i$, and $N_w$ is the number of generated weighted event within the user cuts. Alternatively, the cross section can be calculated from the unweighted events using
\begin{align}
    \sigma &= \diff \sigma_\text{max}\cdot\frac{N_{uw}}{N_t}\,,\nonumber\\
    (\Delta\sigma)^2 &= \frac{\sigma ( \diff \sigma_\text{max} - \sigma )}{N_{uw}}\,,
\end{align}
where $\diff \sigma_\text{max}$ is the {upper bound used in the unweighting procedure}, $N_{uw}$ is the number of unweighted events. $N_t$ represents the number of trials, which is the number of generated weighted events {generated} during the unweighting procedure.

\section{Results}
\subsection{Luminosity Functions}
\begin{figure}[t!b]
    \centering
    \includegraphics[width=\linewidth]{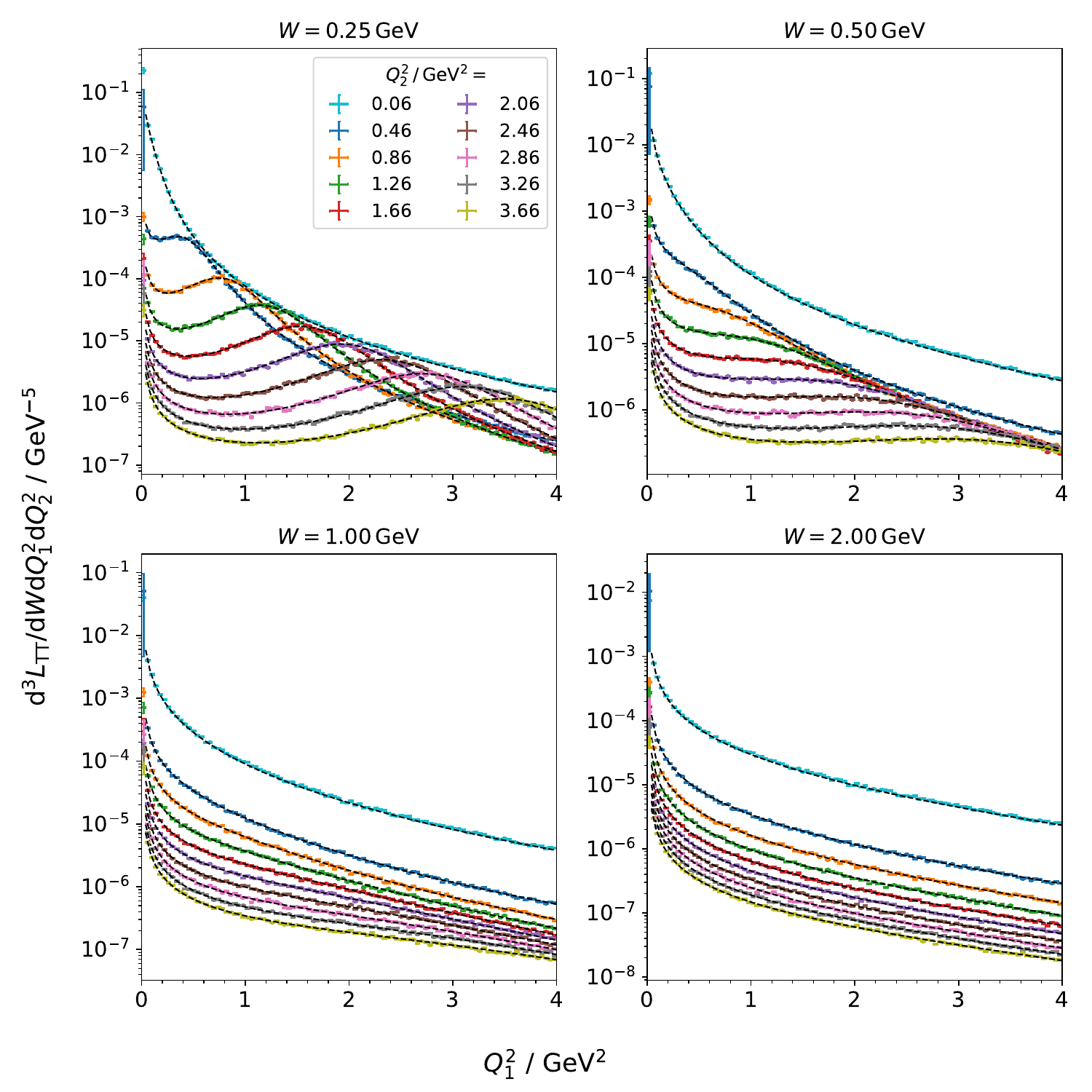}
    \caption{Comparison of the luminosity function of two transversely polarized photons calculated by \HTP (data points) and the analytical result (black dashed lines) from Ref.~\cite{Bonneau:1973kg}. The panels correspond to different two-photon center-of-mass energies $W$, while the colors represent different values of $Q_2^2$. All functions are displayed in dependence on $Q_1^2$ and are evaluated at $\sqrt{s}=4\,$GeV.}
    \label{fig:lumi_func_comparison_TT}
\end{figure}

For validation purposes, the luminosity functions calculated using \HTP can be compared with their analytically calculated counterparts. The exact analytical expressions are given in Ref.\cite{Bonneau:1973kg}. Such a comparison is possible only for the luminosity functions associated with the inclusive cross section in Eq.~(\ref{eq:gagacross}), as no analytical calculation exists for the exclusive formula in Eq.~(\ref{eq:gagapipiunpolcross}).

A systematic comparison between the \HTP output and the analytical curves was performed over a wide range of center-of-mass energies, two-photon invariant masses, and momentum transfers, covering all possible combinations of photon polarizations. The agreement was found to be within the statistical fluctuations of the generated samples. Figure~\ref{fig:lumi_func_comparison_TT} shows the comparison of the double-transverse luminosity function as a function of both momentum transfers and the two-photon center-of-mass energy $(W)$, evaluated at $\sqrt{s}=4\,$GeV. The chosen momentum transfers and masses lie within a typical range accessible at this energy. The comparison demonstrates excellent agreement, even at very high statistics. The same level of agreement is observed for the remaining five polarization combinations. All luminosity functions have been tested at a five per-mille level.

\subsection{Fixed Two-Photon Cross Sections and Luminosity Functions}
\begin{figure}[t!b]
    \centering
    \includegraphics[width=0.49\linewidth]{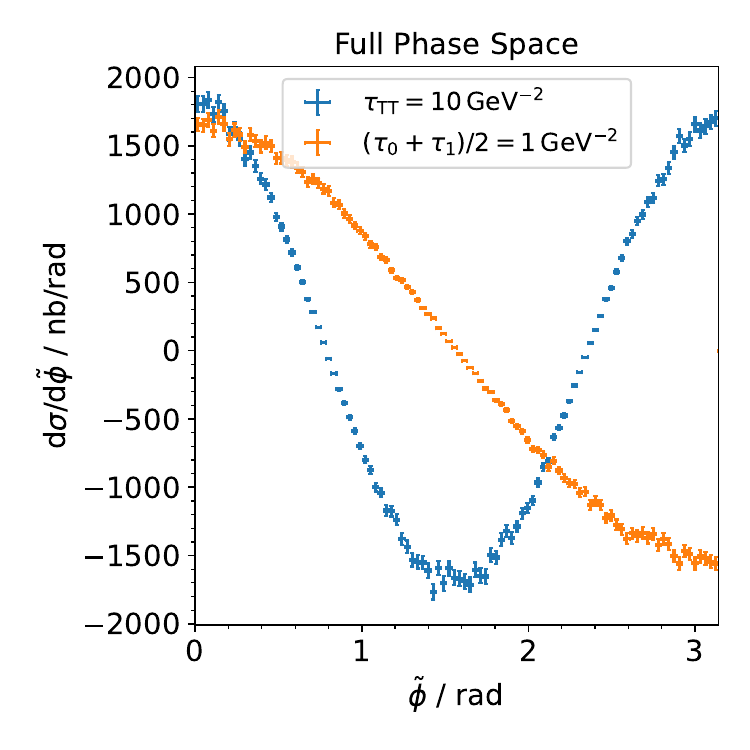}
    \includegraphics[width=0.49\linewidth]{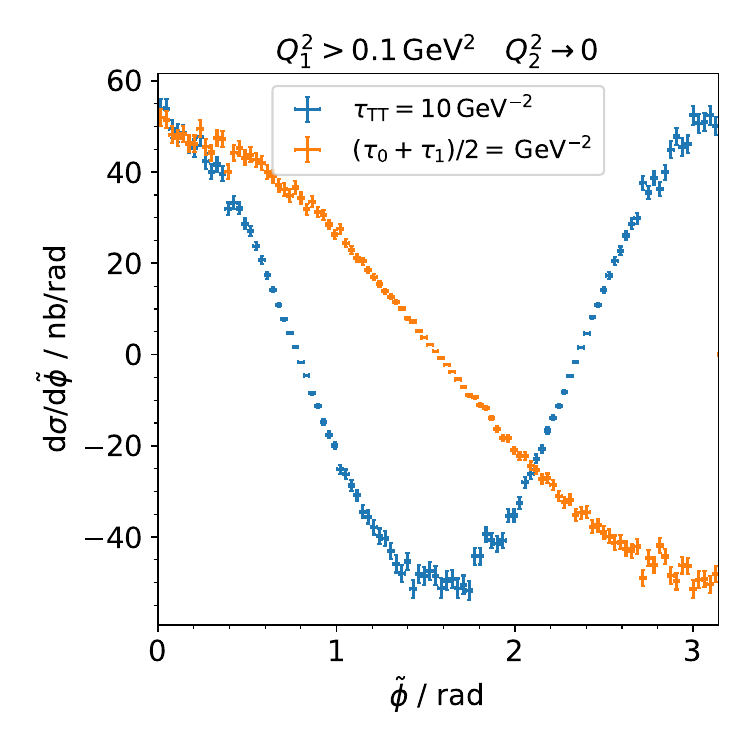}
    \includegraphics[width=0.49\linewidth]{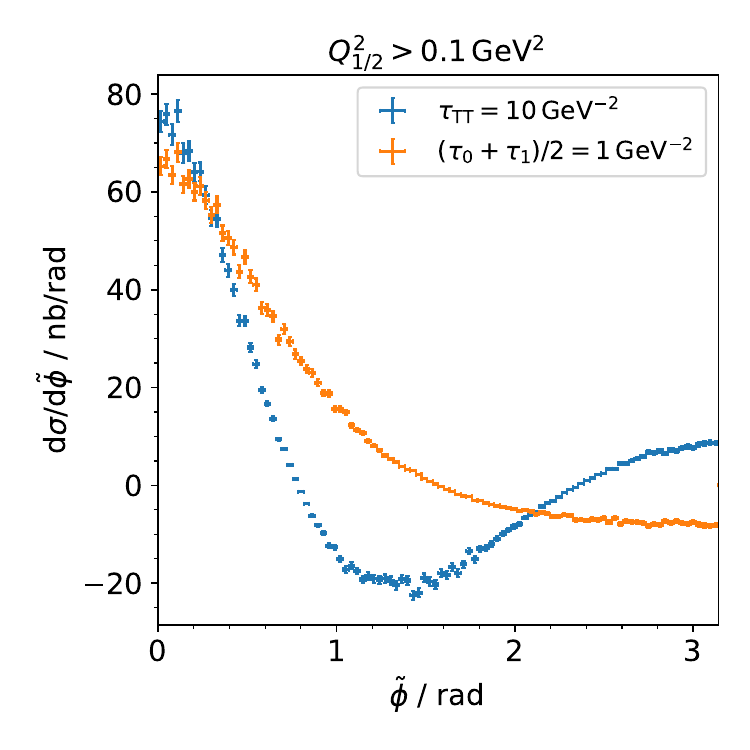}
    \caption{ Differential cross section as a function of the modulation angle $\tilde\phi$. The top-left panel shows the cross section when only $\tau_\text{TT}$ and $(\tau_0 + \tau_1)/2$ are set to non-zero values. The top-right panel shows the same in the single-tag configuration, and the bottom panel in the double-tag setup. All plots are produced at $\sqrt{s}=4\,$GeV for the production of a stable particle with $W=1\,$GeV, with $(\tau_0 + \tau_1)/2=1\,$GeV$^{-2}$ and (for better visibility) $\tau_\text{TT}=10\,$GeV$^{-2}$. }
    \label{fig:fix_cross_section}
\end{figure}

Calculating the $e^+e^-\to e^+e^-X$ cross section with constant two-photon cross sections not only enables a direct determination of the luminosity functions, but also cleanly isolates the lepton-side kinematics that govern the modulation angles $\tilde\phi$, $\tilde{\phi}_1$, and $\tilde{\phi}_2$. 
 As argued at the end of Section 1.1.2, by analyzing the angular-harmonic content of the cross section in $\tilde\phi$ (the $1$, $\cos\tilde\phi$ and $\cos2\tilde\phi$ components) one can project out individual response components. In analyses integrating over the modulation angles, the cosine terms are  removed by construction, leaving only the angle-averaged contribution. Practically, accessing $\tilde\phi$ requires double-tagging (both outgoing leptons reconstructed), which yields lower event rates. 

An additional complication arises, as shown in Fig.~\ref{fig:fix_cross_section}, where the $\tilde\phi$-dependent parts of the cross section are plotted as a function of $\tilde\phi$. In the full phase space and a single-tag environment, the distributions exhibit the expected $\cos\tilde\phi$ and $\cos2\tilde\phi$ behavior. In the double-tag case, however, the shapes are significantly distorted. This distortion stems from the non-trivial dependence of $\tilde\phi$ on the full set of Lorentz invariants, as given in Ref.~\cite{SCHULER1998279}. Therefore, to exploit the modulation-angle dependence, it is not sufficient to simply fit the functional dependencies on these modulation angles, as given in Eqs.~(\ref{eq:gagacross}) and (\ref{eq:gagapipiunpolcross}), to the experimentally extracted differential cross section. Instead, one must fit the full kinematic dependence, as calculated by \HTP, to the experimental data.

\subsection{Cross Sections of Two-Hadron Final States}

\begin{figure}[t!hb]
    \centering
    \includegraphics[width=0.49\linewidth]{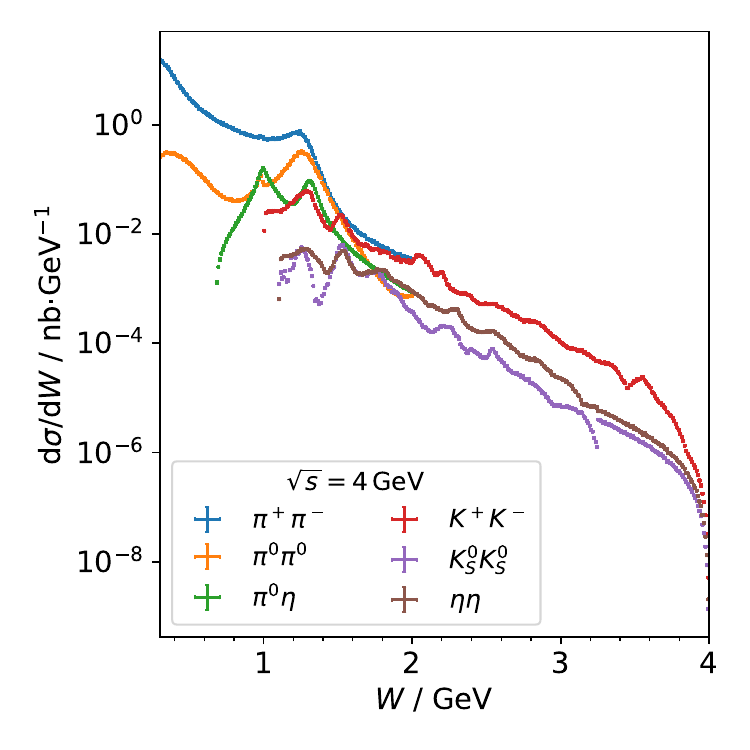}
    \includegraphics[width=0.49\linewidth]{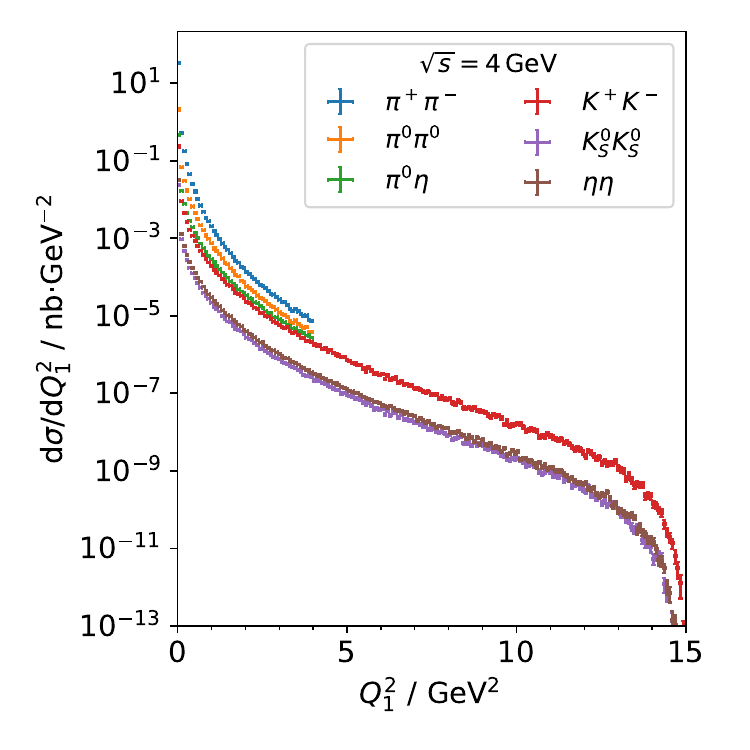}
    \caption{Two-Photon invariant mass distribution (left) and momentum transfer $Q_1^2$ distribution (right) for simulated two-hadron final states generated with \HTP. All events are produced at $\sqrt{s}=4\,$GeV and cover the $W$ and $Q^2$ ranges described in Secs.~\ref{sec:pipi_mode}, \ref{sec:pieta_mode} and \ref{sec::kk_etaeta_mode}. For the $K^+K^-$,$K_sK_s$, and $\eta\eta$ channels, the $Q^2$ dependence of the two-photon cross section is modeled using a vector-pole parametrization with $M_V$=775$\,$MeV.}
    \label{fig:two_body_W}
\end{figure}

Figure~\ref{fig:two_body_W} presents the two-photon invariant mass distributions and the positron momentum transfer distributions generated by \HTP, when simulating the fully available phase space at $\sqrt{s}=4\,$GeV. A strong decrease of the cross section with increasing masses is clearly visible, as is the rapid drop with increasing $Q^2$. The limited coverage in $Q^2$ for the $\pi\pi$ and $\pi\eta$ final states, which originates from the theoretical input, is also apparent. In contrast, the modeling of the $Q^2$ dependence for other final states extends the coverage to larger momentum transfers, but relies on the simplified models described in Sec.~\ref{sec::custom_mode}.

The fluctuations observed in the $K^+K^-$, $K_S^0K_S^0$, and $\eta\eta$ final states can only partially be attributed to the intermediate resonances. They are primarily caused by the limited statistics of the experimental input, which is particularly evident for the $K^+K^-$ channel. Compared to the neutral-pion case, the charged-pion cross section is roughly twice as large in the $f_2(1270)$ mass region, as expected from isospin symmetry. At lower invariant masses, however, the charged-pion contribution exceeds the neutral-pion one by more than an order of magnitude. This difference arises from the presence of the Born contribution, which is absent for neutral pions.
A similar effect is expected for the charged and neutral kaon states, but the available experimental input does not cover the threshold region, and this feature is therefore not produced in \HTP.
\begin{figure}[t!b]
    \centering
    \includegraphics[width=0.5\linewidth]{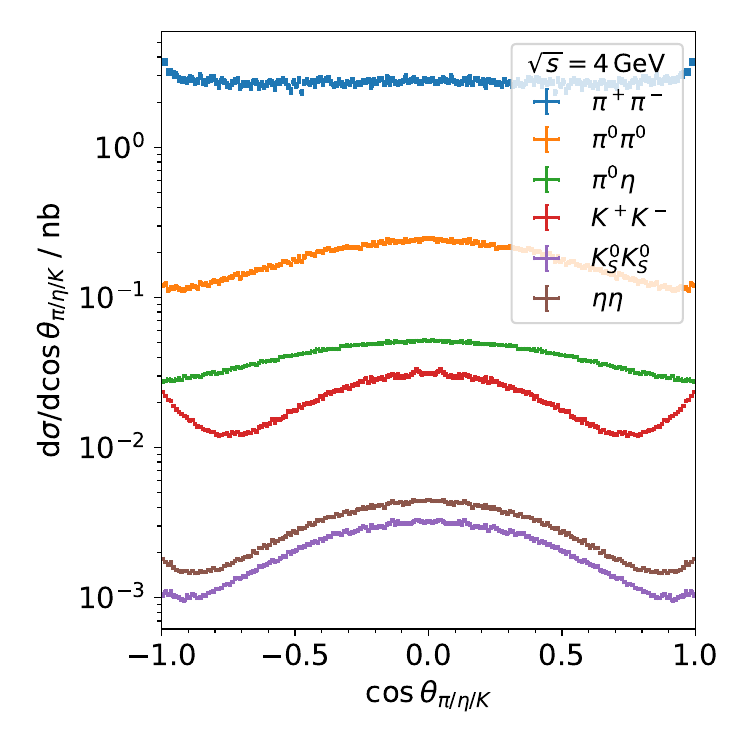}
    \caption{Helicity angle $\cos\theta_{\pi/\eta/K}$ distributions for simulated two-hadron final states generated with \HTP. All events are produced at $\sqrt{s}=4\,$GeV and cover the $W$ and $Q^2$ ranges described in Secs.~\ref{sec:pipi_mode}, \ref{sec:pieta_mode} and \ref{sec::kk_etaeta_mode}, and correspond to those shown in Fig.~\ref{fig:two_body_W}.}
    \label{fig:two_body_z}
\end{figure}
The helicity-angle distributions shown in Fig.~\ref{fig:two_body_z} illustrate the characteristic angular patterns of the final states, which depend on the spin and helicity configuration of the intermediate resonance. For the charged channels, the pronounced forward and backward peaking originates mainly from the Born contribution.

\subsubsection{Cross Sections of the $e^+e^-\to e^+e^-\pi\pi$ Process}
The authors of the forthcoming \textsc{Ekhara3.2} generator have kindly provided early access to the code, allowing a comparison between the two generators for the process $e^+e^-\to e^+e^-\pi^0\pi^0$. A corresponding comparison for the charged-pion channel is not feasible, since \textsc{Ekhara3.2} also includes pion-pair production from a virtual photon radiated in Bhabha scattering.

\begin{figure}[t!bh]
    \centering
    \includegraphics[width=0.49\linewidth]{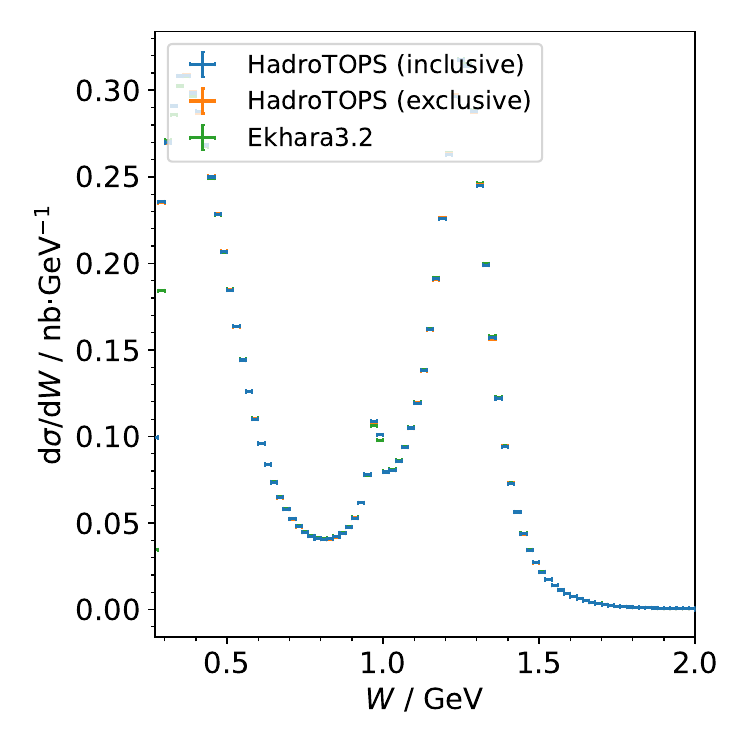}
    \includegraphics[width=0.49\linewidth]{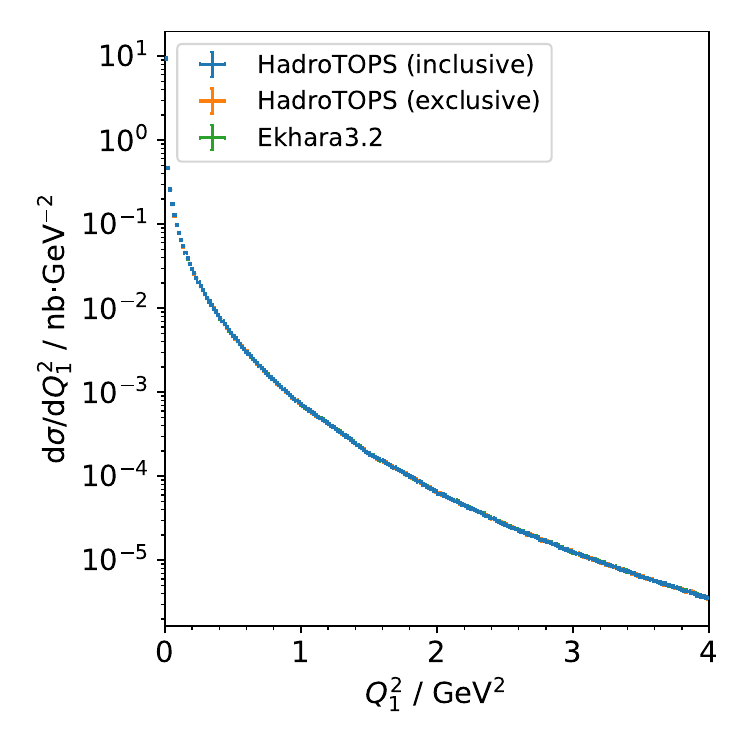}
    \includegraphics[width=0.49\linewidth]{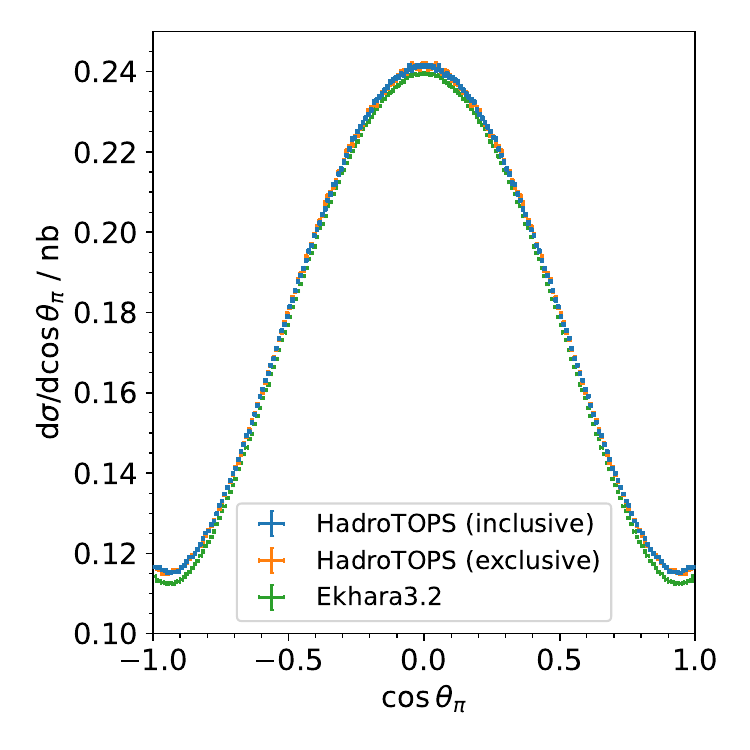}
    \includegraphics[width=0.49\linewidth]{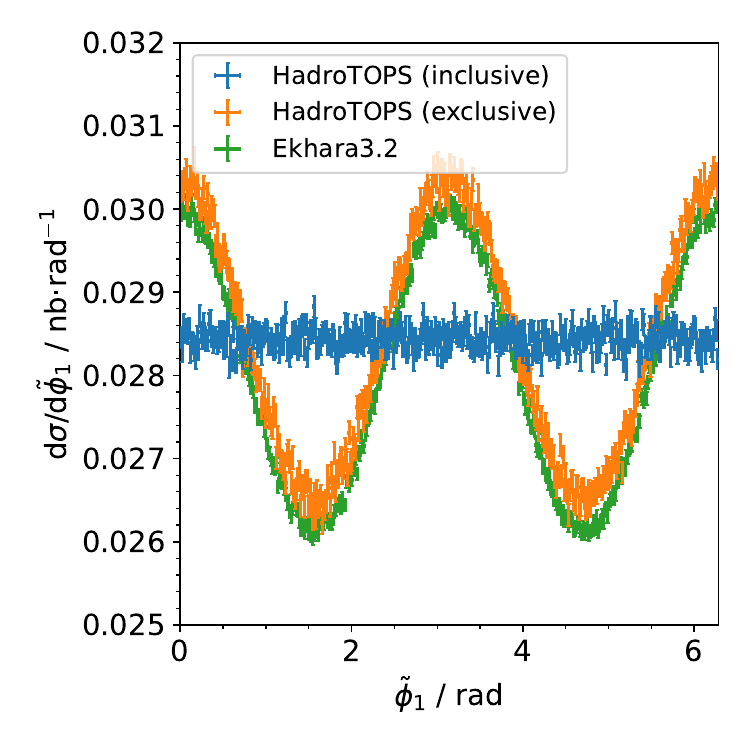}
    \caption{Comparison of the predicted cross sections for the process $e^+e^-\to e^+e^-\pi^0\pi^0$ calculated with \HTP, using the inclusive (blue) and exclusive (orange) cross-section formulas, and with \textsc{Ekhara3.2} (green). The distributions are shown as functions of the two-pion invariant mass (upper left), positron momentum transfer (upper right), pion polar angle $\cos\theta_\pi$ in the two-photon c.m. frame (lower left), and the modulation angle $\tilde\phi_1$ (lower right). The center-of-mass energy is fixed at $\sqrt{s} = 4$\,GeV, with $W$ limited to $W = 0-2$\,GeV, and the momentum transfers to $Q^2_{1,2} = 0-4$\,GeV$^2$.}
    \label{fig:pi0pi0_comparisons_ekhara}
\end{figure}

A comparison between \HTP and \textsc{Ekhara3.2} at $\sqrt{s}=4\,$GeV is shown in Fig.~\ref{fig:pi0pi0_comparisons_ekhara}, as a function of the kinematic variables $W$, $Q_1^2$, $\cos\theta_\pi$, and $\tilde\phi_1$. For \HTP, two separate samples are generated using the cross-section formulas for the inclusive process Eq.~(\ref{eq:gagacross}) and the exclusive process Eq.~(\ref{eq:gagapipiunpolcross}). The two resulting \HTP distributions agree perfectly across all variables except for the modulation angle $\tilde\phi_1$ between one of the lepton scattering planes and the hadron production plane. This is expected since this dependence only enters through the exclusive process formula. 

Overall, the results from \HTP are in good agreement with those from \textsc{Ekhara3.2} apart from some minor differences. Both generators rely on the same theoretical input, but these discrepancies originate from differences in how the differential cross section is evaluated. Unlike \textsc{HadroTOPS}, where the cross section is analytically computed from Eqs.~(\ref{eq:gagacross}) and (\ref{eq:gagapipiunpolcross}) using the two-photon cross sections and response functions, \textsc{Ekhara3.2} evaluates the full $e^+e^-\to e^+e^-\pi\pi$ cross section directly from the matrix element in Eq.~(\ref{eq:pipi_matrix_element}).
Consequently, \textsc{Ekhara3.2} uses the helicity amplitudes as direct input, whereas \HTP interpolates the combination of helicity amplitudes and the kinematic factors arising from the two-photon dynamics. In contrast, \textsc{Ekhara3.2} interpolates only the helicity amplitudes and evaluates all remaining kinematic factors at the exact point in phase space where the event is generated. With a very fine (or infinitely dense) grid for the numerical theory input, both generators are expected to yield identical results.

\begin{figure}[tb]
    \centering
    \includegraphics[width=0.5\linewidth]{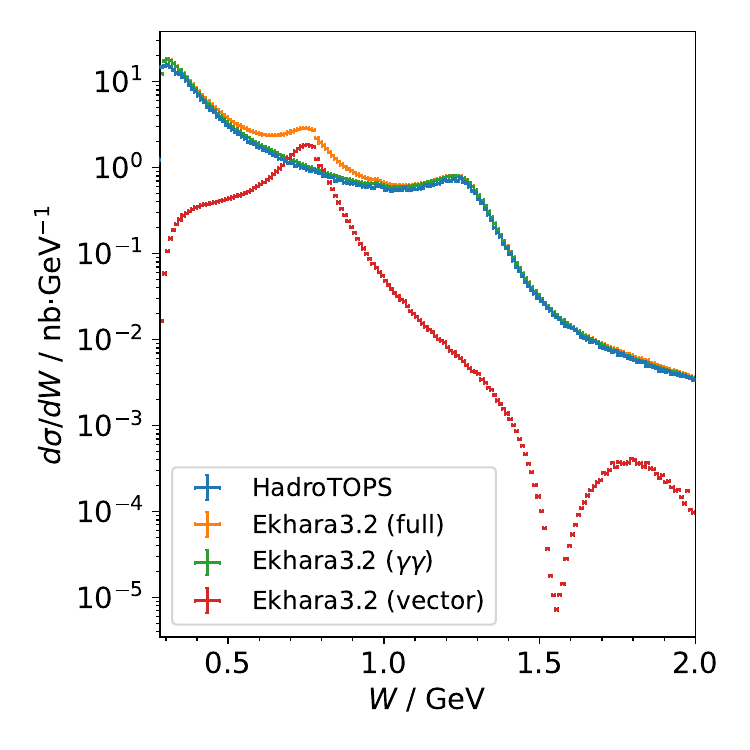}
    \caption{Comparison of the \HTP (blue) and Ekhara3.2 (orange, green, and red) cross-section predictions for the $e^+e^-\to e^+e^-\pi^+\pi^-$ process as a function of the two-pion invariant mass $W$ at a center-of-mass energy of $\sqrt{s}=4$\,GeV. The full Ekhara3.2 prediction is shown in orange, the production from two-photon collisions in green, and the production from a hard photon radiated off a Bhabha-like event in red.  }
    \label{fig:pippim_comparison}
\end{figure}

Both generators can estimate the cross section of the $e^+e^-\to e^+e^-\pi^+\pi^-$ process. While \HTP incorporates pion production solely from the two-photon process, Ekhara3.2 provides an additional estimate for pion production originating from hard-photon radiation in a Bhabha-like process, along with its interference effects with the two-photon process. The cross-section estimates from both generators are presented in Fig.~\ref{fig:pippim_comparison}. The Ekhara3.2 estimate shows a significant deviation from HadroTOPS due to this background process; as a result, the mass distribution shows a clear indication of the $\rho-\omega$ interference pattern entering through the pion vector form factor. At masses larger than $\approx1$\,GeV the two generators are in agreement because of the rapid decrease of the form factor. Consequently, \HTP provides a sensible description from the $f_0(980)$ mass range and above. Future upgrades to \HTP are planned to include these background contributions for the $e^+e^-\pi^+\pi^-$ and $e^+e^-K^+K^-$ final states. 

\subsection{Cross Sections of the $e^+e^-\to e^+e^-f_1(1285)\to e^+e^- \eta\pi^+\pi^-$ Process}

\begin{figure}[!tbh]
    \centering
    \includegraphics[width=\linewidth]{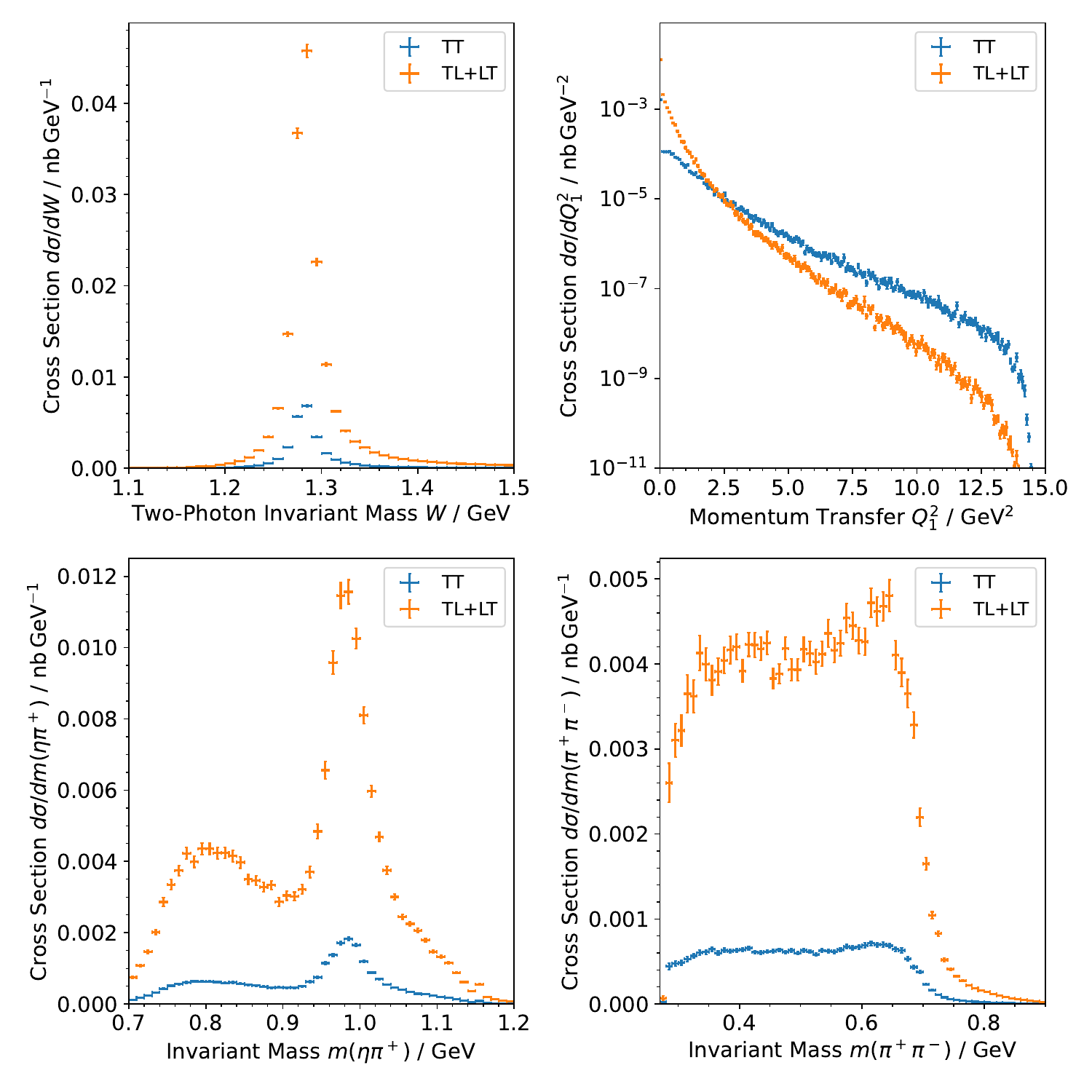}
    \caption{Differential cross sections of the $e^+e^-\to e^+e^- f_1(1285)\to e^+e^- \eta \pi^+\pi^-$ in dependence on two-photon invariant mass $W$ (top left), positron momentum transfer $Q_1^2$ (top right), $\eta\pi^+$ invariant mass (bottom left) and $\pi^+\pi^-$ invariant mass (bottom right) for the case of two transversely polarized photons (blue) and one transversely and one longitudinally polarized photon (orange). All parameters are set as in Ref.~\cite{Ren:2024uui} for a value $\sqrt{s}=4\,$GeV. }
    \label{fig:f1_xs_plots}
\end{figure}

Figure~\ref{fig:f1_xs_plots} shows example distributions of the \HTP prediction for the two-photon production cross section of the $f_1(1285)$ and its decay into $\eta\pi^+\pi^-$. The $W$ dependent cross section is found to be around an order of magnitude smaller than the two-pion modes and comparable to the $K\Bar{K}$ and $\eta\eta$ channels, as seen from Fig.~\ref{fig:two_body_W}. This behavior is expected, since the production of a spin-1 state by two real photons is forbidden by the Landau–Yang theorem \cite{Landau:1948kw,PhysRev.77.242}, and a nonzero cross section arises only for photons with finite virtualities.

The doubly-transverse polarized cross section (TT) decreases more slowly with $Q^2$ than the transverse–longitudinal (TL/LT) one. At small momentum transfers, typical of BESIII at $\sqrt{s}=4\,$GeV, the TL+LT contribution is numerically larger and dominates the cross section. At higher $\sqrt{s}$, as at BaBar or Belle II, where the average $Q^2$ is larger, the TT component becomes relatively more important.

In the $\eta\pi^+$ mass spectrum, the $a_0(980)$ resonance from the $f_1(1285)\to a_0(980)^\pm\pi^\mp$ decay and its kinematic reflection are clearly visible. There is no distinct mass separation between the $a_0(980)^\pm\pi^\mp$ and $f_0(500)\eta$ final states. 
Consequently, a full PWA will be needed in future measurements.
In the $\pi^+\pi^-$ mass spectrum, a clear destructive interference pattern is observed. The interference can be controlled by the relative phase angle, which in this case is set to $\phi = 180^\circ$. The destructive interference disappears for $\phi = 0^\circ$. Experimental input is necessary to determine the correct interference between the two intermediate states.

{
\subsection{Event Generation Time and Unweighting Efficiencies}
For benchmarking purposes, the event generation time and unweighting efficiencies are evaluated for several final-state modes. The benchmarking utilizes a system with AMD Ryzen~7~5700G processor with 16\,GB of memory running Fedora Linux~42. The code is compiled using \texttt{GCC~15.2.1} (Red Hat 15.2.1-1) with histogram routines enabled and quadruple-precision floating-point calculations.

All samples are generated at a center-of-mass energy of $\sqrt{s} = 4\,$GeV without additional phase space restrictions. The two-body processes $e^+e^- \to e^+e^-\pi\pi$ and $e^+e^- \to e^+e^-\pi^0\eta$ are simulated according to Eq.~(\ref{eq:gagapipiunpolcross}), while all other modes use Eq.~\eqref{eq:gagacross}. Numerical inputs are processed using linear spline interpolation. The logarithmic mapping for the momentum transfer generation is used, while a flat mapping is utilized for the mass generation. The upper bound enlargements factor is set to 1.2. The specific job options used for each sample are included in the distributed source files. The average generation times for both weighted and unweighted events, as well as the corresponding unweighting efficiencies, are summarized in Table~\ref{tab:HTP_performance}. The performance checks for simulating the $\tau_{TT}$ and $\tfrac{1}{2}(\tau_0 + \tau_1)$ contributions in two-photon production of a single resonance do not include unweighting, as this is not possible due to the presence of negative weights.

\begin{table}[tbh]
    \caption{Average event generation time for weighted ($\bar{t}_\text{weighted}$) and unweighted events ($\bar{t}_\text{unweighted}$) and the unweighting efficiency, representing the number of generated weighted events per generated unweight event for different final states.}
    \begin{tabular}{lrrr}
    \hline
    Process & $\bar{t}_\text{weighted}$ / ms & $\bar{t}_\text{unweighted}$ / ms & Unweighting Eff. \\
    \hline
    $e^+e^-\pi^+\pi^-$& 0.076 & 7.537 & 99.46:1 \\
    $e^+e^-\pi^0\pi^0$& 0.076 & 2.506 & 33.04:1 \\
    $e^+e^-\pi^0\eta$& 0.076 & 2.640 & 34.62:1 \\
    $e^+e^-K^+K^-$& 0.065 & 4.201 & 64.05:1 \\
    $e^+e^-K_SK_S$& 0.066 & 3.767 & 57.26:1 \\
    $e^+e^-\eta\eta$& 0.067 & 1.973 & 29.19:1 \\
    $e^+e^-f_1(1285) (\to \eta\pi^+\pi^-)$ (TT) & 0.077 & 206.37 & 2675.66:1 \\
    $e^+e^-f_1(1285) (\to \eta\pi^+\pi^-)$ (LT+TL) & 0.078 & 97.33 & 1261.43:1 \\
    $e^+e^-X$ ($\tau_{TT}$, $W=1\,$GeV)& 0.06 & --- & --- \\
    $e^+e^-X$ ($\frac12(\tau_0+\tau_1)$, $W=1\,$GeV)& 0.08 & --- & --- \\
    \hline
    \end{tabular}
    \label{tab:HTP_performance}
\end{table}

It can be concluded that the generation time for weighted events remains consistently below 0.1\,ms, enabling the production of large datasets even on desktop-grade computers. However, the unweighting efficiency depends significantly on the specific final state. For relatively narrow resonances such as the $f_1(1285)$, the generator's efficiency decreases  notably, making the production of large unweighted samples computationally demanding on standard hardware. It should be noted that both the unweighting efficiency and the unweighted event generation time are sensitive to the chosen kinematic cuts and the upper-bound enlargement factor; consequently, the values reported here should be regarded as approximate estimates.

}

\section{Summary}
We present a Monte Carlo event generator capable of computing two-photon luminosity functions and simulating samples of the inclusive process $e^+e^-\to e^+e^- X$ {at leading order QED,} assuming a flat phase-space distribution. The framework is suitable for supporting PWA in two-photon reactions and for estimating reconstruction efficiencies. Furthermore, the code is also extended to simulate the exclusive processes $e^+e^-\to e^+e^- \pi^+\pi^-$, $e^+e^-\to e^+e^- \pi^0\pi^0$, $e^+e^-\to e^+e^- \pi^0\eta$, $e^+e^-\to e^+e^- K^+K^-$, $e^+e^-\to e^+e^- K^0_SK^0_S$, $e^+e^-\to e^+e^- \eta\eta$, and $e^+e^-\to e^+e^- f_1(1285)\to e^+e^-\eta\pi^+\pi^-$ via the two-photon production mechanism. For the exclusive two-meson production processes $e^+e^-\to e^+e^- M_1 M_2$ the full cross section formula is implemented in terms of $\gamma^* \gamma^* \to M_1 M_2$ response functions. The exclusive formulation retains the full azimuthal-angle modulations, enabling studies of interference response functions that are usually washed out by integration over modulation angles. For the $\pi \pi$ and $\pi \eta$ channels, the estimate of the $\gamma^* \gamma^* \to M_1 M_2$ response functions relies on dispersion theory input to describe the dependence on energy and photon virtualities. For $K^+K^-$, $K^0_SK^0_S$, and $\eta\eta$ the input is data-driven from untagged measurements (TT), with a user-configurable form-factor model to describe the $Q^2$ fall-off when needed. The $f_1(1285)$ mode, accessed through the $e^+e^-\to e^+e^- f_1(1285)\to e^+e^-\eta\pi^+\pi^-$ process, is implemented based on an effective-Lagrangian description and includes LT and TT contributions (and TL by symmetry) with a small but nonzero second virtuality. 

The present implementation should be viewed as a baseline based on the best currently available theoretical and experimental input. As improved amplitudes, form factors, and new measurements become available, the generator can be refined iteratively and extended to additional hadronic final states accordingly. Future updates will incorporate the irreducible background contributions from hard-photon radiation in Bhabha-like events for the $e^+e^-\pi^+\pi^-$ and $e^+e^-K^+K^-$ channels to provide a complete simulation framework for these processes.

\section*{Acknowledgements}
We thank Meike K{\"u}\ss{}ner for the discussion of the BESIII results and for providing of the experimental two-photon cross sections and Henryk Czy\.z for discussing results and providing the updated \textsc{Ekhara} Monte Carlo Event Generator. 

This work was supported by the Deutsche Forschungsgemeinschaft (DFG, German Research Foundation) through the Research
Unit FOR 5327 (Photon-photon interactions in the Standard Model and beyond, Projektnummer 458854507).

\bibliographystyle{elsarticle-num}
\bibliography{bibl.bib}

\appendix

\section{Polarized cross section for the exclusive $e^+ e^- \to e^+ e^- \pi_1 \pi_2$ process}
The polarized differential cross section for the exclusive process $e^+ e^- \to e^+ e^- \pi_1 \pi_2$ is expressed as:
\begin{equation}
    d \sigma_{h_1,h_2}= d \sigma^{(0)}+h_1\, d \sigma^{(1)}+h_2\, d \sigma^{(2)}+h_1\,h_2\, d \sigma^{(12)}
\end{equation}
where $h_{1,2}=\pm 1$ denote the helicities (in units $\hbar/2$) of the incoming leptons, and $d\sigma^{(0)}$ represents the unpolarized cross section, which is provided in the main text. The terms $d\sigma^{(1)}$, $d\sigma^{(2)}$, and $d\sigma^{(12)}$ correspond to the polarization-dependent contributions and are given by:\small
\begin{eqnarray}
d \sigma^{(12)}
&=& \frac{\alpha^2}{8 \pi^4 \, Q_1^2 \, Q_2^2} \, \frac{\sqrt{X}}{s (1 - 4 m^2 / s)^{1/2}}  
\cdot \frac{d^3 \vec p_1^{\, \prime}}{E_1^{\prime}} 
\cdot \frac{d^3 \vec p_2^{\, \prime}}{E_2^\prime} d\cos\theta_\pi
\frac{4}{(1 - \varepsilon_1)(1 - \varepsilon_2)}\nonumber \\
&\times& \left\{ 
 \left[ 1 - \varepsilon^2_1 + \frac{4 m^2}{Q_1^2} (1 - \varepsilon_1 )^2 \right]^{1/2} 
\left[ 1 - \varepsilon^2_2 + \frac{4 m^2}{Q_2^2} (1 - \varepsilon_2 )^2 \right]^{1/2} 
\frac{1}{2} \left( \frac{d \sigma_{0}}{d \cos \theta_\pi} - \frac{d \sigma_2}{d \cos \theta_\pi} \right) 
 \right.  \nonumber \\
 &+& \left[ \varepsilon_1 (1 - \varepsilon_1) \right]^{1/2} 
\left[ \varepsilon_2 (1 - \varepsilon_2) \right]^{1/2}
\left[ \cos  (\tilde \phi_2 - \tilde \phi_1) 
\left( \frac{d \tau_{0}}{d \cos \theta_\pi} - \frac{d \tau_1}{d \cos \theta_\pi} \right)  \right. \nonumber \\
&& \hspace{3.5cm}\left. + \cos  (\tilde \phi_1 + \tilde \phi_2) 
\left( \frac{d \tau_{1}}{d \cos \theta_\pi} + \frac{d \tau_{L2}}{d \cos \theta_\pi} \right) 
\right] \nonumber \\
&-&  \left[ \varepsilon_1 (1 - \varepsilon_1) \right]^{1/2} 
\left[ 1 - \varepsilon^2_2 + \frac{4 m^2}{Q_2^2} (1 - \varepsilon_2 )^2 \right]^{1/2} 
\cos  \tilde \phi_1 
\left( \frac{d \tau_{-12}}{d \cos \theta_\pi} + \frac{d \tau_{-1T}}{d \cos \theta_\pi} \right) 
\nonumber \\
&-&\left.   
\left[ 1 - \varepsilon^2_1 + \frac{4 m^2}{Q_1^2} (1 - \varepsilon_1 )^2 \right]^{1/2}
 \left[ \varepsilon_2 (1 - \varepsilon_2) \right]^{1/2} 
\cos  \tilde \phi_2 
\left( \frac{d \tau_{12}}{d \cos \theta_\pi} + \frac{d \tau_{1T}}{d \cos \theta_\pi} \right) 
\right\}, \quad \quad \quad
\end{eqnarray}
and
\begin{eqnarray}
d \sigma^{(1)}
&=& \frac{\alpha^2}{8 \pi^4 \, Q_1^2 \, Q_2^2} \, \frac{\sqrt{X}}{s (1 - 4 m^2 / s)^{1/2}}  
\cdot \frac{d^3 \vec p_1^{\, \prime}}{E_1^{\prime}} 
\cdot \frac{d^3 \vec p_2^{\, \prime}}{E_2^\prime} d\cos\theta_\pi
\frac{4}{(1 - \varepsilon_1)(1 - \varepsilon_2)}\nonumber \\
&\times&\left\{ \left[ \varepsilon_1 (1 - \varepsilon_1) \right]^{1/2}   
\left[ \varepsilon_2 (1 + \varepsilon_2) + \frac{4 m^2}{Q_2^2} \varepsilon_2 (1 - \varepsilon_2 ) \right]^{1/2} 
\right. \nonumber \\
&&\hspace{.75cm} \times \left[ \sin  (\tilde \phi_2 - \tilde \phi_1) 
\left( \frac{d \bar \tau_{0}}{d \cos \theta_\pi} - \frac{d \bar \tau_1}{d \cos \theta_\pi} \right) 
+ \sin  (\tilde \phi_1 + \tilde \phi_2) 
\left( \frac{d \bar \tau_{1}}{d \cos \theta_\pi} - \frac{d \bar \tau_{L2}}{d \cos \theta_\pi} \right) 
\right] 
 \nonumber \\
&-&  
\left[ \varepsilon_1 (1 + \varepsilon_1) + \frac{4 m^2}{Q_1^2} \varepsilon_1 (1 - \varepsilon_1 ) \right]^{1/2} 
\varepsilon_2 \sin (2 \tilde \phi_2) \frac{d \bar \tau_{T2}}{d \cos \theta_\pi}
\nonumber \\
&+& \left[ \varepsilon_1 (1 - \varepsilon_1) \right]^{1/2}   
\left[ \sin  \tilde \phi_1 
\left( \frac{d \bar \tau_{-12}}{d \cos \theta_\pi} - \frac{d \bar \tau_{-1T}}{d \cos \theta_\pi} \right) 
+ 2 \left[ \varepsilon_2 + \frac{2 m^2}{Q_2^2} (1 - \varepsilon_2 ) \right]^{1/2} 
\sin  \tilde \phi_1 \frac{d \bar \tau_{1L}}{d \cos \theta_\pi}
\right. \nonumber \\
&&\hspace{2cm} +\left. 
 \varepsilon_2  \sin  (\tilde \phi_1 + 2 \tilde \phi_2) \frac{d \bar \tau_{-12}}{d \cos \theta_\pi}
+ \varepsilon_2  \sin  (2 \tilde \phi_2 - \tilde \phi_1) \frac{d \bar \tau_{-1T}}{d \cos \theta_\pi}
\right] \nonumber \\ 
&+&\left.   
\left[ 1 - \varepsilon^2_1 + \frac{4 m^2}{Q_1^2} (1 - \varepsilon_1 )^2 \right]^{1/2}  
\left[ \varepsilon_2 (1 + \varepsilon_2) + \frac{4 m^2}{Q_2^2} \varepsilon_2 (1 - \varepsilon_2 ) \right]^{1/2} 
\sin  \tilde \phi_2 \left( \frac{d \bar \tau_{12}}{d \cos \theta_\pi} + \frac{d \bar \tau_{1T}}{d \cos \theta_\pi} \right) 
\right\} , \quad \quad \quad 
\label{eq:gagapipipolcross1} 
\end{eqnarray}

\begin{eqnarray}
d \sigma^{(2)}
&=& \frac{\alpha^2}{8 \pi^4 \, Q_1^2 \, Q_2^2} \, \frac{\sqrt{X}}{s (1 - 4 m^2 / s)^{1/2}}  
\cdot \frac{d^3 \vec p_1^{\, \prime}}{E_1^{\prime}} 
\cdot \frac{d^3 \vec p_2^{\, \prime}}{E_2^\prime}d\cos\theta_\pi
\frac{4}{(1 - \varepsilon_1)(1 - \varepsilon_2)}\nonumber \\
&\times& 
\left\{ 
\left[ \varepsilon_1 (1 + \varepsilon_1) + \frac{4 m^2}{Q_1^2} \varepsilon_1 (1 - \varepsilon_1 ) \right]^{1/2} 
\left[ \varepsilon_2 (1 - \varepsilon_2) \right]^{1/2}   \right. \nonumber \\
&&\hspace{.75cm} \times \left[ \sin  (\tilde \phi_2 - \tilde \phi_1) 
\left( \frac{d \bar \tau_{0}}{d \cos \theta_\pi} + \frac{d \bar \tau_1}{d \cos \theta_\pi} \right) 
+ \sin  (\tilde \phi_1 + \tilde \phi_2) 
\left( \frac{d \bar \tau_{1}}{d \cos \theta_\pi} + \frac{d \bar \tau_{L2}}{d \cos \theta_\pi} \right) 
\right] 
 \nonumber \\
&+&
\varepsilon_1
\left[ \varepsilon_2 (1 + \varepsilon_2) + \frac{4 m^2}{Q_2^2} \varepsilon_2 (1 - \varepsilon_2 ) \right]^{1/2} 
 \sin (2 \tilde \phi_1) \frac{d \bar \tau_{T2}}{d \cos \theta_\pi}
\nonumber \\
&-& \left[ \varepsilon_2 (1 - \varepsilon_2) \right]^{1/2}   
\left[ \sin  \tilde \phi_2 
\left( \frac{d \bar \tau_{12}}{d \cos \theta_\pi} - \frac{d \bar \tau_{1T}}{d \cos \theta_\pi} \right) 
+ 2 \left[ \varepsilon_1 + \frac{2 m^2}{Q_1^2} (1 - \varepsilon_1 ) \right]^{1/2} 
\sin  \tilde \phi_2 \frac{d \bar \tau_{-1L}}{d \cos \theta_\pi}
\right. \nonumber \\
&&\hspace{2cm} \left. 
+ \varepsilon_1  \sin  (2 \tilde \phi_1 + \tilde \phi_2) \frac{d \bar \tau_{12}}{d \cos \theta_\pi}
+ \varepsilon_1  \sin  (2 \tilde \phi_1 - \tilde \phi_2) \frac{d \bar \tau_{1T}}{d \cos \theta_\pi}
\right]  \nonumber \\ 
&-& \left.   
\left[ \varepsilon_1 (1 + \varepsilon_1) + \frac{4 m^2}{Q_1^2} \varepsilon_1 (1 - \varepsilon_1 ) \right]^{1/2} \left[ 1 - \varepsilon^2_2 + \frac{4 m^2}{Q_2^2} (1 - \varepsilon_2 )^2 \right]^{1/2}   
\sin  \tilde \phi_1 \left( \frac{d \bar \tau_{-12}}{d \cos \theta_\pi} + \frac{d \bar \tau_{-1T}}{d \cos \theta_\pi} \right) 
\right\}. \quad \quad \quad 
\label{eq:gagapipipolcross2} 
\end{eqnarray}

\end{document}